\documentclass[a4paper,11pt]{article}

\usepackage{jcappub}
\allowdisplaybreaks
\usepackage{graphicx}
\usepackage{epstopdf}

\newcommand{\be}{\begin{equation}}
\newcommand{\ee}{\end{equation}}
\newcommand{\bea}{\begin{eqnarray}}
\newcommand{\eea}{\end{eqnarray}}
\newcommand{\nn}{\nonumber}
\def\nl{\nonumber \\ &}

\title{Next-to-next-to-leading order gravitational spin-orbit coupling via 
the effective field theory for spinning objects in the post-Newtonian scheme}

\author[a,b]{Mich\`ele Levi}
\author[c,d]{and Jan Steinhoff}

\affiliation[a]{Universit\'e Pierre et Marie Curie, CNRS-UMR 7095, 
Institut d'Astrophysique de Paris,\\ 98 bis Boulevard Arago, 75014 Paris, France} 
\affiliation[b]{Sorbonne Universit\'es, Institut Lagrange de Paris,\\ 
98 bis Boulevard Arago, 75014 Paris, France} 
\affiliation[c]{Max-Planck-Institute for Gravitational Physics 
(Albert-Einstein-Institute),\\ Am M{\"u}hlenberg 1, 14476 Potsdam-Golm, Germany}
\affiliation[d]{Centro Multidisciplinar de Astrofisica, 
Instituto Superior Tecnico, Universidade de Lisboa,\\ 
Avenida Rovisco Pais 1, 1049-001 Lisboa, Portugal}

\emailAdd{michele.levi@upmc.fr}
\emailAdd{jan.steinhoff@aei.mpg.de}

\abstract{We implement the effective field theory for gravitating spinning objects 
in the post-Newtonian scheme at the next-to-next-to-leading order level to derive 
the gravitational spin-orbit interaction potential at the third and a half post-Newtonian order 
for rapidly rotating compact objects. From the next-to-next-to-leading order interaction 
potential, which we obtain here in a Lagrangian form for the first time, we derive 
straightforwardly the corresponding Hamiltonian. The spin-orbit sector constitutes the most 
elaborate spin dependent sector at each order, and accordingly we encounter a 
proliferation of the relevant Feynman diagrams, and a significant increase of the 
computational complexity. We present in detail the evaluation of the interaction 
potential, going over all contributing Feynman diagrams. The computation is carried out 
in terms of the ``nonrelativistic gravitational'' fields, which are advantageous also 
in spin dependent sectors, together with the various gauge choices included in the effective 
field theory for gravitating spinning objects, which also optimize the calculation. In addition, 
we automatize the effective field theory computations, and carry out the automated computations 
in parallel. Such automated effective field theory computations would be most useful to obtain 
higher order post-Newtonian corrections. We compare our Hamiltonian to the ADM Hamiltonian, and 
arrive at a complete agreement between the ADM and effective field theory results. Finally, we 
provide Hamiltonians in the center of mass frame, and complete gauge invariant relations among 
the binding energy, angular momentum, and orbital frequency of an inspiralling binary with 
generic compact spinning components to third and a half post-Newtonian order. The derivation 
presented here is essential to obtain further higher order post-Newtonian corrections, and 
to reach the accuracy level required for the successful detection of gravitational radiation.} 
 
\begin{document}

\maketitle

\flushbottom

\section{Introduction} 

In light of the upcoming operation of second-generation ground-based 
interferometers worldwide, such as Advanced LIGO in the US \cite{Ligo}, Advanced 
Virgo in Europe \cite{Virgo}, and KAGRA in Japan \cite{Kagra}, we may witness a 
direct detection of gravitational waves (GW) within the end of the decade, which 
will open a new era of observational gravitational wave astronomy. Inspiralling 
binaries of compact objects are of the most promising sources in the accessible 
frequency band of these experiments, where they can be treated analytically 
within the post-Newtonian (PN) approximation of General Relativity 
\cite{Blanchet:2013haa}. As the search for the GW signals employs the matched 
filtering technique, there is a pressing need to obtain accurate waveform 
templates, using the Effective-One-Body (EOB) formulation to model the continuous 
signal \cite{Buonanno:1998gg}. 

Recently the fourth PN (4PN) order correction has been completed for the non 
spinning case \cite{Damour:2014jta}, and it is necessary to reach a similar 
accuracy level for the spin dependent case, as such compact objects are expected 
to be rapidly rotating. Of the spin dependent sectors the spin-orbit sector contributes 
the leading order (LO) spin-dependent PN correction, and represents the most dominant 
spin effects. The LO spin-orbit correction at the 1.5PN order has been obtained 
by Tulczyjew already at 1959 \cite{Tulczyjew:1959}, see erratum therein,
and later in the form of a higher-order Lagrangian used here by Damour \cite{Damour:1982}.
Yet, the next-to-leading order 
(NLO) spin-orbit interaction has been approached much later, first at the level of 
the equations of motion (EOM) in \cite{Tagoshi:2000zg}, and \cite{Faye:2006gx}, 
and then within the ADM Hamiltonian formalism \cite{Damour:2007nc}. The 
next-to-next-to-leading order (NNLO) spin-orbit interaction was 
first derived in Hamiltonian form in \cite{Hartung:2011te,Hartung:2013dza}, 
building on \cite{Steinhoff:2008zr,Steinhoff:2009ei}, and then at the level of 
the EOM in \cite{Marsat:2012fn,Bohe:2012mr}.

In this work we apply the effective field theory (EFT) for gravitating spinning 
objects in the PN scheme \cite{Levi:2015msa} at the NNLO level, which was already 
tackled in the spin dependent sector using EFT techniques in \cite{Levi:2011eq}. 
We derive the NNLO gravitational spin-orbit interaction potential at the 3.5PN 
order for rapidly rotating compact objects. The EFT for gravitating spinning objects 
\cite{Levi:2015msa} builds on the novel, self-contained EFT approach for the binary 
inspiral, which was introduced in \cite{Goldberger:2004jt,Goldberger:2007hy}. The 
seminal works in \cite{Hanson:1974qy, Bailey:1975fe} already treated spin in an action 
approach in flat and curved spacetime, respectively. Then an extension to spinning 
objects of EFT techniques was approached in \cite{Porto:2005ac}, which later adopted 
a Routhian approach from \cite{Yee:1993ya}, and leaves the temporal components of the 
spin tensor in the final results. Yet, these components depend on field modes at the 
orbital scale, and they must be eliminated in order to obtain physical observables. 
The EFT for gravitating spinning objects \cite{Levi:2015msa} actually obtains an 
effective action at the orbital scale by integrating out \textit{all} the field modes 
below this scale. Moreover, it actually enables the relation to physical observables: 
the physical EOM are directly obtained via a proper variation of the effective action 
\cite{Levi:2014sba,Levi:2015msa}. Furthermore, it also enables to obtain the 
corresponding Hamiltonians in a straightforward manner from the potentials derived 
via this formulation \cite{Levi:2015msa}. Indeed, from the potential, which we obtain 
here in a Lagrangian form, we derive the corresponding NNLO spin-orbit Hamiltonian, 
and then compare our result to the ADM Hamiltonian in \cite{Hartung:2011te,Hartung:2013dza}. 
We arrive at a complete agreement between the ADM and EFT results. 

The spin-orbit sector constitutes the most elaborate spin dependent 
sector at each order, see \cite{Levi:2015msa} for the LO and NLO levels, and 
\cite{Levi:2011eq, Levi:2015ixa} for the other sectors at the NNLO level. 
Accordingly, we encounter here a proliferation of the
relevant Feynman diagrams, where there are 132 diagrams contributing to this
sector, and a significant increase of the computational complexity, e.g.~there 
are 32 two-loop diagrams here. We also recall that as the spin is derivative-coupled, 
higher-order tensor expressions are required for all integrals involved in the 
calculations, compared to the non spinning case. Yet, the computation is carried 
out in terms of the 
``nonrelativistic gravitational'' (NRG) fields \cite{Kol:2007bc,Kol:2010ze}, which 
are advantageous also in spin dependent sectors, as was first shown in 
\cite{Levi:2008nh}, and later also in 
\cite{Levi:2010zu,Levi:2011eq,Levi:2014gsa,Levi:2015msa}. We also apply the 
various gauge choices included in the EFT for gravitating spinning objects 
\cite{Levi:2015msa}, which also optimize the calculation. 
In addition, we automatize the EFT computations here, and carry out the
automated computations in parallel, where we have 
used the suite of free packages xAct 
with the Mathematica software \cite{xAct, Martin-Garcia:2008aa}. 
Such automated EFT computations would be most useful to obtain higher order 
PN corrections. It should be stressed that in order to obtain further higher 
order results, all lower order results are required, consistently within one 
formalism, and so also in that respect the derivation presented in this work 
is essential. Finally, we provide Hamiltonians in the center of mass frame, 
and complete gauge invariant relations among the binding energy, angular 
momentum, and orbital frequency of an inspiralling binary with generic compact 
spinning components to 3.5PN order. 

The outline of the paper is as follows. In section \ref{frules} we briefly 
review the EFT for gravitating spinning objects in the PN scheme, and present 
the Feynman rules required for the EFT computation. In section \ref{fdiagrams} 
we present the evaluation of the NNLO spin-orbit interaction potential, going 
over all contributing Feynman diagrams, and giving the value of each diagram. 
In section \ref{VtoH} we present the NNLO spin-orbit potential EFT result, 
and from it we obtain the corresponding EFT Hamiltonian. We then compare our 
result to the ADM Hamiltonian, where we resolve the difference between the 
Hamiltonians, using higher order PN canonical transformations, and arrive at 
a complete agreement between the ADM and EFT results. We also present all 
relevant Hamiltonians to 3.5PN order in the center of mass frame. In section 
\ref{GI} we provide the complete gauge invariant relations among the binding 
energy, angular momentum, and orbital frequency of an inspiralling binary with 
generic compact spinning components to 3.5PN order. In section \ref{theendmyfriend} 
we summarize our main conclusions. Finally, in appendix \ref{reduce} we provide 
the additional irreducible two-loop tensor integrals required for this work.

\section{The EFT for gravitating spinning objects in the PN scheme} \label{frules}

In this section we present the effective action, and the Feynman rules, which are 
derived from it, and are required for the EFT computation of the NNLO spin-orbit 
interaction. We employ here the ``NRG'' fields, as applied with spin in 
\cite{Levi:2008nh,Levi:2010zu,Levi:2011eq,Levi:2014gsa,Levi:2015msa}. 
Here, we briefly review and build on 
\cite{Levi:2008nh,Levi:2010zu,Levi:2011eq,Levi:2014sba,Levi:2015msa}, following 
similar notations and conventions as those that were used there. 
Hence we use $c\equiv1$, 
$\eta_{\mu\nu}\equiv \text{Diag}[1,-1,-1,-1]$,
and the convention for the Riemann tensor is 
$R^\mu{}_{\nu\alpha\beta}\equiv\partial_\alpha\Gamma^\mu_{\nu\beta}
-\partial_\beta\Gamma^\mu_{\nu\alpha}
+\Gamma^\mu_{\lambda\alpha}\Gamma^\lambda_{\nu\beta}
-\Gamma^\mu_{\lambda\beta}\Gamma^\lambda_{\nu\alpha}$. 
The scalar triple product appears here with no brackets, 
i.e.~$\vec{a}\times\vec{b}\cdot\vec{c}\equiv(\vec{a}\times\vec{b})\cdot\vec{c}$. 
The notation $\int_{\vec{k}} \equiv \int \frac{d^d\vec{k}}{(2\pi)^d}$ is used for abbreviation. In fact, the generic $d$-dimensional dependence can be and is suppressed in what follows, and $d$ can be set to 3, except for computations  which involve loops, where only the $d$ dependence from the generic $d$-dimensional Feynman integrals, see appendix~A in \cite{Levi:2011eq}, should be considered.

First, in terms of the ``NRG'' fields the metric reads
\begin{align} \label{eq:gkk}
g_{\mu\nu}&=
\left(\begin{array}{cc} 
e^{2\phi}      & \quad -e^{2\phi} A_j \\
-e^{2\phi} A_i & \quad -e^{-2\phi}\gamma_{ij}+e^{2\phi} A_i A_j
\end{array}\right)\nn\\
&\simeq
\left(\begin{array}{cc} 
1+2\phi+2\phi^2+\frac{4}{3}\phi^3 & \quad-A_j-2A_j\phi-2A_j\phi^2 \\
-A_i-2A_i\phi-2A_j\phi^2   & \quad-\delta_{ij}+2\phi\delta_{ij}
-\sigma_{ij}-2\phi^2\delta_{ij}
+2\phi\sigma_{ij}+A_iA_j+\frac{4}{3}\phi^3\delta_{ij}
\end{array}\right), 
\end{align}
where we have written the approximate metric in the weak-field limit 
up to the orders in the fields that are required for this sector. 

We recall that the effective action, describing the binary system, is given by
\be \label{totact}
S=S_{\text{g}}+\sum_{I=1}^{2}S_{\text{(I)pp}},
\ee
where $S_{\text{g}}$ is the pure gravitational action, and $S_{\text{(I)pp}}$ is 
the worldline point particle action for each of the two particles in the binary. 
The gravitational action is the usual Einstein-Hilbert action plus 
a gauge-fixing term, which we choose as the fully harmonic gauge, 
such that we have
\be
S_{\text{g}}=S_{\text{EH}} + S_{\text{GF}} 
= -\frac{1}{16\pi G} \int d^4x \sqrt{g} \,R + \frac{1}{32\pi G} \int d^4x\sqrt{g}
\,g_{\mu\nu}\Gamma^\mu\Gamma^\nu, 
\ee
where $\Gamma^\mu\equiv\Gamma^\mu_{\rho\sigma}g^{\rho\sigma}$.

From the gravitational action we derive the propagators, and the self-gravitational 
vertices. The ``NRG'' scalar, vector, and tensor field propagators in the harmonic 
gauge are then given by 
\begin{align}
\label{eq:prphi} \parbox{18mm}{\includegraphics[scale=0.6]{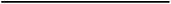}}
 =& \langle{~\phi(x_1)}~~{\phi(x_2)~}\rangle = ~~~~4\pi G~~~ \int_{\vec{k}} 
 \frac{e^{i \vec{k}\cdot\left(\vec{x}_1 - \vec{x}_2\right)}}
 {k^2}~\delta(t_1-t_2),\\ 
\label{eq:prA} \parbox{18mm}{\includegraphics[scale=0.6]{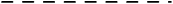}}
 =& \langle{A_i(x_1)}~{A_j(x_2)}\rangle = -16\pi G~\delta_{ij} \int_{\vec{k}} 
 \frac{e^{i\vec{k}\cdot\left(\vec{x}_1 - \vec{x}_2\right)}}
 {k^2}~\delta(t_1-t_2),\\ 
\label{eq:prsigma} \parbox{18mm}{\includegraphics[scale=0.6]{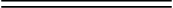}}
 =& \langle{\sigma_{ij}(x_1)}{\sigma_{kl}(x_2)}\rangle = ~~32\pi G~P_{ij;kl}
 \int_{\vec{k}}\frac{e^{i\vec{k}\cdot\left(\vec{x}_1 - \vec{x}_2\right)}}
 {k^2}~\delta(t_1-t_2),
\end{align}
where $P_{ij;kl}\equiv\frac{1}{2}\left(\delta_{ik}\delta_{jl}
+\delta_{il}\delta_{jk}-2\delta_{ij}\delta_{kl}\right)$.

The Feynman rules for the propagator correction vertices are given by
\begin{align}
\label{eq:prtphi}  \parbox{18mm}{\includegraphics[scale=0.6]{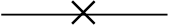}}
 =& ~~\frac{1}{8\pi G}~~\int d^4x~\left(\partial_t\phi\right)^2, \\ 
\label{eq:prtA}   \parbox{18mm}{\includegraphics[scale=0.6]{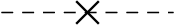}}
 =& -\frac{1}{32\pi G} \int d^4x~\left(\partial_tA_i\right)^2, \\ 
\label{eq:prtsigma} \parbox{18mm}{\includegraphics[scale=0.6]{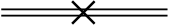}}
 =& \frac{1}{128\pi G} \int d^4x~\left[2(\partial_t\sigma_{ij})^2
 -(\partial_t\sigma_{ii})^2\right], 
\end{align}
where the crosses represent the self-gravitational quadratic vertices, which 
contain two time derivatives.

The Feynman rules for the three-graviton vertices required for the NNLO of the 
spin-orbit interaction are given by
\begin{align}
\label{eq:phiA2} \parbox{18mm}{\includegraphics[scale=0.6]{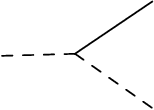}}
 =& \frac{1}{8\pi G}\int d^4x~\phi\left(\partial_iA_j\left(\partial_iA_j
 -\partial_jA_i\right)
 +\left(\partial_iA_i\right)^2\right), \\ 
\label{eq:A2sigma} \parbox{18mm}{\includegraphics[scale=0.6]{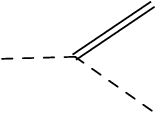}}
 =& -\frac{1}{64\pi G}\int d^4x\Big[2\sigma_{ij}
 \left(\partial_iA_k\partial_jA_k
 +\partial_kA_i\partial_kA_j-2\partial_kA_i\partial_jA_k+2\partial_iA_j
 \partial_kA_k\right)\nn \\ 
 &\left. \qquad\qquad\qquad\quad
 -\sigma_{kk}\left(\partial_iA_j\left(\partial_iA_j-\partial_jA_i\right)
 +\left(\partial_iA_i\right)^2\right)\right], \\ 
\label{eq:sigmaphi2} \parbox{18mm}{\includegraphics[scale=0.6]{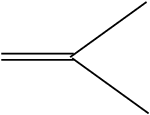}}
 =& ~ \frac{1}{16\pi G}
 \int  d^4x~\left[\left(2\sigma_{ij}\partial_i\phi\partial_j\phi
 -\sigma_{jj}\partial_i\phi\partial_i\phi\right) 
 + \sigma_{ii}\left(\partial_t\phi\right)^2\right],\\
\label{eq:Aphi2dt} \parbox{18mm}{\includegraphics[scale=0.6]{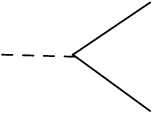}}
 =& -\frac{1}{4\pi G}\int d^4x~\left(A_i\partial_i\phi\partial_t\phi\right),\\
\label{eq:phiAsigmadt} \parbox{18mm}{\includegraphics[scale=0.6]{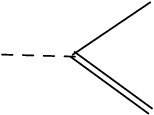}}
 =& ~ \frac{1}{8\pi G}\int d^4x~\left[2\sigma_{ij}\left(\partial_i\phi\partial_tA_j
-\partial_t\phi\partial_iA_j\right)-\sigma_{jj}\left(\partial_i\phi\partial_tA_i
-\partial_t\phi\partial_iA_i\right)\right],\\ 
\label{eq:A3t} \parbox{18mm}{\includegraphics[scale=0.6]{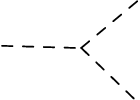}}
 =& \frac{1}{16\pi G}\int d^4x~\left(A_i\partial_iA_j\partial_tA_j\right),\\
\label{eq:phi3dt2} \parbox{18mm}{\includegraphics[scale=0.6]{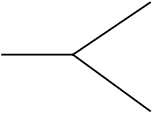}}
 =& -\frac{1}{2\pi G}\int d^4x~\left[\phi\left(\partial_t\phi\right)^2\right],
\end{align}
where the first two vertices are stationary, and can be read off from the 
stationary Kaluza-Klein part of the gravitational action. The next five vertices are time 
dependent, and contain up to two time derivatives.

The Feynman rule for the four-graviton vertex required to the order considered 
here is given by
\begin{align}
\label{eq:phi^2A^2} \parbox{18mm}{\includegraphics[scale=0.6]{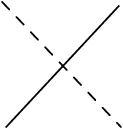}}
 =& \frac{1}{4\pi G}\int d^4x~\phi^2
 \left(\partial_iA_j\left(\partial_iA_j-\partial_jA_i\right)
 +\left(\partial_iA_i\right)^2\right),
\end{align}
where this vertex is stationary.

Next, we recall that the minimal coupling part of the point particle action 
of each of the particles with spins is given by  
\begin{align} \label{mcact}
S_{\text{pp}}=&\int 
d\lambda\left[-m \sqrt{u^2}-\frac{1}{2} \hat{S}_{\mu\nu} \hat{\Omega}^{\mu\nu}
	-\frac{\hat{S}^{\mu\nu} p_{\nu}}{p^2} \frac{D p_{\mu}}{D \lambda}\right],
\end{align}
which is covariant, as well as invariant under rotational variables gauge 
transformations \cite{Levi:2015msa}. Here $\lambda$ is the affine parameter, 
$u^{\mu}\equiv dx^\mu/d\lambda$ is the 
4-velocity, and $\Omega^{\mu\nu}$, $S_{\mu\nu}$ are the angular velocity and spin 
tensors of the particle, respectively \cite{Levi:2015msa}. We parametrize the 
worldline using the coordinate time $t=x^0$, i.e.~$\lambda=t$, so that we have 
for $u_\mu\equiv dx^\mu/d\lambda$: $u^0=1$, $u^i=dx^i/dt\equiv v^i$. Since the 
spin-orbit interaction is linear in the spins, only the minimal coupling part of 
the action, i.e.~that which appears in eq.~\eqref{mcact} is required. We stress 
that in the spin-orbit sector both mass and spin couplings play central roles in 
the interaction.

Let us then present first the mass couplings required for this sector. The Feynman 
rules of the one-graviton couplings to the worldline mass required for the NNLO 
spin-orbit interaction are given by
\begin{align}
\label{eq:mphi} \parbox{12mm}{\includegraphics[scale=0.6]{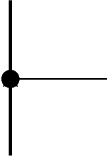}}
 =& - m \int dt~\phi~\left[1+\frac{3}{2}v^2+\frac{7}{8}v^4+\cdots\right], \\ 
\label{eq:mA} \parbox{12mm}{\includegraphics[scale=0.6]{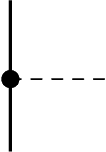}}
 =& ~m \int dt~A_iv^i~\left[1+\frac{1}{2}v^2+\frac{3}{8}v^4+\cdots\right], \\ 
\label{eq:msigma} \parbox{12mm}{\includegraphics[scale=0.6]{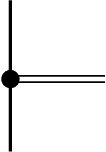}}
 =& \frac{1}{2}m \int dt~\sigma_{ij}v^iv^j~\left[1+\frac{1}{2}v^2+\cdots\right],
\end{align}
where the heavy solid lines represent the worldlines, and the spherical black 
blobs represent the masses on the worldline. 
The ellipsis denotes higher orders in $v$, beyond the order considered here. 

For the two-graviton couplings to the worldline mass required here, we have the following Feynman rules: 
\begin{align}
\label{eq:mphi^2}  \parbox{12mm}{\includegraphics[scale=0.6]{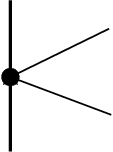}}
 =& -\frac{1}{2}m \int dt~\phi^2~\left[1-\frac{9}{2}v^2+\cdots\right], \\ 
\label{eq:mphiA}   \parbox{12mm}{\includegraphics[scale=0.6]{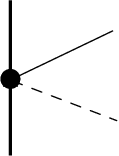}}
 =& ~~m \int dt~\phi A_iv^i~\biggl[1-\frac{3}{2}v^2+\cdots\biggr],\\
\label{eq:mphisigma}  \parbox{12mm}{\includegraphics[scale=0.6]{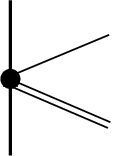}}
 =& -\frac{3}{2}m \int dt~\phi\sigma_{ij}v^iv^j~\left[1+\cdots\right].
\end{align}

Finally, for the three-graviton couplings to the worldline mass required here, we have the following Feynman rules: 
\begin{align}
\label{eq:mphi^3}  \parbox{12mm}{\includegraphics[scale=0.6]{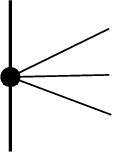}}
 =& -\frac{1}{6}m \int dt~\phi^3~\left[1+\cdots\right],\\
\label{eq:mphi^2A}  \parbox{12mm}{\includegraphics[scale=0.6]{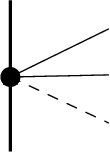}}
 =& \frac{1}{2}m \int dt~\phi^2A_iv^i~\left[1+\cdots\right].
\end{align}

Let us go on to the spin couplings required for this sector. These are given here 
in terms of the physical spatial components of the local spin variable in the 
canonical gauge \cite{Levi:2015msa}. First, we have contributions from kinematic 
terms involving spin without field coupling \cite{Levi:2015msa}. To the order we 
are considering, these are given by
\be \label{frskin}
L_{\text{kin}}=-\vec{S}\cdot \vec{\Omega} + \frac{1}{2} \vec{S}\cdot\vec{v}\times\vec{a}\left(1+\frac{3}{4}v^2
+\frac{5}{8}v^4\right) , 
\ee
where $S_{ij}=\epsilon_{ijk}S_k$, $\Omega_{ij}=\epsilon_{ijk}\Omega_k$, 
$\epsilon_{ijk}$ is the 3-dimensional Levi-Civita symbol, and 
$a^i\equiv\dot{v}^i$. We recall that all indices are Euclidean. 

The required Feynman rules of the one-graviton couplings to the worldline spin 
are thus
\begin{align}
\label{eq:sA}  \parbox{12mm}{\includegraphics[scale=0.6]{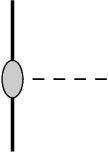}}
 =& \int dt \,\,\left[\epsilon_{ijk}S_{k}\left(\frac{1}{2}\partial_i A_j 
 + \frac{1}{4} v^iv^l \left(\partial_l A_j-\partial_j A_l\right)
 \left(3+\frac{7}{4}v^2\right) 
 + v^i\partial_t A_j\left(1+\frac{1}{2}v^2\right)\right.\right.\nn\\
 & \qquad\qquad\qquad\quad+v^i a^j A_l v^l\bigg)\bigg], \\ 
\label{eq:sphi}   \parbox{12mm}{\includegraphics[scale=0.6]{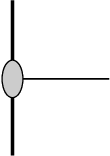}}
 =& \int dt \,\,\left[\epsilon_{ijk}S_{k}v^i\left(
 \partial_j\phi\left(2 + v^2 +\frac{3}{4}v^4\right) 
 -a^j\phi\left(2+3v^2\right)\right)\right], \\  
\label{eq:ssigma}   \parbox{12mm}{\includegraphics[scale=0.6]{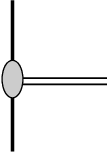}}
 =& \int dt \,\,\left[\frac{1}{2}\epsilon_{ijk}S_{k}\left(\partial_i\sigma_{jl} \,v^l 
 +\left(\frac{1}{2}\partial_i\sigma_{lm}-\partial_l\sigma_{im}\right) v^j v^l v^m 
 -\frac{3}{2}\partial_t\sigma_{il}\,v^j v^l\right.\right.\nn\\
 &\left.\left.\qquad\qquad\qquad\quad 
 -\frac{1}{2}\sigma_{il}\left(v^j a^l - v^l a^j \right)\right)\right],
\end{align}
where the (gray) oval blobs represent the spins on the worldlines.  

For the two-graviton couplings to the worldline spin, the Feynman rules required 
here are: 
\begin{align}
\label{eq:sphiA}  \parbox{12mm}{\includegraphics[scale=0.6]{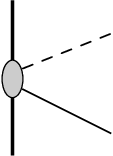}}
 =& \int dt \left[2\epsilon_{ijk}S_{k}\left(\partial_i A_j\,\phi 
 + A_j\,\partial_t\phi\, v^i \right)\right], \\ 
\label{eq:ssigmaA}  \parbox{12mm}{\includegraphics[scale=0.6]{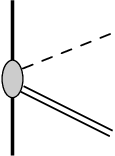}}
 =& \int dt~\left[\frac{1}{4}\epsilon_{ijk}S_{k}\sigma_{il} 
 \left(\partial_j A_l -\partial_l A_j\right)\right], \\ 
\label{eq:sA^2}   \parbox{12mm}{\includegraphics[scale=0.6]{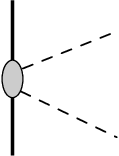}}
 =& \int dt \left[\frac{1}{2}\epsilon_{ijk}S_{k} \left(\partial_jA_i\, A_l\, v^l 
 + A_i\,\partial_t A_j\right)\right],\\ 
\label{eq:sphi^2}   \parbox{12mm}{\includegraphics[scale=0.6]{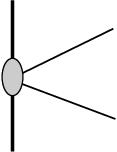}} 
 =& \int dt\left[4\epsilon_{ijk}S_{k} \phi\left(\partial_i\phi\, v^j v^2 + v^i a^j 
 \phi\right) \right] , \\
\label{eq:sphisigma}   \parbox{12mm}{\includegraphics[scale=0.6]{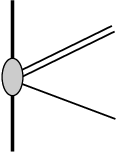}}
 =& \int dt \,\,\left[\epsilon_{ijk}S_{k}\sigma_{il} \left(\partial_j\phi \,v^l
 +\partial_l\phi\, v^j\right)\right].  
\end{align}

Finally, we also have to include three-graviton spin couplings. For three-graviton 
couplings to the worldline spin, the only Feynman rule required here is
\begin{align}
\label{eq:sphi^2A}  \parbox{12mm}{\includegraphics[scale=0.6]{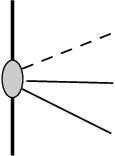}}
 =& \int dt \left[4\epsilon_{ijk}S_{k}\partial_iA_j\, \phi^2\right].
\end{align}
Note that similarly to the NLO, where the two-scalar spin coupling is absent, here 
at NNLO the three-scalar spin coupling vanishes due to the use of the ``NRG'' fields, 
and our gauge of the rotational spin variables. Its appearance is then deferred to 
higher PN orders, which is advantageous.

\section{Next-to-next-to-leading order spin-orbit interaction} \label{fdiagrams}

In this section we evaluate the relevant two-body effective action by its 
diagrammatic expansion. As explained in \cite{Levi:2011eq}, in the NNLO spin-orbit 
potential, which is evaluated at 3.5PN order, we have diagram contributions up to order 
$G^3$, coming from all 12 possible topologies appearing at these orders, as displayed in figures \ref{fig:sonnlo1g}-\ref{fig:sonnlog3twoloop} below: one 
topology at $O(G)$, two at $O(G^2)$, and nine topologies at $O(G^3)$. For the 
construction of the Feynman diagrams we use the Feynman rules presented in the 
previous section, see \cite{Levi:2011eq} for more detail. We have a total of 132 diagrams contributing to the NNLO 
spin-orbit interaction. Eventually, we 
will also have contribution from terms with higher order time derivatives coming 
from the point-mass 2PN order \cite{Gilmore:2008gq}, and up to NLO spin-orbit \cite{Levi:2015msa} sectors.
		
We denote ${\vec{r}}\equiv{\vec{x}}_1-{\vec{x}}_2$, $r\equiv\left|\vec{r}\right|$, 
and ${\vec{n}}\equiv\frac{{\vec{r}}} {r}$. Labels 1 and 2 are used for the left 
and right worldlines in the figures, respectively. All of the diagrams should be included together with their mirror images. Accordingly, the ($1\leftrightarrow2$) 
notation stands for a term, whose value is obtained under the interchange of 
particles labels. Finally, a multiplicative factor of $\int dt$ is 
omitted from all diagram values.

\subsection{One-graviton exchange}

\begin{figure}[t]
\includegraphics[scale=0.93]{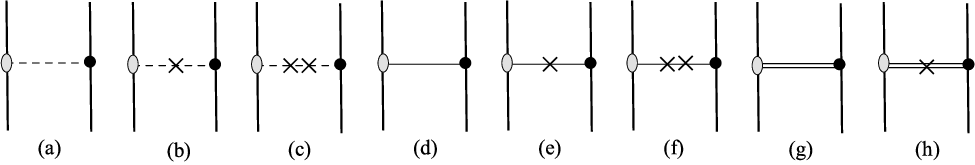}
\caption{NNLO spin-orbit Feynman diagrams of one-graviton exchange.}
\label{fig:sonnlo1g}
\end{figure}

For the NNLO spin-orbit interaction we have 8 one-graviton exchange diagrams as 
shown in figure \ref{fig:sonnlo1g}. In addition to the one-graviton exchange 
diagrams, which already appeared in the NLO spin-orbit sector 
\cite{Levi:2010zu,Levi:2015msa}, new diagrams are added here by inserting further 
propagator correction vertices. Tensor Fourier integrals of up to order 5 are 
required here, due to the derivative-coupling of spin, which makes the 
computations heavier, see \cite{Levi:2011eq}, and appendix~A there.

At the NNLO level we inevitably obtain terms with higher order time derivative, 
i.e.~with accelerations and precessions, and these are all kept until they are 
treated rigorously in the resulting action \cite{Levi:2014sba,Levi:2015msa}, see 
section \ref{VtoH} below. Finally, we recall that there are several ways to 
evaluate diagrams with time derivatives, which differ only by total time 
derivatives. Our convention for their evaluation is, that time derivatives from 
the spin couplings are taken on their respective worldlines, whereas those from 
the propagator correction vertices are taken symmetrically on the worldlines.

The values of the one-graviton exchange diagrams are given as follows: 
\begin{align}
\text{Fig.~1(a)} =&
\frac{2 G m_{2}}{r^2} \vec{S}_{1}\times\vec{n}\cdot\vec{v}_{2}
+ \frac{G m_{2}}{r^2} \Big[ 3 \vec{S}_{1}\times\vec{n}\cdot\vec{v}_{1} \vec{v}_{1}\cdot\vec{v}_{2}
	 + \vec{S}_{1}\times\vec{n}\cdot\vec{v}_{2} v_{2}^2
	 - \vec{S}_{1}\times\vec{v}_{1}\cdot\vec{v}_{2} \vec{v}_{1}\cdot\vec{n} \Big] \nl
+ \frac{4 G m_{2}}{r} \vec{S}_{1}\times\vec{a}_{1}\cdot\vec{v}_{2}
+ \frac{4 G m_{2}}{r} \dot{\vec{S}}_{1}\times\vec{v}_{1}\cdot\vec{v}_{2}
+ \frac{G m_{2}}{4 r^2} \Big[ \vec{S}_{1}\times\vec{n}\cdot\vec{v}_{1} \big( 7 v_{1}^2 \vec{v}_{1}\cdot\vec{v}_{2} \nl
	 + 6 \vec{v}_{1}\cdot\vec{v}_{2} v_{2}^2 \big)
	 + 3 \vec{S}_{1}\times\vec{n}\cdot\vec{v}_{2} v_{2}^{4}
	 - \vec{S}_{1}\times\vec{v}_{1}\cdot\vec{v}_{2} \big( \vec{v}_{1}\cdot\vec{n} v_{1}^2
	 + 2 \vec{v}_{1}\cdot\vec{n} v_{2}^2 \big) \Big] \nl
- \frac{2 G m_{2}}{r} \Big[ 2 \vec{S}_{1}\times\vec{v}_{1}\cdot\vec{a}_{1} \vec{v}_{1}\cdot\vec{v}_{2}
	 - 2 \vec{S}_{1}\times\vec{v}_{1}\cdot\vec{v}_{2} \vec{v}_{1}\cdot\vec{a}_{1}
	 - \vec{S}_{1}\times\vec{a}_{1}\cdot\vec{v}_{2} \big( v_{1}^2
	 + v_{2}^2 \big) \Big] \nl
+ \frac{2 G m_{2}}{r} \dot{\vec{S}}_{1}\times\vec{v}_{1}\cdot\vec{v}_{2} \big( v_{1}^2
	 + v_{2}^2 \big)\\
\text{Fig.~1(b)} =&
\frac{G m_{2}}{r^2} \Big[ \vec{S}_{1}\times\vec{n}\cdot\vec{v}_{2} \big( \vec{v}_{1}\cdot\vec{v}_{2} -3 \vec{v}_{1}\cdot\vec{n} \vec{v}_{2}\cdot\vec{n} \big)
	 + \vec{S}_{1}\times\vec{v}_{1}\cdot\vec{v}_{2} \vec{v}_{2}\cdot\vec{n} \Big] \nl
- \frac{G m_{2}}{r} \Big[ \vec{S}_{1}\times\vec{n}\cdot\vec{a}_{2} \vec{v}_{1}\cdot\vec{n}
	 - \vec{S}_{1}\times\vec{v}_{1}\cdot\vec{a}_{2} \Big]
+ \frac{G m_{2}}{r} \vec{v}_{2}\cdot\vec{n} \dot{\vec{S}}_{1}\times\vec{n}\cdot\vec{v}_{2} \nl
+ G m_{2} \dot{\vec{S}}_{1}\times\vec{n}\cdot\vec{a}_{2}
+ \frac{G m_{2}}{2 r^2} \Big[ 3 \vec{S}_{1}\times\vec{n}\cdot\vec{v}_{1} \big( ( \vec{v}_{1}\cdot\vec{v}_{2} )^{2} -3 \vec{v}_{1}\cdot\vec{n} \vec{v}_{2}\cdot\vec{n} \vec{v}_{1}\cdot\vec{v}_{2} \big) \nl
	 + \vec{S}_{1}\times\vec{n}\cdot\vec{v}_{2} \big( \vec{v}_{1}\cdot\vec{v}_{2} v_{2}^2 -3 \vec{v}_{1}\cdot\vec{n} \vec{v}_{2}\cdot\vec{n} v_{2}^2 \big)
	 - \vec{S}_{1}\times\vec{v}_{1}\cdot\vec{v}_{2} \big( v_{1}^2 \vec{v}_{2}\cdot\vec{n}
	 + 5 \vec{v}_{1}\cdot\vec{n} \vec{v}_{1}\cdot\vec{v}_{2} \nl
	 - \vec{v}_{2}\cdot\vec{n} v_{2}^2 -3 \vec{v}_{2}\cdot\vec{n} ( \vec{v}_{1}\cdot\vec{n} )^{2} \big) \Big]
+ \frac{G m_{2}}{2 r} \Big[ 3 \vec{S}_{1}\times\vec{n}\cdot\vec{v}_{1} \big( \vec{v}_{2}\cdot\vec{n} \vec{a}_{1}\cdot\vec{v}_{2}
	 - \vec{v}_{1}\cdot\vec{n} \vec{v}_{1}\cdot\vec{a}_{2} \big) \nl
	 + 3 \vec{S}_{1}\times\vec{n}\cdot\vec{a}_{1} \vec{v}_{2}\cdot\vec{n} \vec{v}_{1}\cdot\vec{v}_{2}
	 - 2 \vec{S}_{1}\times\vec{n}\cdot\vec{v}_{2} \vec{v}_{1}\cdot\vec{n} \vec{v}_{2}\cdot\vec{a}_{2}
	 + \vec{S}_{1}\times\vec{v}_{1}\cdot\vec{v}_{2} \big( 4 \vec{a}_{1}\cdot\vec{v}_{2} \nl
	 + 2 \vec{v}_{2}\cdot\vec{a}_{2}
	 - \vec{a}_{1}\cdot\vec{n} \vec{v}_{2}\cdot\vec{n} \big)
	 + \vec{S}_{1}\times\vec{a}_{1}\cdot\vec{v}_{2} \big( 8 \vec{v}_{1}\cdot\vec{v}_{2} -5 \vec{v}_{1}\cdot\vec{n} \vec{v}_{2}\cdot\vec{n} \big) \nl
	 - \vec{S}_{1}\times\vec{n}\cdot\vec{a}_{2} \vec{v}_{1}\cdot\vec{n} v_{2}^2
	 - \vec{S}_{1}\times\vec{v}_{1}\cdot\vec{a}_{2} \big( v_{1}^2
	 - v_{2}^2
	 - ( \vec{v}_{1}\cdot\vec{n} )^{2} \big) \Big] \nl
+ \frac{G m_{2}}{2 r} \Big[ 3 \dot{\vec{S}}_{1}\times\vec{n}\cdot\vec{v}_{1} \vec{v}_{2}\cdot\vec{n} \vec{v}_{1}\cdot\vec{v}_{2}
	 + \dot{\vec{S}}_{1}\times\vec{n}\cdot\vec{v}_{2} \vec{v}_{2}\cdot\vec{n} v_{2}^2 
	 + \dot{\vec{S}}_{1}\times\vec{v}_{1}\cdot\vec{v}_{2} \big( 8 \vec{v}_{1}\cdot\vec{v}_{2} \nl
                    -5 \vec{v}_{1}\cdot\vec{n} \vec{v}_{2}\cdot\vec{n} \big) \Big]
+ \frac{1}{2} G m_{2} \Big[ 4 \vec{S}_{1}\times\dot{\vec{a}}_{1}\cdot\vec{v}_{2} \vec{v}_{2}\cdot\vec{n}
	 + 4 \ddot{\vec{S}}_{1}\times\vec{v}_{1}\cdot\vec{v}_{2} \vec{v}_{2}\cdot\vec{n} \nl
	 + (3 \vec{S}_{1}\times\vec{n}\cdot\vec{v}_{1} \vec{a}_{1}\cdot\vec{a}_{2}
	 + 3 \vec{S}_{1}\times\vec{n}\cdot\vec{a}_{1} \vec{v}_{1}\cdot\vec{a}_{2}
	 - \vec{S}_{1}\times\vec{v}_{1}\cdot\vec{a}_{2} \vec{a}_{1}\cdot\vec{n} \nl
	 - 5 \vec{S}_{1}\times\vec{a}_{1}\cdot\vec{a}_{2} \vec{v}_{1}\cdot\vec{n} )
	 + (3 \dot{\vec{S}}_{1}\times\vec{n}\cdot\vec{v}_{1} \vec{v}_{1}\cdot\vec{a}_{2}
	 + 2 \dot{\vec{S}}_{1}\times\vec{n}\cdot\vec{v}_{2} \vec{v}_{2}\cdot\vec{a}_{2} \nl
	 + 8 \dot{\vec{S}}_{1}\times\vec{a}_{1}\cdot\vec{v}_{2} \vec{v}_{2}\cdot\vec{n}
	 + \dot{\vec{S}}_{1}\times\vec{n}\cdot\vec{a}_{2} v_{2}^2
	 - 5 \dot{\vec{S}}_{1}\times\vec{v}_{1}\cdot\vec{a}_{2} \vec{v}_{1}\cdot\vec{n} ) \Big] \nl
- 2 G m_{2} r \Big[ \vec{S}_{1}\times\dot{\vec{a}}_{1}\cdot\vec{a}_{2}
	 + \ddot{\vec{S}}_{1}\times\vec{v}_{1}\cdot\vec{a}_{2}
	 + 2 \dot{\vec{S}}_{1}\times\vec{a}_{1}\cdot\vec{a}_{2} \Big]\\
\text{Fig.~1(c)} =&
\frac{G m_{2}}{4 r^2} \Big[ \vec{S}_{1}\times\vec{n}\cdot\vec{v}_{2} \big( v_{1}^2 v_{2}^2
	 + 2 ( \vec{v}_{1}\cdot\vec{v}_{2} )^{2} -12 \vec{v}_{1}\cdot\vec{n} \vec{v}_{2}\cdot\vec{n} \vec{v}_{1}\cdot\vec{v}_{2}
	 - 3 v_{2}^2 ( \vec{v}_{1}\cdot\vec{n} )^{2} \nl
	 - 3 v_{1}^2 ( \vec{v}_{2}\cdot\vec{n} )^{2}
	 + 15 ( \vec{v}_{1}\cdot\vec{n} )^{2} ( \vec{v}_{2}\cdot\vec{n} )^{2} \big)
	 + 2 \vec{S}_{1}\times\vec{v}_{1}\cdot\vec{v}_{2} \big( 2 \vec{v}_{2}\cdot\vec{n} \vec{v}_{1}\cdot\vec{v}_{2}
	 + \vec{v}_{1}\cdot\vec{n} v_{2}^2 \nl
                    -3 \vec{v}_{1}\cdot\vec{n} ( \vec{v}_{2}\cdot\vec{n} )^{2} \big) \Big]
+ \frac{G m_{2}}{4 r} \Big[ \vec{S}_{1}\times\vec{n}\cdot\vec{v}_{2} \big( 2 \vec{v}_{2}\cdot\vec{n} \vec{a}_{1}\cdot\vec{v}_{2}
	 + \vec{a}_{1}\cdot\vec{n} v_{2}^2
	 - v_{1}^2 \vec{a}_{2}\cdot\vec{n} \nl
	 - 2 \vec{v}_{1}\cdot\vec{n} \vec{v}_{1}\cdot\vec{a}_{2}
	 + 3 \vec{a}_{2}\cdot\vec{n} ( \vec{v}_{1}\cdot\vec{n} )^{2}
	 - 3 \vec{a}_{1}\cdot\vec{n} ( \vec{v}_{2}\cdot\vec{n} )^{2} \big)
	 + 2 \vec{S}_{1}\times\vec{v}_{1}\cdot\vec{v}_{2} \big( \vec{v}_{1}\cdot\vec{a}_{2} \nl
	 - \vec{v}_{1}\cdot\vec{n} \vec{a}_{2}\cdot\vec{n} \big)
	 - \vec{S}_{1}\times\vec{a}_{1}\cdot\vec{v}_{2} \big( v_{2}^2
	 - ( \vec{v}_{2}\cdot\vec{n} )^{2} \big)
	 - 2 \vec{S}_{1}\times\vec{n}\cdot\vec{a}_{2} \big( v_{1}^2 \vec{v}_{2}\cdot\vec{n} \nl
	 + 2 \vec{v}_{1}\cdot\vec{n} \vec{v}_{1}\cdot\vec{v}_{2} -3 \vec{v}_{2}\cdot\vec{n} ( \vec{v}_{1}\cdot\vec{n} )^{2} \big)
	 + 4 \vec{S}_{1}\times\vec{v}_{1}\cdot\vec{a}_{2} \big( \vec{v}_{1}\cdot\vec{v}_{2}
	 - \vec{v}_{1}\cdot\vec{n} \vec{v}_{2}\cdot\vec{n} \big) \nl
	 + \vec{S}_{1}\times\vec{v}_{2}\cdot\vec{a}_{2} \big( v_{1}^2
	 - ( \vec{v}_{1}\cdot\vec{n} )^{2} \big) \Big]
+ \frac{G m_{2}}{2 r} \Big[ \dot{\vec{S}}_{1}\times\vec{n}\cdot\vec{v}_{2} \big( 2 \vec{v}_{2}\cdot\vec{n} \vec{v}_{1}\cdot\vec{v}_{2}
	 + \vec{v}_{1}\cdot\vec{n} v_{2}^2 \nl
                    -3 \vec{v}_{1}\cdot\vec{n} ( \vec{v}_{2}\cdot\vec{n} )^{2} \big)
	 - \dot{\vec{S}}_{1}\times\vec{v}_{1}\cdot\vec{v}_{2} \big( v_{2}^2
	 - ( \vec{v}_{2}\cdot\vec{n} )^{2} \big) \Big]
- \frac{1}{4} G m_{2} \Big[ ( \vec{S}_{1}\times\vec{n}\cdot\dot{\vec{a}}_{2} \big( v_{1}^2 \nl
	 - ( \vec{v}_{1}\cdot\vec{n} )^{2} \big)
	 + 2 \vec{S}_{1}\times\vec{v}_{1}\cdot\dot{\vec{a}}_{2} \vec{v}_{1}\cdot\vec{n} )
	 - ( \vec{S}_{1}\times\vec{n}\cdot\vec{v}_{2} \big( \vec{a}_{1}\cdot\vec{a}_{2}
	 - \vec{a}_{1}\cdot\vec{n} \vec{a}_{2}\cdot\vec{n} \big) \nl
	 + \vec{S}_{1}\times\vec{a}_{1}\cdot\vec{v}_{2} \vec{a}_{2}\cdot\vec{n}
	 + 2 \vec{S}_{1}\times\vec{n}\cdot\vec{a}_{2} \big( \vec{a}_{1}\cdot\vec{v}_{2}
	 - \vec{a}_{1}\cdot\vec{n} \vec{v}_{2}\cdot\vec{n} \big)
	 + 2 \vec{S}_{1}\times\vec{a}_{1}\cdot\vec{a}_{2} \vec{v}_{2}\cdot\vec{n} \nl
	 + \vec{S}_{1}\times\vec{v}_{2}\cdot\vec{a}_{2} \vec{a}_{1}\cdot\vec{n} )
	 - (2 \dot{\vec{S}}_{1}\times\vec{n}\cdot\vec{v}_{2} \big( \vec{v}_{1}\cdot\vec{a}_{2}
	 - \vec{v}_{1}\cdot\vec{n} \vec{a}_{2}\cdot\vec{n} \big)
	 + 2 \dot{\vec{S}}_{1}\times\vec{v}_{1}\cdot\vec{v}_{2} \vec{a}_{2}\cdot\vec{n} \nl
	 + 4 \dot{\vec{S}}_{1}\times\vec{n}\cdot\vec{a}_{2} \big( \vec{v}_{1}\cdot\vec{v}_{2}
	 - \vec{v}_{1}\cdot\vec{n} \vec{v}_{2}\cdot\vec{n} \big)
	 + 4 \dot{\vec{S}}_{1}\times\vec{v}_{1}\cdot\vec{a}_{2} \vec{v}_{2}\cdot\vec{n}
	 + 2 \dot{\vec{S}}_{1}\times\vec{v}_{2}\cdot\vec{a}_{2} \vec{v}_{1}\cdot\vec{n} ) \nl
	 + \ddot{\vec{S}}_{1}\times\vec{n}\cdot\vec{v}_{2} \big( v_{2}^2
	 - ( \vec{v}_{2}\cdot\vec{n} )^{2} \big) \Big]
- \frac{1}{4} G m_{2} r \Big[ ( \vec{S}_{1}\times\vec{n}\cdot\dot{\vec{a}}_{2} \vec{a}_{1}\cdot\vec{n}
	 + \vec{S}_{1}\times\vec{a}_{1}\cdot\dot{\vec{a}}_{2} ) \nl
	 + (2 \dot{\vec{S}}_{1}\times\vec{n}\cdot\dot{\vec{a}}_{2} \vec{v}_{1}\cdot\vec{n}
	 + 2 \dot{\vec{S}}_{1}\times\vec{v}_{1}\cdot\dot{\vec{a}}_{2} )
	 - ( \ddot{\vec{S}}_{1}\times\vec{n}\cdot\vec{v}_{2} \vec{a}_{2}\cdot\vec{n}
	 + 2 \ddot{\vec{S}}_{1}\times\vec{n}\cdot\vec{a}_{2} \vec{v}_{2}\cdot\vec{n} \nl
	 + \ddot{\vec{S}}_{1}\times\vec{v}_{2}\cdot\vec{a}_{2} ) \Big]
- \frac{1}{4} G m_{2} r^2 \ddot{\vec{S}}_{1}\times\vec{n}\cdot\dot{\vec{a}}_{2} \\
\text{Fig.~1(d)} =&
- \frac{2 G m_{2}}{r^2} \vec{S}_{1}\times\vec{n}\cdot\vec{v}_{1}
- \frac{G m_{2}}{r^2} \vec{S}_{1}\times\vec{n}\cdot\vec{v}_{1} \big( v_{1}^2
	 + 3 v_{2}^2 \big)
+ \frac{2 G m_{2}}{r} \vec{S}_{1}\times\vec{v}_{1}\cdot\vec{a}_{1} \nl
- \frac{G m_{2}}{4 r^2} \vec{S}_{1}\times\vec{n}\cdot\vec{v}_{1} \big( 6 v_{1}^2 v_{2}^2
	 + 3 v_{1}^{4}
	 + 7 v_{2}^{4} \big)
+ \frac{3 G m_{2}}{r} \vec{S}_{1}\times\vec{v}_{1}\cdot\vec{a}_{1} \big( v_{1}^2
	 + v_{2}^2 \big)\\
\text{Fig.~1(e)} =&
- \frac{G m_{2}}{r^2} \Big[ \vec{S}_{1}\times\vec{n}\cdot\vec{v}_{1} \big( \vec{v}_{1}\cdot\vec{v}_{2} -3 \vec{v}_{1}\cdot\vec{n} \vec{v}_{2}\cdot\vec{n} \big)
	 - \vec{S}_{1}\times\vec{v}_{1}\cdot\vec{v}_{2} \vec{v}_{1}\cdot\vec{n} \Big] \nl
- \frac{G m_{2}}{r} \Big[ \vec{S}_{1}\times\vec{n}\cdot\vec{a}_{1} \vec{v}_{2}\cdot\vec{n}
	 + \vec{S}_{1}\times\vec{a}_{1}\cdot\vec{v}_{2} \Big]
- \frac{G m_{2}}{r} \Big[ \dot{\vec{S}}_{1}\times\vec{n}\cdot\vec{v}_{1} \vec{v}_{2}\cdot\vec{n}
	 + \dot{\vec{S}}_{1}\times\vec{v}_{1}\cdot\vec{v}_{2} \Big] \nl
- \frac{G m_{2}}{2 r^2} \Big[ \vec{S}_{1}\times\vec{n}\cdot\vec{v}_{1} \big( v_{1}^2 \vec{v}_{1}\cdot\vec{v}_{2}
	 + 3 \vec{v}_{1}\cdot\vec{v}_{2} v_{2}^2 -3 \vec{v}_{1}\cdot\vec{n} v_{1}^2 \vec{v}_{2}\cdot\vec{n}
	 - 9 \vec{v}_{1}\cdot\vec{n} \vec{v}_{2}\cdot\vec{n} v_{2}^2 \big) \nl
	 - \vec{S}_{1}\times\vec{v}_{1}\cdot\vec{v}_{2} \big( \vec{v}_{1}\cdot\vec{n} v_{1}^2
	 + 3 \vec{v}_{1}\cdot\vec{n} v_{2}^2 \big) \Big]
- \frac{G m_{2}}{2 r} \Big[ 2 \vec{S}_{1}\times\vec{n}\cdot\vec{v}_{1} \big( \vec{v}_{1}\cdot\vec{a}_{1} \vec{v}_{2}\cdot\vec{n} \nl
	 - 3 \vec{v}_{1}\cdot\vec{n} \vec{v}_{2}\cdot\vec{a}_{2} \big)
	 + \vec{S}_{1}\times\vec{n}\cdot\vec{a}_{1} \big( v_{1}^2 \vec{v}_{2}\cdot\vec{n}
	 + 3 \vec{v}_{2}\cdot\vec{n} v_{2}^2 \big)
	 - 2 \vec{S}_{1}\times\vec{v}_{1}\cdot\vec{a}_{1} \big( \vec{v}_{1}\cdot\vec{v}_{2} \nl
	 - \vec{v}_{1}\cdot\vec{n} \vec{v}_{2}\cdot\vec{n} \big)
	 + 2 \vec{S}_{1}\times\vec{v}_{1}\cdot\vec{v}_{2} \vec{v}_{1}\cdot\vec{a}_{1}
	 + \vec{S}_{1}\times\vec{a}_{1}\cdot\vec{v}_{2} \big( v_{1}^2
	 + 3 v_{2}^2 \big) \Big] \nl
- \frac{G m_{2}}{2 r} \Big[ \dot{\vec{S}}_{1}\times\vec{n}\cdot\vec{v}_{1} \big( v_{1}^2 \vec{v}_{2}\cdot\vec{n}
	 + 3 \vec{v}_{2}\cdot\vec{n} v_{2}^2 \big)
	 + \dot{\vec{S}}_{1}\times\vec{v}_{1}\cdot\vec{v}_{2} \big( v_{1}^2
	 + 3 v_{2}^2 \big) \Big] \nl
- G m_{2} \Big[ 3 \vec{S}_{1}\times\vec{n}\cdot\vec{a}_{1} \vec{v}_{2}\cdot\vec{a}_{2}
	 + (3 \dot{\vec{S}}_{1}\times\vec{n}\cdot\vec{v}_{1} \vec{v}_{2}\cdot\vec{a}_{2}
	 - \dot{\vec{S}}_{1}\times\vec{v}_{1}\cdot\vec{a}_{1} \vec{v}_{2}\cdot\vec{n} ) \nl
	 - \vec{S}_{1}\times\vec{v}_{1}\cdot\dot{\vec{a}}_{1} \vec{v}_{2}\cdot\vec{n} \Big]\\
\text{Fig.~1(f)} =&
- \frac{G m_{2}}{4 r^2} \Big[ \vec{S}_{1}\times\vec{n}\cdot\vec{v}_{1} \big( v_{1}^2 v_{2}^2
	 + 2 ( \vec{v}_{1}\cdot\vec{v}_{2} )^{2} -12 \vec{v}_{1}\cdot\vec{n} \vec{v}_{2}\cdot\vec{n} \vec{v}_{1}\cdot\vec{v}_{2}
	 - 3 v_{2}^2 ( \vec{v}_{1}\cdot\vec{n} )^{2} \nl
	 - 3 v_{1}^2 ( \vec{v}_{2}\cdot\vec{n} )^{2}
	 + 15 ( \vec{v}_{1}\cdot\vec{n} )^{2} ( \vec{v}_{2}\cdot\vec{n} )^{2} \big)
	 - 2 \vec{S}_{1}\times\vec{v}_{1}\cdot\vec{v}_{2} \big( v_{1}^2 \vec{v}_{2}\cdot\vec{n}
	 + 2 \vec{v}_{1}\cdot\vec{n} \vec{v}_{1}\cdot\vec{v}_{2} \nl
                    -3 \vec{v}_{2}\cdot\vec{n} ( \vec{v}_{1}\cdot\vec{n} )^{2} \big) \Big]
- \frac{G m_{2}}{4 r} \Big[ \vec{S}_{1}\times\vec{n}\cdot\vec{v}_{1} \big( 2 \vec{v}_{2}\cdot\vec{n} \vec{a}_{1}\cdot\vec{v}_{2}
	 + \vec{a}_{1}\cdot\vec{n} v_{2}^2
	 - v_{1}^2 \vec{a}_{2}\cdot\vec{n} \nl
	 - 2 \vec{v}_{1}\cdot\vec{n} \vec{v}_{1}\cdot\vec{a}_{2}
	 + 3 \vec{a}_{2}\cdot\vec{n} ( \vec{v}_{1}\cdot\vec{n} )^{2}
	 - 3 \vec{a}_{1}\cdot\vec{n} ( \vec{v}_{2}\cdot\vec{n} )^{2} \big)
	 + 2 \vec{S}_{1}\times\vec{n}\cdot\vec{a}_{1} \big( 2 \vec{v}_{2}\cdot\vec{n} \vec{v}_{1}\cdot\vec{v}_{2} \nl
	 + \vec{v}_{1}\cdot\vec{n} v_{2}^2 -3 \vec{v}_{1}\cdot\vec{n} ( \vec{v}_{2}\cdot\vec{n} )^{2} \big)
	 - \vec{S}_{1}\times\vec{v}_{1}\cdot\vec{a}_{1} \big( v_{2}^2
	 - ( \vec{v}_{2}\cdot\vec{n} )^{2} \big)
	 + 2 \vec{S}_{1}\times\vec{v}_{1}\cdot\vec{v}_{2} \big( \vec{a}_{1}\cdot\vec{v}_{2} \nl
	 - \vec{a}_{1}\cdot\vec{n} \vec{v}_{2}\cdot\vec{n} \big)
	 + 4 \vec{S}_{1}\times\vec{a}_{1}\cdot\vec{v}_{2} \big( \vec{v}_{1}\cdot\vec{v}_{2}
	 - \vec{v}_{1}\cdot\vec{n} \vec{v}_{2}\cdot\vec{n} \big)
	 - \vec{S}_{1}\times\vec{v}_{1}\cdot\vec{a}_{2} \big( v_{1}^2
	 - ( \vec{v}_{1}\cdot\vec{n} )^{2} \big) \Big] \nl
- \frac{G m_{2}}{2 r} \Big[ \dot{\vec{S}}_{1}\times\vec{n}\cdot\vec{v}_{1} \big( 2 \vec{v}_{2}\cdot\vec{n} \vec{v}_{1}\cdot\vec{v}_{2}
	 + \vec{v}_{1}\cdot\vec{n} v_{2}^2 -3 \vec{v}_{1}\cdot\vec{n} ( \vec{v}_{2}\cdot\vec{n} )^{2} \big) \nl
	 + 2 \dot{\vec{S}}_{1}\times\vec{v}_{1}\cdot\vec{v}_{2} \big( \vec{v}_{1}\cdot\vec{v}_{2}
	 - \vec{v}_{1}\cdot\vec{n} \vec{v}_{2}\cdot\vec{n} \big) \Big]
+ \frac{1}{4} G m_{2} \Big[ ( \vec{S}_{1}\times\vec{n}\cdot\dot{\vec{a}}_{1} \big( v_{2}^2
	 - ( \vec{v}_{2}\cdot\vec{n} )^{2} \big) \nl
	 - 2 \vec{S}_{1}\times\dot{\vec{a}}_{1}\cdot\vec{v}_{2} \vec{v}_{2}\cdot\vec{n} )
	 + ( \ddot{\vec{S}}_{1}\times\vec{n}\cdot\vec{v}_{1} \big( v_{2}^2
	 - ( \vec{v}_{2}\cdot\vec{n} )^{2} \big)
	 - 2 \ddot{\vec{S}}_{1}\times\vec{v}_{1}\cdot\vec{v}_{2} \vec{v}_{2}\cdot\vec{n} ) \nl
	 - ( \vec{S}_{1}\times\vec{n}\cdot\vec{v}_{1} \big( \vec{a}_{1}\cdot\vec{a}_{2}
	 - \vec{a}_{1}\cdot\vec{n} \vec{a}_{2}\cdot\vec{n} \big)
	 + 2 \vec{S}_{1}\times\vec{n}\cdot\vec{a}_{1} \big( \vec{v}_{1}\cdot\vec{a}_{2}
	 - \vec{v}_{1}\cdot\vec{n} \vec{a}_{2}\cdot\vec{n} \big) \nl
	 + \vec{S}_{1}\times\vec{v}_{1}\cdot\vec{a}_{1} \vec{a}_{2}\cdot\vec{n}
	 - \vec{S}_{1}\times\vec{v}_{1}\cdot\vec{a}_{2} \vec{a}_{1}\cdot\vec{n}
	 - 2 \vec{S}_{1}\times\vec{a}_{1}\cdot\vec{a}_{2} \vec{v}_{1}\cdot\vec{n} ) \nl
	 - (2 \dot{\vec{S}}_{1}\times\vec{n}\cdot\vec{v}_{1} \big( \vec{v}_{1}\cdot\vec{a}_{2}
	 - \vec{v}_{1}\cdot\vec{n} \vec{a}_{2}\cdot\vec{n} \big)
	 - 2 \dot{\vec{S}}_{1}\times\vec{n}\cdot\vec{a}_{1} \big( v_{2}^2
	 - ( \vec{v}_{2}\cdot\vec{n} )^{2} \big) \nl
	 + 4 \dot{\vec{S}}_{1}\times\vec{a}_{1}\cdot\vec{v}_{2} \vec{v}_{2}\cdot\vec{n}
	 - 2 \dot{\vec{S}}_{1}\times\vec{v}_{1}\cdot\vec{a}_{2} \vec{v}_{1}\cdot\vec{n} ) \Big]
- \frac{1}{4} G m_{2} r \Big[ ( \vec{S}_{1}\times\vec{n}\cdot\dot{\vec{a}}_{1} \vec{a}_{2}\cdot\vec{n} \nl
	 - \vec{S}_{1}\times\dot{\vec{a}}_{1}\cdot\vec{a}_{2} )
	 + ( \ddot{\vec{S}}_{1}\times\vec{n}\cdot\vec{v}_{1} \vec{a}_{2}\cdot\vec{n}
	 - \ddot{\vec{S}}_{1}\times\vec{v}_{1}\cdot\vec{a}_{2} )
	 + (2 \dot{\vec{S}}_{1}\times\vec{n}\cdot\vec{a}_{1} \vec{a}_{2}\cdot\vec{n} \nl
	 - 2 \dot{\vec{S}}_{1}\times\vec{a}_{1}\cdot\vec{a}_{2} ) \Big]\\
\text{Fig.~1(g)} =&
\frac{2 G m_{2}}{r^2} \Big[ \vec{S}_{1}\times\vec{n}\cdot\vec{v}_{1} v_{2}^2
	 - \vec{S}_{1}\times\vec{n}\cdot\vec{v}_{2} \vec{v}_{1}\cdot\vec{v}_{2} \Big]
+ \frac{G m_{2}}{r^2} \Big[ \vec{S}_{1}\times\vec{n}\cdot\vec{v}_{1} \big( v_{1}^2 v_{2}^2
	 - ( \vec{v}_{1}\cdot\vec{v}_{2} )^{2} \nl
	 + v_{2}^{4} \big)
	 - \vec{S}_{1}\times\vec{n}\cdot\vec{v}_{2} \vec{v}_{1}\cdot\vec{v}_{2} v_{2}^2
	 + \vec{S}_{1}\times\vec{v}_{1}\cdot\vec{v}_{2} \vec{v}_{1}\cdot\vec{n} \vec{v}_{1}\cdot\vec{v}_{2} \Big]
- \frac{2 G m_{2}}{r} \Big[ \vec{S}_{1}\times\vec{v}_{1}\cdot\vec{a}_{1} v_{2}^2 \nl
	 + \vec{S}_{1}\times\vec{v}_{1}\cdot\vec{v}_{2} \vec{a}_{1}\cdot\vec{v}_{2}
	 + 2 \vec{S}_{1}\times\vec{a}_{1}\cdot\vec{v}_{2} \vec{v}_{1}\cdot\vec{v}_{2} \Big]
- \frac{3 G m_{2}}{r} \vec{v}_{1}\cdot\vec{v}_{2} \dot{\vec{S}}_{1}\times\vec{v}_{1}\cdot\vec{v}_{2} \\
\text{Fig.~1(h)} =&
\frac{G m_{2}}{r^2} \Big[ \vec{S}_{1}\times\vec{n}\cdot\vec{v}_{1} \big( \vec{v}_{1}\cdot\vec{v}_{2} v_{2}^2 -3 \vec{v}_{1}\cdot\vec{n} \vec{v}_{2}\cdot\vec{n} v_{2}^2 \big)
	 - \vec{S}_{1}\times\vec{n}\cdot\vec{v}_{2} \big( ( \vec{v}_{1}\cdot\vec{v}_{2} )^{2} \nl
                    -3 \vec{v}_{1}\cdot\vec{n} \vec{v}_{2}\cdot\vec{n} \vec{v}_{1}\cdot\vec{v}_{2} \big)
	 - \vec{S}_{1}\times\vec{v}_{1}\cdot\vec{v}_{2} \big( \vec{v}_{2}\cdot\vec{n} \vec{v}_{1}\cdot\vec{v}_{2}
	 + \vec{v}_{1}\cdot\vec{n} v_{2}^2 \big) \Big] \nl
- \frac{G m_{2}}{r} \Big[ 2 \vec{S}_{1}\times\vec{n}\cdot\vec{v}_{1} \vec{v}_{1}\cdot\vec{n} \vec{v}_{2}\cdot\vec{a}_{2}
	 - \vec{S}_{1}\times\vec{n}\cdot\vec{a}_{1} \vec{v}_{2}\cdot\vec{n} v_{2}^2
	 + \vec{S}_{1}\times\vec{n}\cdot\vec{v}_{2} \big( \vec{v}_{2}\cdot\vec{n} \vec{a}_{1}\cdot\vec{v}_{2} \nl
	 - \vec{v}_{1}\cdot\vec{n} \vec{v}_{1}\cdot\vec{a}_{2} \big)
	 + \vec{S}_{1}\times\vec{v}_{1}\cdot\vec{v}_{2} \vec{v}_{1}\cdot\vec{a}_{2}
	 - \vec{S}_{1}\times\vec{a}_{1}\cdot\vec{v}_{2} v_{2}^2
	 - \vec{S}_{1}\times\vec{n}\cdot\vec{a}_{2} \vec{v}_{1}\cdot\vec{n} \vec{v}_{1}\cdot\vec{v}_{2} \nl
	 + \vec{S}_{1}\times\vec{v}_{1}\cdot\vec{a}_{2} \vec{v}_{1}\cdot\vec{v}_{2} \Big]
+ \frac{G m_{2}}{r} \Big[ \dot{\vec{S}}_{1}\times\vec{n}\cdot\vec{v}_{1} \vec{v}_{2}\cdot\vec{n} v_{2}^2
	 - \dot{\vec{S}}_{1}\times\vec{n}\cdot\vec{v}_{2} \vec{v}_{2}\cdot\vec{n} \vec{v}_{1}\cdot\vec{v}_{2} \nl
	 + \dot{\vec{S}}_{1}\times\vec{v}_{1}\cdot\vec{v}_{2} v_{2}^2 \Big]
+ G m_{2} \Big[ (2 \vec{S}_{1}\times\vec{n}\cdot\vec{a}_{1} \vec{v}_{2}\cdot\vec{a}_{2}
	 - \vec{S}_{1}\times\vec{n}\cdot\vec{v}_{2} \vec{a}_{1}\cdot\vec{a}_{2} \nl
	 - \vec{S}_{1}\times\vec{n}\cdot\vec{a}_{2} \vec{a}_{1}\cdot\vec{v}_{2} )
	 + (2 \dot{\vec{S}}_{1}\times\vec{n}\cdot\vec{v}_{1} \vec{v}_{2}\cdot\vec{a}_{2}
	 - \dot{\vec{S}}_{1}\times\vec{n}\cdot\vec{v}_{2} \vec{v}_{1}\cdot\vec{a}_{2} \nl
	 - \dot{\vec{S}}_{1}\times\vec{n}\cdot\vec{a}_{2} \vec{v}_{1}\cdot\vec{v}_{2} ) \Big]
\end{align}

\subsection{Two-graviton exchange and cubic self-interaction}

\subsubsection{Two-graviton exchange}

\begin{figure}[t]
\includegraphics[scale=0.93]{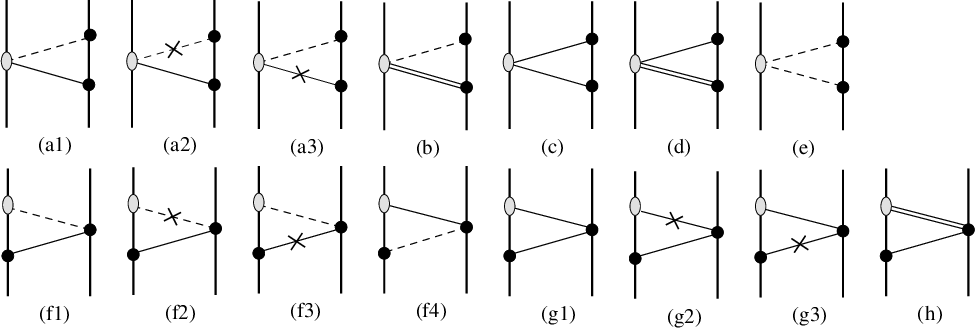}
\caption{NNLO spin-orbit Feynman diagrams of two-graviton exchange.}
\label{fig:sonnlo2g}
\end{figure}

For the NNLO spin-orbit interaction we have 15 two-graviton exchange diagrams, 
as shown in figure \ref{fig:sonnlo2g}, where they either contain a two-graviton 
spin or mass coupling.

The two-graviton exchange diagrams also require only tensor Fourier integrals. 
We encounter here two-graviton exchange diagrams, which involve time derivatives, 
either from the spin couplings or from propagator correction vertices. 

The values of the two-graviton exchange diagrams are given in the following:
\begin{align}
\text{Fig.~2(a1)} =&
- \frac{8 G^2 m_{2}^2}{r^3} \vec{S}_{1}\times\vec{n}\cdot\vec{v}_{2}
- \frac{8 G^2 m_{2}^2}{r^3} \Big[ 2 \vec{S}_{1}\times\vec{n}\cdot\vec{v}_{2} v_{2}^2
	 - \vec{S}_{1}\times\vec{v}_{1}\cdot\vec{v}_{2} \vec{v}_{2}\cdot\vec{n} \Big]\\
\text{Fig.~2(a2)} =&
- \frac{4 G^2 m_{2}^2}{r^3} \Big[ \vec{S}_{1}\times\vec{n}\cdot\vec{v}_{2} \big( \vec{v}_{1}\cdot\vec{v}_{2} -4 \vec{v}_{1}\cdot\vec{n} \vec{v}_{2}\cdot\vec{n}
	 + ( \vec{v}_{2}\cdot\vec{n} )^{2} \big)
	 + \vec{S}_{1}\times\vec{v}_{1}\cdot\vec{v}_{2} \vec{v}_{2}\cdot\vec{n} \Big] \nl
+ \frac{4 G^2 m_{2}^2}{r^2} \Big[ \vec{S}_{1}\times\vec{n}\cdot\vec{a}_{2} \big( 2 \vec{v}_{1}\cdot\vec{n}
	 - \vec{v}_{2}\cdot\vec{n} \big)
	 - \vec{S}_{1}\times\vec{v}_{1}\cdot\vec{a}_{2} \Big]
- \frac{4 G^2 m_{2}^2}{r^2} \vec{v}_{2}\cdot\vec{n} \dot{\vec{S}}_{1}\times\vec{n}\cdot\vec{v}_{2} \nl
- \frac{4 G^2 m_{2}^2}{r} \dot{\vec{S}}_{1}\times\vec{n}\cdot\vec{a}_{2} \\
\text{Fig.~2(a3)} =&
- \frac{4 G^2 m_{2}^2}{r^3} \Big[ \vec{S}_{1}\times\vec{n}\cdot\vec{v}_{2} \big( \vec{v}_{1}\cdot\vec{v}_{2} -4 \vec{v}_{1}\cdot\vec{n} \vec{v}_{2}\cdot\vec{n}
	 + 3 ( \vec{v}_{2}\cdot\vec{n} )^{2} \big)
	 + \vec{S}_{1}\times\vec{v}_{1}\cdot\vec{v}_{2} \vec{v}_{2}\cdot\vec{n} \Big] \nl
- \frac{4 G^2 m_{2}^2}{r^2} \vec{v}_{2}\cdot\vec{n} \vec{S}_{1}\times\vec{n}\cdot\vec{a}_{2}
- \frac{4 G^2 m_{2}^2}{r^2} \vec{v}_{2}\cdot\vec{n} \dot{\vec{S}}_{1}\times\vec{n}\cdot\vec{v}_{2} \\
\text{Fig.~2(b)} =&
\frac{4 G^2 m_{2}^2}{r^3} v_{2}^2 \vec{S}_{1}\times\vec{n}\cdot\vec{v}_{2} \\
\text{Fig.~2(c)} =&
- \frac{4 G^2 m_{2}^2}{r^3} v_{1}^2 \vec{S}_{1}\times\vec{n}\cdot\vec{v}_{1}
+ \frac{4 G^2 m_{2}^2}{r^2} \vec{S}_{1}\times\vec{v}_{1}\cdot\vec{a}_{1} \\
\text{Fig.~2(d)} =&
- \frac{4 G^2 m_{2}^2}{r^3} \Big[ \vec{S}_{1}\times\vec{n}\cdot\vec{v}_{2} \vec{v}_{1}\cdot\vec{v}_{2}
	 + \vec{S}_{1}\times\vec{v}_{1}\cdot\vec{v}_{2} \vec{v}_{2}\cdot\vec{n} \Big]\\
\text{Fig.~2(e)} =&
\frac{8 G^2 m_{2}^2}{r^3} \vec{v}_{1}\cdot\vec{v}_{2} \vec{S}_{1}\times\vec{n}\cdot\vec{v}_{2}
+ \frac{8 G^2 m_{2}^2}{r^2} \vec{S}_{1}\times\vec{v}_{2}\cdot\vec{a}_{2} \\
\text{Fig.~2(f1)} =&
- \frac{2 G^2 m_{1} m_{2}}{r^3} \vec{S}_{1}\times\vec{n}\cdot\vec{v}_{2}
- \frac{G^2 m_{1} m_{2}}{r^3} \Big[ 3 \vec{S}_{1}\times\vec{n}\cdot\vec{v}_{1} \vec{v}_{1}\cdot\vec{v}_{2}
	 + 3 \vec{S}_{1}\times\vec{n}\cdot\vec{v}_{2} \big( v_{1}^2
	 - v_{2}^2 \big) \nl
	 - \vec{S}_{1}\times\vec{v}_{1}\cdot\vec{v}_{2} \vec{v}_{1}\cdot\vec{n} \Big]
- \frac{4 G^2 m_{1} m_{2}}{r^2} \vec{S}_{1}\times\vec{a}_{1}\cdot\vec{v}_{2}
- \frac{4 G^2 m_{1} m_{2}}{r^2} \dot{\vec{S}}_{1}\times\vec{v}_{1}\cdot\vec{v}_{2} \\
\text{Fig.~2(f2)} =&
- \frac{G^2 m_{1} m_{2}}{r^3} \Big[ \vec{S}_{1}\times\vec{n}\cdot\vec{v}_{2} \big( \vec{v}_{1}\cdot\vec{v}_{2} -4 \vec{v}_{1}\cdot\vec{n} \vec{v}_{2}\cdot\vec{n}
	 + ( \vec{v}_{1}\cdot\vec{n} )^{2} \big)
	 - \vec{S}_{1}\times\vec{v}_{1}\cdot\vec{v}_{2} \big( \vec{v}_{1}\cdot\vec{n} \nl
	 - 2 \vec{v}_{2}\cdot\vec{n} \big) \Big]
+ \frac{G^2 m_{1} m_{2}}{r^2} \Big[ \vec{S}_{1}\times\vec{n}\cdot\vec{a}_{2} \vec{v}_{1}\cdot\vec{n}
	 - \vec{S}_{1}\times\vec{v}_{1}\cdot\vec{a}_{2} \Big] \nl
+ \frac{G^2 m_{1} m_{2}}{r^2} \dot{\vec{S}}_{1}\times\vec{n}\cdot\vec{v}_{2} \big( \vec{v}_{1}\cdot\vec{n}
	 - 2 \vec{v}_{2}\cdot\vec{n} \big)
- \frac{G^2 m_{1} m_{2}}{r} \dot{\vec{S}}_{1}\times\vec{n}\cdot\vec{a}_{2} \\
\text{Fig.~2(f3)} =&
- \frac{G^2 m_{1} m_{2}}{r^3} \Big[ \vec{S}_{1}\times\vec{n}\cdot\vec{v}_{2} \big( \vec{v}_{1}\cdot\vec{v}_{2} -4 \vec{v}_{1}\cdot\vec{n} \vec{v}_{2}\cdot\vec{n}
	 + 3 ( \vec{v}_{1}\cdot\vec{n} )^{2} \big)
	 - \vec{S}_{1}\times\vec{v}_{1}\cdot\vec{v}_{2} \vec{v}_{1}\cdot\vec{n} \Big] \nl
+ \frac{G^2 m_{1} m_{2}}{r^2} \vec{v}_{1}\cdot\vec{n} \vec{S}_{1}\times\vec{n}\cdot\vec{a}_{2}
+ \frac{G^2 m_{1} m_{2}}{r^2} \vec{v}_{1}\cdot\vec{n} \dot{\vec{S}}_{1}\times\vec{n}\cdot\vec{v}_{2} \\
\text{Fig.~2(f4)} =&
- \frac{8 G^2 m_{1} m_{2}}{r^3} \vec{S}_{1}\times\vec{n}\cdot\vec{v}_{1} \vec{v}_{1}\cdot\vec{v}_{2} \\
\text{Fig.~2(g1)} =&
\frac{2 G^2 m_{1} m_{2}}{r^3} \vec{S}_{1}\times\vec{n}\cdot\vec{v}_{1}
+ \frac{G^2 m_{1} m_{2}}{r^3} \vec{S}_{1}\times\vec{n}\cdot\vec{v}_{1} \big( 4 v_{1}^2
	 - 9 v_{2}^2 \big)
- \frac{2 G^2 m_{1} m_{2}}{r^2} \vec{S}_{1}\times\vec{v}_{1}\cdot\vec{a}_{1} \\
\text{Fig.~2(g2)} =&
\frac{G^2 m_{1} m_{2}}{r^3} \Big[ \vec{S}_{1}\times\vec{n}\cdot\vec{v}_{1} \big( \vec{v}_{1}\cdot\vec{v}_{2} -4 \vec{v}_{1}\cdot\vec{n} \vec{v}_{2}\cdot\vec{n}
	 + ( \vec{v}_{1}\cdot\vec{n} )^{2} \big)
	 - \vec{S}_{1}\times\vec{v}_{1}\cdot\vec{v}_{2} \vec{v}_{1}\cdot\vec{n} \Big] \nl
- \frac{G^2 m_{1} m_{2}}{r^2} \Big[ \vec{S}_{1}\times\vec{n}\cdot\vec{a}_{1} \big( \vec{v}_{1}\cdot\vec{n}
	 - 2 \vec{v}_{2}\cdot\vec{n} \big)
	 - \vec{S}_{1}\times\vec{a}_{1}\cdot\vec{v}_{2} \Big] \nl
- \frac{G^2 m_{1} m_{2}}{r^2} \Big[ \dot{\vec{S}}_{1}\times\vec{n}\cdot\vec{v}_{1} \big( \vec{v}_{1}\cdot\vec{n}
	 - 2 \vec{v}_{2}\cdot\vec{n} \big)
	 - \dot{\vec{S}}_{1}\times\vec{v}_{1}\cdot\vec{v}_{2} \Big]\\
\text{Fig.~2(g3)} =&
\frac{G^2 m_{1} m_{2}}{r^3} \Big[ \vec{S}_{1}\times\vec{n}\cdot\vec{v}_{1} \big( \vec{v}_{1}\cdot\vec{v}_{2} -4 \vec{v}_{1}\cdot\vec{n} \vec{v}_{2}\cdot\vec{n}
	 + 3 ( \vec{v}_{1}\cdot\vec{n} )^{2} \big)
	 - \vec{S}_{1}\times\vec{v}_{1}\cdot\vec{v}_{2} \vec{v}_{1}\cdot\vec{n} \Big] \nl
- \frac{G^2 m_{1} m_{2}}{r^2} \vec{v}_{1}\cdot\vec{n} \vec{S}_{1}\times\vec{n}\cdot\vec{a}_{1}
- \frac{G^2 m_{1} m_{2}}{r^2} \vec{v}_{1}\cdot\vec{n} \dot{\vec{S}}_{1}\times\vec{n}\cdot\vec{v}_{1} \\
\text{Fig.~2(h)} =&
\frac{6 G^2 m_{1} m_{2}}{r^3} \Big[ \vec{S}_{1}\times\vec{n}\cdot\vec{v}_{1} v_{2}^2
	 - \vec{S}_{1}\times\vec{n}\cdot\vec{v}_{2} \vec{v}_{1}\cdot\vec{v}_{2} \Big]
\end{align}

\subsubsection{Cubic self-interaction} 

\begin{figure}[p]
	\begin{center}
		\includegraphics[scale=0.92]{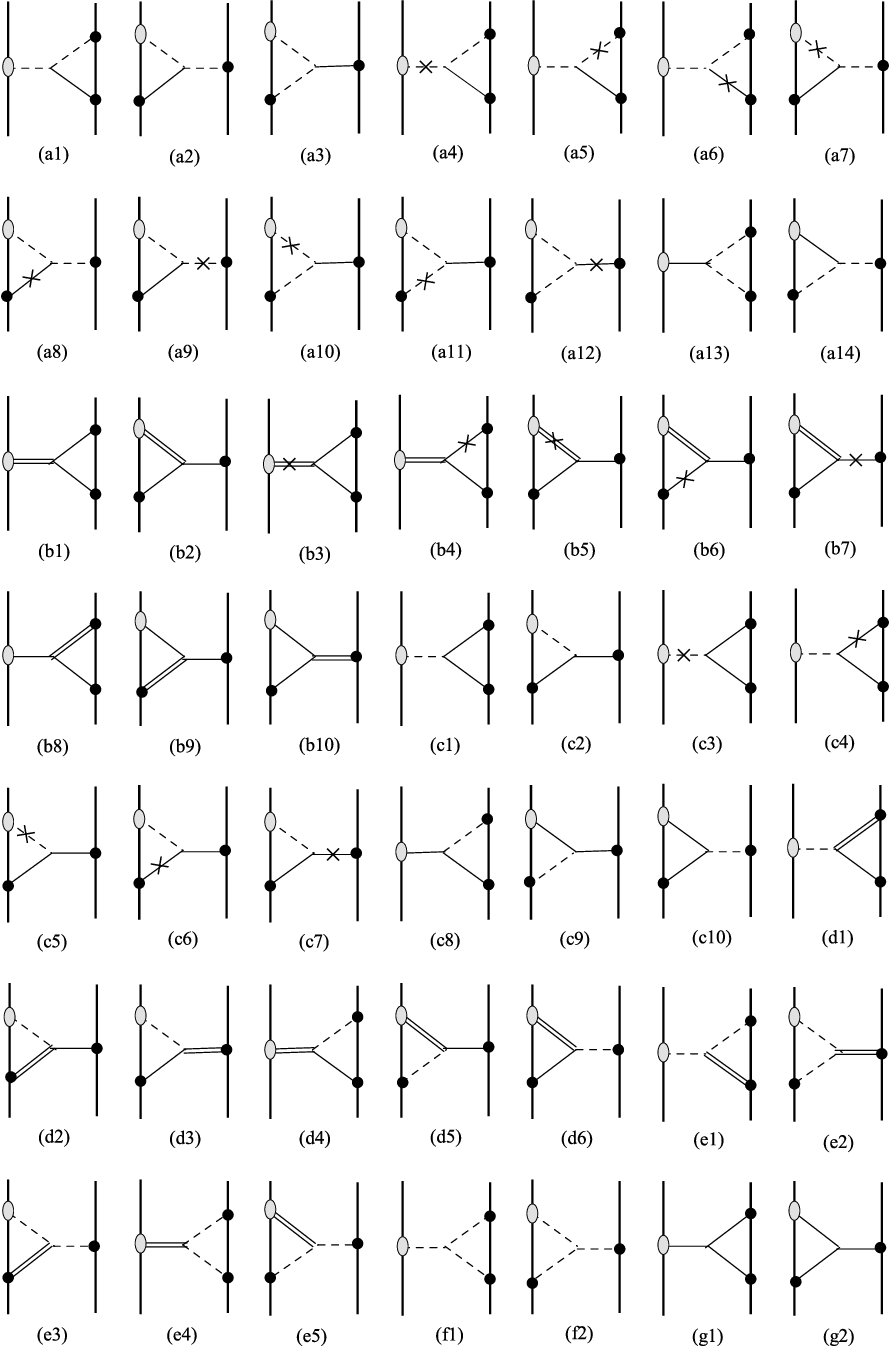}%
	\end{center}
	\caption{NNLO spin-orbit Feynman diagrams of cubic self-interaction.}
	\label{fig:sonnlo1loop}
\end{figure}

For the NNLO spin-orbit interaction we have 49 cubic self-interaction diagrams, 
as shown in figure \ref{fig:sonnlo1loop}, where the cubic vertices contain up to 
two time derivatives.

The cubic self-gravitational interaction diagrams require first the application 
of one-loop tensor integrals up to order 3, in addition to the Fourier tensor 
integrals, see appendix~A in \cite{Levi:2011eq}. Here, we encounter time 
derivatives from the spin couplings, the propagator correction vertices, and the 
time dependent cubic self-gravitational vertices.

The values of the cubic self-interaction diagrams are given as follows:
\begin{align}
\text{Fig.~3(a1)} =&
\frac{8 G^2 m_{2}^2}{r^3} \vec{S}_{1}\times\vec{n}\cdot\vec{v}_{2}
+ \frac{4 G^2 m_{2}^2}{r^3} \Big[ 3 \vec{S}_{1}\times\vec{n}\cdot\vec{v}_{1} \vec{v}_{1}\cdot\vec{v}_{2}
	 + 4 \vec{S}_{1}\times\vec{n}\cdot\vec{v}_{2} v_{2}^2 \nl
	 - \vec{S}_{1}\times\vec{v}_{1}\cdot\vec{v}_{2} \vec{v}_{1}\cdot\vec{n} \Big]
+ \frac{8 G^2 m_{2}^2}{r^2} \vec{S}_{1}\times\vec{a}_{1}\cdot\vec{v}_{2}
+ \frac{8 G^2 m_{2}^2}{r^2} \dot{\vec{S}}_{1}\times\vec{v}_{1}\cdot\vec{v}_{2} \\
\text{Fig.~3(a2)} =&
\frac{2 G^2 m_{1} m_{2}}{r^3} \vec{S}_{1}\times\vec{n}\cdot\vec{v}_{2}
+ \frac{G^2 m_{1} m_{2}}{r^3} \Big[ 7 \vec{S}_{1}\times\vec{n}\cdot\vec{v}_{1} \vec{v}_{1}\cdot\vec{v}_{2}
	 + \vec{S}_{1}\times\vec{n}\cdot\vec{v}_{2} \big( 3 v_{1}^2
	 + v_{2}^2 \big) \nl
	 - 5 \vec{S}_{1}\times\vec{v}_{1}\cdot\vec{v}_{2} \vec{v}_{1}\cdot\vec{n} \Big]
+ \frac{8 G^2 m_{1} m_{2}}{r^2} \vec{S}_{1}\times\vec{a}_{1}\cdot\vec{v}_{2}
+ \frac{8 G^2 m_{1} m_{2}}{r^2} \dot{\vec{S}}_{1}\times\vec{v}_{1}\cdot\vec{v}_{2} 
\end{align}

\clearpage

\begin{align}
\text{Fig.~3(a3)} =&
- \frac{2 G^2 m_{1} m_{2}}{r^3} \vec{S}_{1}\times\vec{n}\cdot\vec{v}_{1}
- \frac{G^2 m_{1} m_{2}}{r^3} \vec{S}_{1}\times\vec{n}\cdot\vec{v}_{1} \big( 8 v_{1}^2
	 + 3 v_{2}^2 \big) \nl
+ \frac{8 G^2 m_{1} m_{2}}{r^2} \vec{S}_{1}\times\vec{v}_{1}\cdot\vec{a}_{1} \\
\text{Fig.~3(a4)} =&
\frac{8 G^2 m_{2}^2}{r^3} \Big[ \vec{S}_{1}\times\vec{n}\cdot\vec{v}_{2} \big( \vec{v}_{1}\cdot\vec{v}_{2} -4 \vec{v}_{1}\cdot\vec{n} \vec{v}_{2}\cdot\vec{n} \big)
	 + \vec{S}_{1}\times\vec{v}_{1}\cdot\vec{v}_{2} \vec{v}_{2}\cdot\vec{n} \Big] \nl
- \frac{4 G^2 m_{2}^2}{r^2} \Big[ 2 \vec{S}_{1}\times\vec{n}\cdot\vec{a}_{2} \vec{v}_{1}\cdot\vec{n}
	 - \vec{S}_{1}\times\vec{v}_{1}\cdot\vec{a}_{2} \Big]
+ \frac{8 G^2 m_{2}^2}{r^2} \vec{v}_{2}\cdot\vec{n} \dot{\vec{S}}_{1}\times\vec{n}\cdot\vec{v}_{2} \nl
+ \frac{4 G^2 m_{2}^2}{r} \dot{\vec{S}}_{1}\times\vec{n}\cdot\vec{a}_{2}\\
\text{Fig.~3(a5)} =&
\frac{4 G^2 m_{2}^2}{r^3} \Big[ \vec{S}_{1}\times\vec{n}\cdot\vec{v}_{2} \big( \vec{v}_{1}\cdot\vec{v}_{2} -4 \vec{v}_{1}\cdot\vec{n} \vec{v}_{2}\cdot\vec{n}
	 + ( \vec{v}_{2}\cdot\vec{n} )^{2} \big)
	 + \vec{S}_{1}\times\vec{v}_{1}\cdot\vec{v}_{2} \vec{v}_{2}\cdot\vec{n} \Big] \nl
+ \frac{2 G^2 m_{2}^2}{r^2} \Big[ 2 \vec{S}_{1}\times\vec{n}\cdot\vec{v}_{2} \vec{a}_{2}\cdot\vec{n}
	 - 4 \vec{S}_{1}\times\vec{n}\cdot\vec{a}_{2} \vec{v}_{1}\cdot\vec{n}
	 + 2 \vec{S}_{1}\times\vec{v}_{1}\cdot\vec{a}_{2} \nl
	 + \vec{S}_{1}\times\vec{v}_{2}\cdot\vec{a}_{2} \Big]
+ \frac{4 G^2 m_{2}^2}{r^2} \vec{v}_{2}\cdot\vec{n} \dot{\vec{S}}_{1}\times\vec{n}\cdot\vec{v}_{2}
+ \frac{4 G^2 m_{2}^2}{r} \dot{\vec{S}}_{1}\times\vec{n}\cdot\vec{a}_{2}
\\
\text{Fig.~3(a6)} =&
\frac{4 G^2 m_{2}^2}{r^3} \Big[ \vec{S}_{1}\times\vec{n}\cdot\vec{v}_{2} \big( \vec{v}_{1}\cdot\vec{v}_{2} -4 \vec{v}_{1}\cdot\vec{n} \vec{v}_{2}\cdot\vec{n}
	 + 3 ( \vec{v}_{2}\cdot\vec{n} )^{2} \big)
	 + \vec{S}_{1}\times\vec{v}_{1}\cdot\vec{v}_{2} \vec{v}_{2}\cdot\vec{n} \Big] \nl
+ \frac{2 G^2 m_{2}^2}{r^2} \Big[ 2 \vec{S}_{1}\times\vec{n}\cdot\vec{v}_{2} \vec{a}_{2}\cdot\vec{n}
	 + \vec{S}_{1}\times\vec{v}_{2}\cdot\vec{a}_{2} \Big]
+ \frac{4 G^2 m_{2}^2}{r^2} \vec{v}_{2}\cdot\vec{n} \dot{\vec{S}}_{1}\times\vec{n}\cdot\vec{v}_{2} \\
\text{Fig.~3(a7)} =&
- \frac{G^2 m_{1} m_{2}}{r^3} \Big[ \vec{S}_{1}\times\vec{n}\cdot\vec{v}_{1} \big( \vec{v}_{1}\cdot\vec{v}_{2}
	 - v_{2}^2
	 + 4 ( \vec{v}_{2}\cdot\vec{n} )^{2} \big)
	 + 2 \vec{S}_{1}\times\vec{n}\cdot\vec{v}_{2} \big( v_{1}^2
	 - 3 \vec{v}_{1}\cdot\vec{v}_{2} \nl
	 + 12 \vec{v}_{1}\cdot\vec{n} \vec{v}_{2}\cdot\vec{n}
	 - 3 ( \vec{v}_{1}\cdot\vec{n} )^{2} \big)
	 + \vec{S}_{1}\times\vec{v}_{1}\cdot\vec{v}_{2} \big( 4 \vec{v}_{1}\cdot\vec{n}
	 - 5 \vec{v}_{2}\cdot\vec{n} \big) \Big] \nl
- \frac{G^2 m_{1} m_{2}}{r^2} \Big[ \vec{S}_{1}\times\vec{n}\cdot\vec{v}_{1} \vec{a}_{2}\cdot\vec{n}
	 + 6 \vec{S}_{1}\times\vec{n}\cdot\vec{a}_{2} \vec{v}_{1}\cdot\vec{n}
	 - 3 \vec{S}_{1}\times\vec{v}_{1}\cdot\vec{a}_{2} \Big] \nl
+ \frac{G^2 m_{1} m_{2}}{r^2} \Big[ \dot{\vec{S}}_{1}\times\vec{n}\cdot\vec{v}_{1} \vec{v}_{2}\cdot\vec{n}
	 - 6 \dot{\vec{S}}_{1}\times\vec{n}\cdot\vec{v}_{2} \big( \vec{v}_{1}\cdot\vec{n}
	 - 2 \vec{v}_{2}\cdot\vec{n} \big)
	 + 3 \dot{\vec{S}}_{1}\times\vec{v}_{1}\cdot\vec{v}_{2} \Big] \nl
+ \frac{6 G^2 m_{1} m_{2}}{r} \dot{\vec{S}}_{1}\times\vec{n}\cdot\vec{a}_{2} \\
\text{Fig.~3(a8)} =&
\frac{G^2 m_{1} m_{2}}{r^3} \Big[ \vec{S}_{1}\times\vec{n}\cdot\vec{v}_{1} \big( \vec{v}_{1}\cdot\vec{v}_{2}
	 - 3 v_{2}^2 -8 \vec{v}_{1}\cdot\vec{n} \vec{v}_{2}\cdot\vec{n}
	 + 12 ( \vec{v}_{2}\cdot\vec{n} )^{2} \big) \nl
	 + 2 \vec{S}_{1}\times\vec{n}\cdot\vec{v}_{2} \big( \vec{v}_{1}\cdot\vec{v}_{2} -4 \vec{v}_{1}\cdot\vec{n} \vec{v}_{2}\cdot\vec{n}
	 + 3 ( \vec{v}_{1}\cdot\vec{n} )^{2} \big)
	 - \vec{S}_{1}\times\vec{v}_{1}\cdot\vec{v}_{2} \big( 4 \vec{v}_{1}\cdot\vec{n} \nl
	 - 5 \vec{v}_{2}\cdot\vec{n} \big) \Big]
+ \frac{G^2 m_{1} m_{2}}{r^2} \Big[ 3 \vec{S}_{1}\times\vec{n}\cdot\vec{v}_{1} \vec{a}_{2}\cdot\vec{n}
	 - 2 \vec{S}_{1}\times\vec{n}\cdot\vec{a}_{2} \vec{v}_{1}\cdot\vec{n}
	 + \vec{S}_{1}\times\vec{v}_{1}\cdot\vec{a}_{2} \Big] \nl
+ \frac{G^2 m_{1} m_{2}}{r^2} \Big[ 3 \dot{\vec{S}}_{1}\times\vec{n}\cdot\vec{v}_{1} \vec{v}_{2}\cdot\vec{n}
	 - 2 \dot{\vec{S}}_{1}\times\vec{n}\cdot\vec{v}_{2} \vec{v}_{1}\cdot\vec{n}
	 + \dot{\vec{S}}_{1}\times\vec{v}_{1}\cdot\vec{v}_{2} \Big]\\
\text{Fig.~3(a9)} =&
\frac{2 G^2 m_{1} m_{2}}{r^3} \Big[ \vec{S}_{1}\times\vec{n}\cdot\vec{v}_{2} \big( \vec{v}_{1}\cdot\vec{v}_{2} -4 \vec{v}_{1}\cdot\vec{n} \vec{v}_{2}\cdot\vec{n} \big)
	 + \vec{S}_{1}\times\vec{v}_{1}\cdot\vec{v}_{2} \vec{v}_{2}\cdot\vec{n} \Big] \nl
- \frac{G^2 m_{1} m_{2}}{r^2} \Big[ 2 \vec{S}_{1}\times\vec{n}\cdot\vec{a}_{2} \vec{v}_{1}\cdot\vec{n}
	 - \vec{S}_{1}\times\vec{v}_{1}\cdot\vec{a}_{2} \Big]
+ \frac{2 G^2 m_{1} m_{2}}{r^2} \vec{v}_{2}\cdot\vec{n} \dot{\vec{S}}_{1}\times\vec{n}\cdot\vec{v}_{2} \nl
+ \frac{G^2 m_{1} m_{2}}{r} \dot{\vec{S}}_{1}\times\vec{n}\cdot\vec{a}_{2}\\ 
\text{Fig.~3(a10)} =&
\frac{G^2 m_{1} m_{2}}{r^3} \Big[ \vec{S}_{1}\times\vec{n}\cdot\vec{v}_{1} \big( v_{1}^2
	 - \vec{v}_{1}\cdot\vec{v}_{2}
	 + 4 \vec{v}_{1}\cdot\vec{n} \vec{v}_{2}\cdot\vec{n}
	 - 2 ( \vec{v}_{1}\cdot\vec{n} )^{2} \big) \nl
	 + \vec{S}_{1}\times\vec{v}_{1}\cdot\vec{v}_{2} \vec{v}_{1}\cdot\vec{n} \Big]
+ \frac{G^2 m_{1} m_{2}}{r^2} \Big[ \vec{S}_{1}\times\vec{n}\cdot\vec{v}_{1} \vec{a}_{1}\cdot\vec{n}
	 - \vec{S}_{1}\times\vec{v}_{1}\cdot\vec{a}_{1} \Big] \nl
+ \frac{G^2 m_{1} m_{2}}{r^2} \Big[ \dot{\vec{S}}_{1}\times\vec{n}\cdot\vec{v}_{1} \big( 3 \vec{v}_{1}\cdot\vec{n}
	 - 4 \vec{v}_{2}\cdot\vec{n} \big)
	 - 2 \dot{\vec{S}}_{1}\times\vec{v}_{1}\cdot\vec{v}_{2} \Big] \nl
- \frac{2 G^2 m_{1} m_{2}}{r} \dot{\vec{S}}_{1}\times\vec{n}\cdot\vec{a}_{1} \\
\text{Fig.~3(a11)} =&
- \frac{G^2 m_{1} m_{2}}{r^3} \Big[ \vec{S}_{1}\times\vec{n}\cdot\vec{v}_{1} \big( v_{1}^2
	 - \vec{v}_{1}\cdot\vec{v}_{2}
	 + 4 \vec{v}_{1}\cdot\vec{n} \vec{v}_{2}\cdot\vec{n}
	 - 2 ( \vec{v}_{1}\cdot\vec{n} )^{2} \big) \nl
	 + \vec{S}_{1}\times\vec{v}_{1}\cdot\vec{v}_{2} \vec{v}_{1}\cdot\vec{n} \Big]
+ \frac{G^2 m_{1} m_{2}}{r^2} \Big[ 3 \vec{S}_{1}\times\vec{n}\cdot\vec{v}_{1} \vec{a}_{1}\cdot\vec{n}
	 - 2 \vec{S}_{1}\times\vec{n}\cdot\vec{a}_{1} \big( 3 \vec{v}_{1}\cdot\vec{n} \nl
	 - 2 \vec{v}_{2}\cdot\vec{n} \big)
	 + 3 \vec{S}_{1}\times\vec{v}_{1}\cdot\vec{a}_{1}
	 + 2 \vec{S}_{1}\times\vec{a}_{1}\cdot\vec{v}_{2} \Big]
- \frac{G^2 m_{1} m_{2}}{r^2} \vec{v}_{1}\cdot\vec{n} \dot{\vec{S}}_{1}\times\vec{n}\cdot\vec{v}_{1} \nl
+ \frac{2 G^2 m_{1} m_{2}}{r} \dot{\vec{S}}_{1}\times\vec{n}\cdot\vec{a}_{1} \\
\text{Fig.~3(a12)} =&
- \frac{2 G^2 m_{1} m_{2}}{r^3} \Big[ \vec{S}_{1}\times\vec{n}\cdot\vec{v}_{1} \big( \vec{v}_{1}\cdot\vec{v}_{2} -4 \vec{v}_{1}\cdot\vec{n} \vec{v}_{2}\cdot\vec{n} \big)
	 - \vec{S}_{1}\times\vec{v}_{1}\cdot\vec{v}_{2} \vec{v}_{1}\cdot\vec{n} \Big] \nl
- \frac{G^2 m_{1} m_{2}}{r^2} \Big[ 2 \vec{S}_{1}\times\vec{n}\cdot\vec{a}_{1} \vec{v}_{2}\cdot\vec{n}
	 + \vec{S}_{1}\times\vec{a}_{1}\cdot\vec{v}_{2} \Big]
- \frac{G^2 m_{1} m_{2}}{r^2} \Big[ 2 \dot{\vec{S}}_{1}\times\vec{n}\cdot\vec{v}_{1} \vec{v}_{2}\cdot\vec{n} \nl
	 + \dot{\vec{S}}_{1}\times\vec{v}_{1}\cdot\vec{v}_{2} \Big]\\
\text{Fig.~3(a13)} =&
- \frac{16 G^2 m_{2}^2}{r^3} \vec{S}_{1}\times\vec{n}\cdot\vec{v}_{1} v_{2}^2 \\
\text{Fig.~3(a14)} =&
\frac{8 G^2 m_{1} m_{2}}{r^3} \Big[ 2 \vec{S}_{1}\times\vec{n}\cdot\vec{v}_{1} \vec{v}_{1}\cdot\vec{v}_{2}
	 + \vec{S}_{1}\times\vec{v}_{1}\cdot\vec{v}_{2} \vec{v}_{1}\cdot\vec{n} \Big]\\
\text{Fig.~3(b1)} =&
\frac{G^2 m_{2}^2}{2 r^3} \vec{S}_{1}\times\vec{n}\cdot\vec{v}_{1}
+ \frac{G^2 m_{2}^2}{4 r^3} \Big[ \vec{S}_{1}\times\vec{n}\cdot\vec{v}_{1} \big( 3 v_{1}^2
	 + 2 v_{2}^2 -8 ( \vec{v}_{1}\cdot\vec{n} )^{2}
	 + 8 ( \vec{v}_{2}\cdot\vec{n} )^{2} \big) \nl
	 + 4 \vec{S}_{1}\times\vec{v}_{1}\cdot\vec{v}_{2} \vec{v}_{2}\cdot\vec{n} \Big]
+ \frac{G^2 m_{2}^2}{2 r^2} \Big[ \vec{S}_{1}\times\vec{n}\cdot\vec{v}_{1} \vec{a}_{1}\cdot\vec{n}
	 + 2 \vec{S}_{1}\times\vec{n}\cdot\vec{a}_{1} \vec{v}_{1}\cdot\vec{n} \nl
	 - \vec{S}_{1}\times\vec{v}_{1}\cdot\vec{a}_{1} \Big]
+ \frac{3 G^2 m_{2}^2}{4 r^2} \vec{v}_{1}\cdot\vec{n} \dot{\vec{S}}_{1}\times\vec{n}\cdot\vec{v}_{1} \\
\text{Fig.~3(b2)} =&
\frac{G^2 m_{1} m_{2}}{2 r^3} \vec{S}_{1}\times\vec{n}\cdot\vec{v}_{1}
- \frac{G^2 m_{1} m_{2}}{4 r^3} \Big[ \vec{S}_{1}\times\vec{n}\cdot\vec{v}_{1} \big( 2 v_{1}^2
	 - 4 \vec{v}_{1}\cdot\vec{v}_{2}
	 - 3 v_{2}^2 \nl
	 + 16 \vec{v}_{1}\cdot\vec{n} \vec{v}_{2}\cdot\vec{n}
	 - 16 ( \vec{v}_{1}\cdot\vec{n} )^{2} \big)
	 + 4 \vec{S}_{1}\times\vec{v}_{1}\cdot\vec{v}_{2} \vec{v}_{1}\cdot\vec{n} \Big] \nl
- \frac{2 G^2 m_{1} m_{2}}{r^2} \Big[ 2 \vec{S}_{1}\times\vec{n}\cdot\vec{v}_{1} \vec{a}_{1}\cdot\vec{n}
	 + 4 \vec{S}_{1}\times\vec{n}\cdot\vec{a}_{1} \vec{v}_{1}\cdot\vec{n}
	 - \vec{S}_{1}\times\vec{v}_{1}\cdot\vec{a}_{1} \Big] \nl
- \frac{6 G^2 m_{1} m_{2}}{r^2} \vec{v}_{1}\cdot\vec{n} \dot{\vec{S}}_{1}\times\vec{n}\cdot\vec{v}_{1} \\
\text{Fig.~3(b3)} =&
\frac{G^2 m_{2}^2}{2 r^3} \Big[ \vec{S}_{1}\times\vec{n}\cdot\vec{v}_{1} \big( \vec{v}_{1}\cdot\vec{v}_{2} -4 \vec{v}_{1}\cdot\vec{n} \vec{v}_{2}\cdot\vec{n} \big)
	 - \vec{S}_{1}\times\vec{v}_{1}\cdot\vec{v}_{2} \vec{v}_{1}\cdot\vec{n} \Big] \nl
+ \frac{G^2 m_{2}^2}{4 r^2} \Big[ 2 \vec{S}_{1}\times\vec{n}\cdot\vec{a}_{1} \vec{v}_{2}\cdot\vec{n}
	 + \vec{S}_{1}\times\vec{a}_{1}\cdot\vec{v}_{2} \Big] \nl
+ \frac{G^2 m_{2}^2}{4 r^2} \Big[ 2 \dot{\vec{S}}_{1}\times\vec{n}\cdot\vec{v}_{1} \vec{v}_{2}\cdot\vec{n}
	 + \dot{\vec{S}}_{1}\times\vec{v}_{1}\cdot\vec{v}_{2} \Big]\\
\text{Fig.~3(b4)} =&
- \frac{G^2 m_{2}^2}{2 r^3} \Big[ \vec{S}_{1}\times\vec{n}\cdot\vec{v}_{1} \big( \vec{v}_{1}\cdot\vec{v}_{2}
	 - v_{2}^2 -4 \vec{v}_{1}\cdot\vec{n} \vec{v}_{2}\cdot\vec{n}
	 + 2 ( \vec{v}_{2}\cdot\vec{n} )^{2} \big)
	 - \vec{S}_{1}\times\vec{n}\cdot\vec{v}_{2} \big( v_{1}^2 \nl
	 + 4 \vec{v}_{1}\cdot\vec{n} \vec{v}_{2}\cdot\vec{n}
	 - 4 ( \vec{v}_{1}\cdot\vec{n} )^{2} \big)
	 - \vec{S}_{1}\times\vec{v}_{1}\cdot\vec{v}_{2} \big( 3 \vec{v}_{1}\cdot\vec{n}
	 - 2 \vec{v}_{2}\cdot\vec{n} \big) \Big] \nl
- \frac{G^2 m_{2}^2}{2 r^2} \Big[ \vec{S}_{1}\times\vec{n}\cdot\vec{a}_{1} \vec{v}_{2}\cdot\vec{n}
	 - \vec{S}_{1}\times\vec{n}\cdot\vec{v}_{2} \vec{a}_{1}\cdot\vec{n}
	 + \vec{S}_{1}\times\vec{a}_{1}\cdot\vec{v}_{2} \Big] \nl
- \frac{G^2 m_{2}^2}{2 r^2} \Big[ \dot{\vec{S}}_{1}\times\vec{n}\cdot\vec{v}_{1} \vec{v}_{2}\cdot\vec{n}
	 - \dot{\vec{S}}_{1}\times\vec{n}\cdot\vec{v}_{2} \vec{v}_{1}\cdot\vec{n}
	 + \dot{\vec{S}}_{1}\times\vec{v}_{1}\cdot\vec{v}_{2} \Big]\\
\text{Fig.~3(b5)} =&
- \frac{G^2 m_{1} m_{2}}{2 r^3} \Big[ \vec{S}_{1}\times\vec{n}\cdot\vec{v}_{1} \big( v_{1}^2
	 - 3 \vec{v}_{1}\cdot\vec{v}_{2}
	 + 12 \vec{v}_{1}\cdot\vec{n} \vec{v}_{2}\cdot\vec{n}
	 - 7 ( \vec{v}_{1}\cdot\vec{n} )^{2} \big) \nl
	 + 3 \vec{S}_{1}\times\vec{v}_{1}\cdot\vec{v}_{2} \vec{v}_{1}\cdot\vec{n} \Big]
+ \frac{G^2 m_{1} m_{2}}{4 r^2} \Big[ 7 \vec{S}_{1}\times\vec{n}\cdot\vec{v}_{1} \vec{a}_{1}\cdot\vec{n}
	 + \vec{S}_{1}\times\vec{n}\cdot\vec{a}_{1} \big( 3 \vec{v}_{1}\cdot\vec{n} \nl
	 - 4 \vec{v}_{2}\cdot\vec{n} \big)
	 + 2 \vec{S}_{1}\times\vec{v}_{1}\cdot\vec{a}_{1}
	 - 2 \vec{S}_{1}\times\vec{a}_{1}\cdot\vec{v}_{2} \Big]
+ \frac{G^2 m_{1} m_{2}}{2 r^2} \Big[ \dot{\vec{S}}_{1}\times\vec{n}\cdot\vec{v}_{1} \big( 5 \vec{v}_{1}\cdot\vec{n} \nl
	 - 2 \vec{v}_{2}\cdot\vec{n} \big)
	 - \dot{\vec{S}}_{1}\times\vec{v}_{1}\cdot\vec{v}_{2} \Big]\\
\text{Fig.~3(b6)} =&
\frac{G^2 m_{1} m_{2}}{2 r^3} \Big[ \vec{S}_{1}\times\vec{n}\cdot\vec{v}_{1} \big( 2 v_{1}^2
	 - 3 \vec{v}_{1}\cdot\vec{v}_{2}
	 + 12 \vec{v}_{1}\cdot\vec{n} \vec{v}_{2}\cdot\vec{n}
	 - 5 ( \vec{v}_{1}\cdot\vec{n} )^{2} \big) \nl
	 + 3 \vec{S}_{1}\times\vec{v}_{1}\cdot\vec{v}_{2} \vec{v}_{1}\cdot\vec{n} \Big]
+ \frac{G^2 m_{1} m_{2}}{4 r^2} \Big[ 5 \vec{S}_{1}\times\vec{n}\cdot\vec{v}_{1} \vec{a}_{1}\cdot\vec{n}
	 + \vec{S}_{1}\times\vec{n}\cdot\vec{a}_{1} \vec{v}_{1}\cdot\vec{n} \nl
	 + 2 \vec{S}_{1}\times\vec{v}_{1}\cdot\vec{a}_{1} \Big]
+ \frac{3 G^2 m_{1} m_{2}}{2 r^2} \vec{v}_{1}\cdot\vec{n} \dot{\vec{S}}_{1}\times\vec{n}\cdot\vec{v}_{1} \\
\text{Fig.~3(b7)} =&
\frac{G^2 m_{1} m_{2}}{2 r^3} \Big[ \vec{S}_{1}\times\vec{n}\cdot\vec{v}_{1} \big( \vec{v}_{1}\cdot\vec{v}_{2} -4 \vec{v}_{1}\cdot\vec{n} \vec{v}_{2}\cdot\vec{n} \big)
	 - \vec{S}_{1}\times\vec{v}_{1}\cdot\vec{v}_{2} \vec{v}_{1}\cdot\vec{n} \Big] \nl
+ \frac{G^2 m_{1} m_{2}}{4 r^2} \Big[ 2 \vec{S}_{1}\times\vec{n}\cdot\vec{a}_{1} \vec{v}_{2}\cdot\vec{n}
	 + \vec{S}_{1}\times\vec{a}_{1}\cdot\vec{v}_{2} \Big]
+ \frac{G^2 m_{1} m_{2}}{4 r^2} \Big[ 2 \dot{\vec{S}}_{1}\times\vec{n}\cdot\vec{v}_{1} \vec{v}_{2}\cdot\vec{n} \nl
	 + \dot{\vec{S}}_{1}\times\vec{v}_{1}\cdot\vec{v}_{2} \Big]\\
\text{Fig.~3(b8)} =&
- \frac{8 G^2 m_{2}^2}{r^3} \Big[ \vec{S}_{1}\times\vec{n}\cdot\vec{v}_{1} \big( v_{2}^2 -4 ( \vec{v}_{2}\cdot\vec{n} )^{2} \big)
	 - 2 \vec{S}_{1}\times\vec{v}_{1}\cdot\vec{v}_{2} \vec{v}_{2}\cdot\vec{n} \Big]\\
\text{Fig.~3(b9)} =&
- \frac{6 G^2 m_{1} m_{2}}{r^3} \vec{S}_{1}\times\vec{n}\cdot\vec{v}_{1} \big( v_{1}^2 -4 ( \vec{v}_{1}\cdot\vec{n} )^{2} \big)\\
\text{Fig.~3(b10)} =&
\frac{2 G^2 m_{1} m_{2}}{r^3} \Big[ \vec{S}_{1}\times\vec{n}\cdot\vec{v}_{1} \big( v_{2}^2 -2 ( \vec{v}_{2}\cdot\vec{n} )^{2} \big)
	 - \vec{S}_{1}\times\vec{v}_{1}\cdot\vec{v}_{2} \vec{v}_{2}\cdot\vec{n} \Big]\\
\text{Fig.~3(c1)} =&
- \frac{G^2 m_{2}^2}{2 r^3} \vec{S}_{1}\times\vec{n}\cdot\vec{v}_{2}
- \frac{G^2 m_{2}^2}{4 r^3} \Big[ \vec{S}_{1}\times\vec{n}\cdot\vec{v}_{1} \big( 7 \vec{v}_{1}\cdot\vec{v}_{2} -16 \vec{v}_{1}\cdot\vec{n} \vec{v}_{2}\cdot\vec{n} \big) \nl
	 + 6 \vec{S}_{1}\times\vec{n}\cdot\vec{v}_{2} v_{2}^2
	 - 5 \vec{S}_{1}\times\vec{v}_{1}\cdot\vec{v}_{2} \vec{v}_{1}\cdot\vec{n} \Big]
- \frac{G^2 m_{2}^2}{r^2} \Big[ \vec{S}_{1}\times\vec{n}\cdot\vec{a}_{1} \vec{v}_{2}\cdot\vec{n} \nl
	 + \vec{S}_{1}\times\vec{a}_{1}\cdot\vec{v}_{2} \Big]
- \frac{G^2 m_{2}^2}{r^2} \Big[ \dot{\vec{S}}_{1}\times\vec{n}\cdot\vec{v}_{1} \vec{v}_{2}\cdot\vec{n}
	 + \dot{\vec{S}}_{1}\times\vec{v}_{1}\cdot\vec{v}_{2} \Big]\\
\text{Fig.~3(c2)} =&
- \frac{G^2 m_{1} m_{2}}{2 r^3} \vec{S}_{1}\times\vec{n}\cdot\vec{v}_{1}
+ \frac{G^2 m_{1} m_{2}}{4 r^3} \Big[ \vec{S}_{1}\times\vec{n}\cdot\vec{v}_{1} \big( 2 v_{1}^2
	 + 4 \vec{v}_{1}\cdot\vec{v}_{2}
	 - 3 v_{2}^2 \nl
         -16 \vec{v}_{1}\cdot\vec{n} \vec{v}_{2}\cdot\vec{n}
	 - 16 ( \vec{v}_{1}\cdot\vec{n} )^{2} \big)
	 - 4 \vec{S}_{1}\times\vec{v}_{1}\cdot\vec{v}_{2} \vec{v}_{1}\cdot\vec{n} \Big] \nl
+ \frac{2 G^2 m_{1} m_{2}}{r^2} \Big[ 2 \vec{S}_{1}\times\vec{n}\cdot\vec{a}_{1} \big( \vec{v}_{1}\cdot\vec{n}
	 + \vec{v}_{2}\cdot\vec{n} \big)
	 - \vec{S}_{1}\times\vec{v}_{1}\cdot\vec{a}_{1}
	 + \vec{S}_{1}\times\vec{a}_{1}\cdot\vec{v}_{2} \Big] \nl
+ \frac{2 G^2 m_{1} m_{2}}{r^2} \Big[ 2 \dot{\vec{S}}_{1}\times\vec{n}\cdot\vec{v}_{1} \big( \vec{v}_{1}\cdot\vec{n}
	 + \vec{v}_{2}\cdot\vec{n} \big)
	 + \dot{\vec{S}}_{1}\times\vec{v}_{1}\cdot\vec{v}_{2} \Big]\\
\text{Fig.~3(c3)} =&
- \frac{G^2 m_{2}^2}{2 r^3} \Big[ \vec{S}_{1}\times\vec{n}\cdot\vec{v}_{2} \big( \vec{v}_{1}\cdot\vec{v}_{2} -4 \vec{v}_{1}\cdot\vec{n} \vec{v}_{2}\cdot\vec{n} \big)
	 + \vec{S}_{1}\times\vec{v}_{1}\cdot\vec{v}_{2} \vec{v}_{2}\cdot\vec{n} \Big] \nl
+ \frac{G^2 m_{2}^2}{4 r^2} \Big[ 2 \vec{S}_{1}\times\vec{n}\cdot\vec{a}_{2} \vec{v}_{1}\cdot\vec{n}
	 - \vec{S}_{1}\times\vec{v}_{1}\cdot\vec{a}_{2} \Big]
- \frac{G^2 m_{2}^2}{2 r^2} \vec{v}_{2}\cdot\vec{n} \dot{\vec{S}}_{1}\times\vec{n}\cdot\vec{v}_{2} \nl
- \frac{G^2 m_{2}^2}{4 r} \dot{\vec{S}}_{1}\times\vec{n}\cdot\vec{a}_{2} \\
\text{Fig.~3(c4)} =&
- \frac{G^2 m_{2}^2}{2 r^3} \vec{S}_{1}\times\vec{n}\cdot\vec{v}_{2} \big( v_{2}^2
	 + 2 ( \vec{v}_{2}\cdot\vec{n} )^{2} \big)
- \frac{G^2 m_{2}^2}{2 r^2} \Big[ 2 \vec{S}_{1}\times\vec{n}\cdot\vec{v}_{2} \vec{a}_{2}\cdot\vec{n} \nl
	 + 2 \vec{S}_{1}\times\vec{n}\cdot\vec{a}_{2} \big( \vec{v}_{1}\cdot\vec{n}
	 - \vec{v}_{2}\cdot\vec{n} \big)
	 - \vec{S}_{1}\times\vec{v}_{1}\cdot\vec{a}_{2}
	 + 2 \vec{S}_{1}\times\vec{v}_{2}\cdot\vec{a}_{2} \Big] \nl
+ \frac{G^2 m_{2}^2}{2 r} \dot{\vec{S}}_{1}\times\vec{n}\cdot\vec{a}_{2} \\
\text{Fig.~3(c5)} =&
- \frac{G^2 m_{1} m_{2}}{2 r^3} \Big[ \vec{S}_{1}\times\vec{n}\cdot\vec{v}_{1} \big( 2 v_{2}^2
	 + 3 ( \vec{v}_{1}\cdot\vec{n} )^{2}
	 - 8 ( \vec{v}_{2}\cdot\vec{n} )^{2} \big)
	 - 4 \vec{S}_{1}\times\vec{v}_{1}\cdot\vec{v}_{2} \vec{v}_{2}\cdot\vec{n} \Big] \nl
+ \frac{G^2 m_{1} m_{2}}{4 r^2} \Big[ \vec{S}_{1}\times\vec{n}\cdot\vec{v}_{1} \big( 3 \vec{a}_{1}\cdot\vec{n}
	 + 4 \vec{a}_{2}\cdot\vec{n} \big)
	 - \vec{S}_{1}\times\vec{n}\cdot\vec{a}_{1} \vec{v}_{1}\cdot\vec{n}
	 + 2 \vec{S}_{1}\times\vec{v}_{1}\cdot\vec{a}_{1} \nl
	 + 2 \vec{S}_{1}\times\vec{v}_{1}\cdot\vec{a}_{2} \Big]
- \frac{3 G^2 m_{1} m_{2}}{2 r^2} \vec{v}_{1}\cdot\vec{n} \dot{\vec{S}}_{1}\times\vec{n}\cdot\vec{v}_{1}
+ \frac{G^2 m_{1} m_{2}}{2 r} \dot{\vec{S}}_{1}\times\vec{n}\cdot\vec{a}_{1} \\
\text{Fig.~3(c6)} =&
- \frac{G^2 m_{1} m_{2}}{2 r^3} \Big[ \vec{S}_{1}\times\vec{n}\cdot\vec{v}_{1} \big( v_{1}^2
	 - 2 v_{2}^2
	 - ( \vec{v}_{1}\cdot\vec{n} )^{2}
	 + 8 ( \vec{v}_{2}\cdot\vec{n} )^{2} \big)
	 + 4 \vec{S}_{1}\times\vec{v}_{1}\cdot\vec{v}_{2} \vec{v}_{2}\cdot\vec{n} \Big] \nl
+ \frac{G^2 m_{1} m_{2}}{4 r^2} \Big[ \vec{S}_{1}\times\vec{n}\cdot\vec{v}_{1} \big( \vec{a}_{1}\cdot\vec{n}
	 - 4 \vec{a}_{2}\cdot\vec{n} \big)
	 - \vec{S}_{1}\times\vec{n}\cdot\vec{a}_{1} \big( 3 \vec{v}_{1}\cdot\vec{n}
	 - 4 \vec{v}_{2}\cdot\vec{n} \big) \nl
	 + 2 \vec{S}_{1}\times\vec{v}_{1}\cdot\vec{a}_{1}
	 + 2 \vec{S}_{1}\times\vec{a}_{1}\cdot\vec{v}_{2}
	 - 2 \vec{S}_{1}\times\vec{v}_{1}\cdot\vec{a}_{2} \Big]
- \frac{G^2 m_{1} m_{2}}{2 r^2} \Big[ \dot{\vec{S}}_{1}\times\vec{n}\cdot\vec{v}_{1} \big( \vec{v}_{1}\cdot\vec{n} \nl
	 + 2 \vec{v}_{2}\cdot\vec{n} \big)
	 + \dot{\vec{S}}_{1}\times\vec{v}_{1}\cdot\vec{v}_{2} \Big]
+ \frac{G^2 m_{1} m_{2}}{2 r} \dot{\vec{S}}_{1}\times\vec{n}\cdot\vec{a}_{1} \\
\text{Fig.~3(c7)} =&
- \frac{G^2 m_{1} m_{2}}{2 r^3} \Big[ \vec{S}_{1}\times\vec{n}\cdot\vec{v}_{1} \big( \vec{v}_{1}\cdot\vec{v}_{2} -4 \vec{v}_{1}\cdot\vec{n} \vec{v}_{2}\cdot\vec{n} \big)
	 - \vec{S}_{1}\times\vec{v}_{1}\cdot\vec{v}_{2} \vec{v}_{1}\cdot\vec{n} \Big] \nl
- \frac{G^2 m_{1} m_{2}}{4 r^2} \Big[ 2 \vec{S}_{1}\times\vec{n}\cdot\vec{a}_{1} \vec{v}_{2}\cdot\vec{n}
	 + \vec{S}_{1}\times\vec{a}_{1}\cdot\vec{v}_{2} \Big] \nl
- \frac{G^2 m_{1} m_{2}}{4 r^2} \Big[ 2 \dot{\vec{S}}_{1}\times\vec{n}\cdot\vec{v}_{1} \vec{v}_{2}\cdot\vec{n}
	 + \dot{\vec{S}}_{1}\times\vec{v}_{1}\cdot\vec{v}_{2} \Big]\\
\text{Fig.~3(c8)} =&
\frac{8 G^2 m_{2}^2}{r^3} \Big[ \vec{S}_{1}\times\vec{n}\cdot\vec{v}_{1} \big( \vec{v}_{1}\cdot\vec{v}_{2}
	 + v_{2}^2 -4 \vec{v}_{1}\cdot\vec{n} \vec{v}_{2}\cdot\vec{n}
	 - 4 ( \vec{v}_{2}\cdot\vec{n} )^{2} \big)
	 - \vec{S}_{1}\times\vec{v}_{1}\cdot\vec{v}_{2} \big( \vec{v}_{1}\cdot\vec{n} \nl
	 + 2 \vec{v}_{2}\cdot\vec{n} \big) \Big]
+ \frac{4 G^2 m_{2}^2}{r^2} \Big[ 2 \vec{S}_{1}\times\vec{n}\cdot\vec{a}_{1} \vec{v}_{2}\cdot\vec{n}
	 + \vec{S}_{1}\times\vec{a}_{1}\cdot\vec{v}_{2} \Big] \nl
+ \frac{4 G^2 m_{2}^2}{r^2} \Big[ 2 \dot{\vec{S}}_{1}\times\vec{n}\cdot\vec{v}_{1} \vec{v}_{2}\cdot\vec{n}
	 + \dot{\vec{S}}_{1}\times\vec{v}_{1}\cdot\vec{v}_{2} \Big]\\
\text{Fig.~3(c9)} =&
\frac{6 G^2 m_{1} m_{2}}{r^3} \Big[ \vec{S}_{1}\times\vec{n}\cdot\vec{v}_{1} \big( v_{1}^2
	 + \vec{v}_{1}\cdot\vec{v}_{2} -4 \vec{v}_{1}\cdot\vec{n} \vec{v}_{2}\cdot\vec{n}
	 - 4 ( \vec{v}_{1}\cdot\vec{n} )^{2} \big) \nl
	 - \vec{S}_{1}\times\vec{v}_{1}\cdot\vec{v}_{2} \vec{v}_{1}\cdot\vec{n} \Big]
+ \frac{4 G^2 m_{1} m_{2}}{r^2} \Big[ 2 \vec{S}_{1}\times\vec{n}\cdot\vec{a}_{1} \vec{v}_{1}\cdot\vec{n}
	 - \vec{S}_{1}\times\vec{v}_{1}\cdot\vec{a}_{1} \Big] \nl
+ \frac{8 G^2 m_{1} m_{2}}{r^2} \vec{v}_{1}\cdot\vec{n} \dot{\vec{S}}_{1}\times\vec{n}\cdot\vec{v}_{1} \\
\text{Fig.~3(c10)} =&
- \frac{2 G^2 m_{1} m_{2}}{r^3} \Big[ 2 \vec{S}_{1}\times\vec{n}\cdot\vec{v}_{1} \big( \vec{v}_{1}\cdot\vec{v}_{2} -2 \vec{v}_{1}\cdot\vec{n} \vec{v}_{2}\cdot\vec{n} \big)
	 - \vec{S}_{1}\times\vec{v}_{1}\cdot\vec{v}_{2} \vec{v}_{1}\cdot\vec{n} \Big] \nl
- \frac{2 G^2 m_{1} m_{2}}{r^2} \Big[ \vec{S}_{1}\times\vec{n}\cdot\vec{a}_{1} \vec{v}_{2}\cdot\vec{n}
	 + \vec{S}_{1}\times\vec{a}_{1}\cdot\vec{v}_{2} \Big]
- \frac{2 G^2 m_{1} m_{2}}{r^2} \Big[ \dot{\vec{S}}_{1}\times\vec{n}\cdot\vec{v}_{1} \vec{v}_{2}\cdot\vec{n} \nl
	 + \dot{\vec{S}}_{1}\times\vec{v}_{1}\cdot\vec{v}_{2} \Big]\\
\text{Fig.~3(d1)} =&
\frac{8 G^2 m_{2}^2}{r^3} \Big[ \vec{S}_{1}\times\vec{n}\cdot\vec{v}_{2} \big( \vec{v}_{1}\cdot\vec{v}_{2}
	 - v_{2}^2 -4 \vec{v}_{1}\cdot\vec{n} \vec{v}_{2}\cdot\vec{n}
	 + 4 ( \vec{v}_{2}\cdot\vec{n} )^{2} \big) \nl
	 + \vec{S}_{1}\times\vec{v}_{1}\cdot\vec{v}_{2} \vec{v}_{2}\cdot\vec{n} \Big]
+ \frac{8 G^2 m_{2}^2}{r^2} \vec{v}_{2}\cdot\vec{n} \dot{\vec{S}}_{1}\times\vec{n}\cdot\vec{v}_{2} \\
\text{Fig.~3(d2)} =&
\frac{6 G^2 m_{1} m_{2}}{r^3} \Big[ \vec{S}_{1}\times\vec{n}\cdot\vec{v}_{1} \big( v_{1}^2
	 - \vec{v}_{1}\cdot\vec{v}_{2}
	 + 4 \vec{v}_{1}\cdot\vec{n} \vec{v}_{2}\cdot\vec{n}
	 - 4 ( \vec{v}_{1}\cdot\vec{n} )^{2} \big) \nl
	 + \vec{S}_{1}\times\vec{v}_{1}\cdot\vec{v}_{2} \vec{v}_{1}\cdot\vec{n} \Big]
+ \frac{8 G^2 m_{1} m_{2}}{r^2} \vec{v}_{1}\cdot\vec{n} \dot{\vec{S}}_{1}\times\vec{n}\cdot\vec{v}_{1} \\
\text{Fig.~3(d3)} =&
\frac{2 G^2 m_{1} m_{2}}{r^3} \vec{v}_{2}\cdot\vec{n} \vec{S}_{1}\times\vec{v}_{1}\cdot\vec{v}_{2}
- \frac{2 G^2 m_{1} m_{2}}{r^2} \vec{v}_{2}\cdot\vec{n} \dot{\vec{S}}_{1}\times\vec{n}\cdot\vec{v}_{2} \\
\text{Fig.~3(d4)} =&
\frac{4 G^2 m_{2}^2}{r^2} \Big[ 2 \vec{S}_{1}\times\vec{n}\cdot\vec{a}_{2} \vec{v}_{1}\cdot\vec{n}
	 - \vec{S}_{1}\times\vec{v}_{1}\cdot\vec{a}_{2} \Big]\\
\text{Fig.~3(d5)} =&
- \frac{2 G^2 m_{1} m_{2}}{r^3} \Big[ \vec{S}_{1}\times\vec{n}\cdot\vec{v}_{1} \big( \vec{v}_{1}\cdot\vec{v}_{2} -4 \vec{v}_{1}\cdot\vec{n} \vec{v}_{2}\cdot\vec{n}
	 + 4 ( \vec{v}_{1}\cdot\vec{n} )^{2} \big) \nl
	 - \vec{S}_{1}\times\vec{v}_{1}\cdot\vec{v}_{2} \vec{v}_{1}\cdot\vec{n} \Big]
+ \frac{4 G^2 m_{1} m_{2}}{r^2} \Big[ 2 \vec{S}_{1}\times\vec{n}\cdot\vec{a}_{1} \vec{v}_{1}\cdot\vec{n}
	 - \vec{S}_{1}\times\vec{v}_{1}\cdot\vec{a}_{1} \Big]\\
\text{Fig.~3(d6)} =&
\frac{2 G^2 m_{1} m_{2}}{r^3} \Big[ \vec{S}_{1}\times\vec{n}\cdot\vec{v}_{1} \vec{v}_{1}\cdot\vec{v}_{2}
	 - \vec{S}_{1}\times\vec{n}\cdot\vec{v}_{2} \big( v_{1}^2
	 - \vec{v}_{1}\cdot\vec{v}_{2}
	 + 4 \vec{v}_{1}\cdot\vec{n} \vec{v}_{2}\cdot\vec{n} \nl
	 - 4 ( \vec{v}_{1}\cdot\vec{n} )^{2} \big)
	 - \vec{S}_{1}\times\vec{v}_{1}\cdot\vec{v}_{2} \big( 3 \vec{v}_{1}\cdot\vec{n}
	 - 2 \vec{v}_{2}\cdot\vec{n} \big) \Big]
- \frac{2 G^2 m_{1} m_{2}}{r^2} \Big[ \vec{S}_{1}\times\vec{n}\cdot\vec{a}_{2} \vec{v}_{1}\cdot\vec{n} \nl
	 - \vec{S}_{1}\times\vec{v}_{1}\cdot\vec{a}_{2} \Big]\\
\text{Fig.~3(e1)} =&
\frac{8 G^2 m_{2}^2}{r^3} \vec{S}_{1}\times\vec{n}\cdot\vec{v}_{2} \big( v_{2}^2 -4 ( \vec{v}_{2}\cdot\vec{n} )^{2} \big)\\
\text{Fig.~3(e2)} =&
\frac{2 G^2 m_{1} m_{2}}{r^3} \Big[ 2 \vec{S}_{1}\times\vec{n}\cdot\vec{v}_{1} ( \vec{v}_{2}\cdot\vec{n} )^{2}
	 - \vec{S}_{1}\times\vec{n}\cdot\vec{v}_{2} \vec{v}_{1}\cdot\vec{v}_{2} \Big]\\
\text{Fig.~3(e3)} =&
\frac{2 G^2 m_{1} m_{2}}{r^3} \Big[ \vec{S}_{1}\times\vec{n}\cdot\vec{v}_{1} \vec{v}_{1}\cdot\vec{v}_{2}
	 + 2 \vec{S}_{1}\times\vec{n}\cdot\vec{v}_{2} \big( v_{1}^2 -6 ( \vec{v}_{1}\cdot\vec{n} )^{2} \big) \nl
	 + 4 \vec{S}_{1}\times\vec{v}_{1}\cdot\vec{v}_{2} \vec{v}_{1}\cdot\vec{n} \Big]\\
\text{Fig.~3(e4)} =&
\frac{2 G^2 m_{2}^2}{r^3} \Big[ 3 \vec{S}_{1}\times\vec{n}\cdot\vec{v}_{1} v_{2}^2
	 - 4 \vec{S}_{1}\times\vec{n}\cdot\vec{v}_{2} \vec{v}_{1}\cdot\vec{v}_{2} \Big]\\
\text{Fig.~3(e5)} =&
- \frac{2 G^2 m_{1} m_{2}}{r^3} \Big[ 4 \vec{S}_{1}\times\vec{n}\cdot\vec{v}_{1} \vec{v}_{1}\cdot\vec{v}_{2}
	 - 3 \vec{S}_{1}\times\vec{n}\cdot\vec{v}_{2} v_{1}^2
	 + \vec{S}_{1}\times\vec{v}_{1}\cdot\vec{v}_{2} \vec{v}_{1}\cdot\vec{n} \Big]\\
\text{Fig.~3(f1)} =&
- \frac{2 G^2 m_{2}^2}{r^3} \Big[ \vec{S}_{1}\times\vec{n}\cdot\vec{v}_{2} \big( 4 \vec{v}_{1}\cdot\vec{v}_{2}
	 + 3 v_{2}^2 -16 \vec{v}_{1}\cdot\vec{n} \vec{v}_{2}\cdot\vec{n}
	 - 16 ( \vec{v}_{2}\cdot\vec{n} )^{2} \big) \nl
	 + 4 \vec{S}_{1}\times\vec{v}_{1}\cdot\vec{v}_{2} \vec{v}_{2}\cdot\vec{n} \Big]
+ \frac{8 G^2 m_{2}^2}{r^2} \Big[ 2 \vec{S}_{1}\times\vec{n}\cdot\vec{a}_{2} \vec{v}_{2}\cdot\vec{n}
	 - \vec{S}_{1}\times\vec{v}_{2}\cdot\vec{a}_{2} \Big] \nl
- \frac{8 G^2 m_{2}^2}{r^2} \vec{v}_{2}\cdot\vec{n} \dot{\vec{S}}_{1}\times\vec{n}\cdot\vec{v}_{2} \\
\text{Fig.~3(f2)} =&
\frac{2 G^2 m_{1} m_{2}}{r^3} \Big[ \vec{S}_{1}\times\vec{n}\cdot\vec{v}_{1} \big( 3 \vec{v}_{1}\cdot\vec{v}_{2} -4 \vec{v}_{1}\cdot\vec{n} \vec{v}_{2}\cdot\vec{n} \big)
	 - 3 \vec{S}_{1}\times\vec{n}\cdot\vec{v}_{2} \big( v_{1}^2
	 + \vec{v}_{1}\cdot\vec{v}_{2} \nl
                     -4 \vec{v}_{1}\cdot\vec{n} \vec{v}_{2}\cdot\vec{n}
	 - 4 ( \vec{v}_{1}\cdot\vec{n} )^{2} \big)
	 - 2 \vec{S}_{1}\times\vec{v}_{1}\cdot\vec{v}_{2} \big( 3 \vec{v}_{1}\cdot\vec{n}
	 + \vec{v}_{2}\cdot\vec{n} \big) \Big] \nl
+ \frac{2 G^2 m_{1} m_{2}}{r^2} \Big[ 3 \vec{S}_{1}\times\vec{n}\cdot\vec{a}_{1} \vec{v}_{2}\cdot\vec{n}
	 + 3 \vec{S}_{1}\times\vec{n}\cdot\vec{a}_{2} \vec{v}_{1}\cdot\vec{n}
	 + \vec{S}_{1}\times\vec{a}_{1}\cdot\vec{v}_{2} \nl
	 - \vec{S}_{1}\times\vec{v}_{1}\cdot\vec{a}_{2} \Big]
+ \frac{2 G^2 m_{1} m_{2}}{r^2} \Big[ \dot{\vec{S}}_{1}\times\vec{n}\cdot\vec{v}_{1} \vec{v}_{2}\cdot\vec{n}
	 - 4 \dot{\vec{S}}_{1}\times\vec{n}\cdot\vec{v}_{2} \vec{v}_{1}\cdot\vec{n} \nl
	 + 3 \dot{\vec{S}}_{1}\times\vec{v}_{1}\cdot\vec{v}_{2} \Big]\\
\text{Fig.~3(g1)} =&
- \frac{2 G^2 m_{2}^2}{r^3} \Big[ \vec{S}_{1}\times\vec{n}\cdot\vec{v}_{1} \big( 4 \vec{v}_{1}\cdot\vec{v}_{2}
	 - v_{2}^2 -16 \vec{v}_{1}\cdot\vec{n} \vec{v}_{2}\cdot\vec{n}
	 + 2 ( \vec{v}_{2}\cdot\vec{n} )^{2} \big) \nl
	 - \vec{S}_{1}\times\vec{v}_{1}\cdot\vec{v}_{2} \big( 4 \vec{v}_{1}\cdot\vec{n}
	 - \vec{v}_{2}\cdot\vec{n} \big) \Big]
- \frac{4 G^2 m_{2}^2}{r^2} \Big[ 2 \vec{S}_{1}\times\vec{n}\cdot\vec{a}_{1} \vec{v}_{2}\cdot\vec{n}
	 + \vec{S}_{1}\times\vec{a}_{1}\cdot\vec{v}_{2} \Big] \nl
- \frac{4 G^2 m_{2}^2}{r^2} \Big[ 2 \dot{\vec{S}}_{1}\times\vec{n}\cdot\vec{v}_{1} \vec{v}_{2}\cdot\vec{n}
	 + \dot{\vec{S}}_{1}\times\vec{v}_{1}\cdot\vec{v}_{2} \Big]\\
\text{Fig.~3(g2)} =&
\frac{2 G^2 m_{1} m_{2}}{r^3} \Big[ \vec{S}_{1}\times\vec{n}\cdot\vec{v}_{1} \big( v_{1}^2
	 - 4 \vec{v}_{1}\cdot\vec{v}_{2}
	 + 16 \vec{v}_{1}\cdot\vec{n} \vec{v}_{2}\cdot\vec{n}
	 - 2 ( \vec{v}_{1}\cdot\vec{n} )^{2} \big) \nl
	 + 4 \vec{S}_{1}\times\vec{v}_{1}\cdot\vec{v}_{2} \vec{v}_{1}\cdot\vec{n} \Big]
+ \frac{2 G^2 m_{1} m_{2}}{r^2} \Big[ \vec{S}_{1}\times\vec{n}\cdot\vec{a}_{1} \big( \vec{v}_{1}\cdot\vec{n}
	 - 4 \vec{v}_{2}\cdot\vec{n} \big)
	 - \vec{S}_{1}\times\vec{v}_{1}\cdot\vec{a}_{1} \nl
	 - 2 \vec{S}_{1}\times\vec{a}_{1}\cdot\vec{v}_{2} \Big]
+ \frac{2 G^2 m_{1} m_{2}}{r^2} \Big[ \dot{\vec{S}}_{1}\times\vec{n}\cdot\vec{v}_{1} \big( \vec{v}_{1}\cdot\vec{n}
	 - 4 \vec{v}_{2}\cdot\vec{n} \big)
	 - 2 \dot{\vec{S}}_{1}\times\vec{v}_{1}\cdot\vec{v}_{2} \Big]
\end{align}

\subsection{Cubic in G interaction}

\subsubsection{Three-graviton exchange}

\begin{figure}[t]
\includegraphics[scale=0.93]{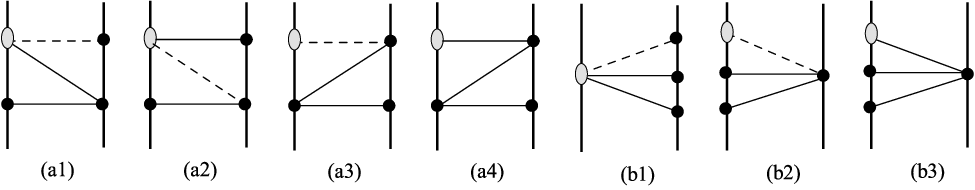}
\caption{NNLO spin-orbit Feynman diagrams of order $G^3$ with no loops.}
\label{fig:sonnlog3nonloop}
\end{figure}

For the NNLO spin-orbit interaction we have 7 diagrams at order $G^3$ with no 
loops, as shown in figure \ref{fig:sonnlog3nonloop}. These three-graviton 
exchange diagrams are constructed with either one-, two-, or three-graviton 
spin couplings.

The values of these diagrams are given by
\begin{align}
\text{Fig.~4(a1)}=& -8\frac{G^3 m_1 m_2^2}{r^4}
\vec{S}_1\cdot\vec{v}_2\times\vec{n},\\
\text{Fig.~4(a2)}=& -8\frac{G^3 m_1 m_2^2}{r^4}
\vec{S}_1\cdot\vec{v}_2\times\vec{n},\\
\text{Fig.~4(a3)}=& -2\frac{G^3 m_1 m_2^2}{r^4}
\vec{S}_1\cdot\vec{v}_2\times\vec{n},\\
\text{Fig.~4(a4)}=& 2\frac{G^3 m_1 m_2^2}{r^4}
\vec{S}_1\cdot\vec{v}_1\times\vec{n},\\
\text{Fig.~4(b1)}=& -16\frac{G^3 m_2^3}{r^4}
\vec{S}_1\cdot\vec{v}_2\times\vec{n},\\
\text{Fig.~4(b2)}=& -\frac{G^3 m_1^2 m_2}{r^4}
\vec{S}_1\cdot\vec{v}_2\times\vec{n},\\
\text{Fig.~4(b3)}=& \frac{G^3 m_1^2 m_2}{r^4}
\vec{S}_1\cdot\vec{v}_1\times\vec{n}.
\end{align}

\subsubsection{Cubic self-interaction with two-graviton exchange}

\begin{figure}[t]
\includegraphics[scale=0.93]{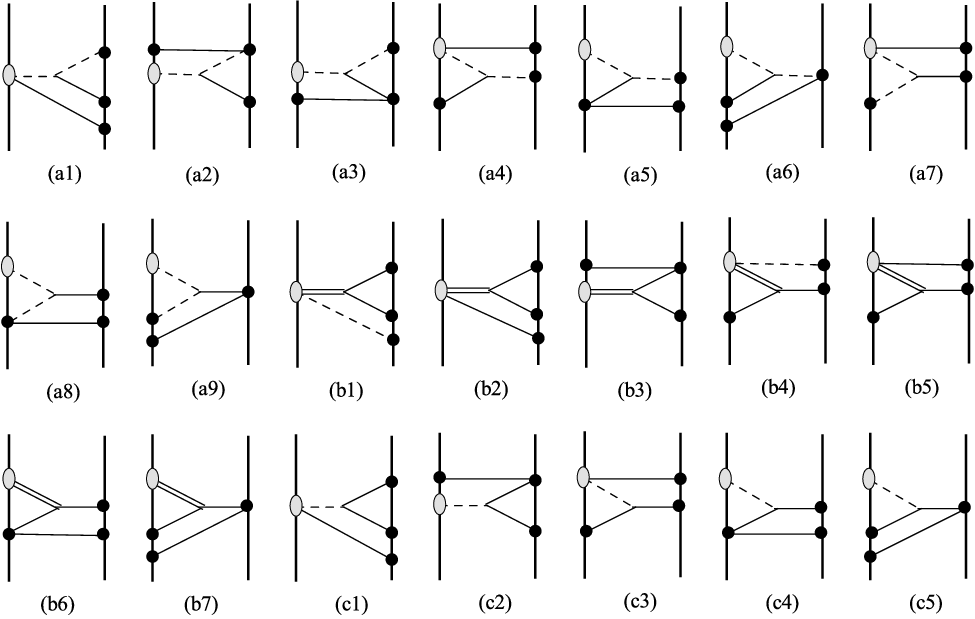}
\caption{NNLO spin-orbit Feynman diagrams of order G$^3$ with one loop.}
\label{fig:sonnlog3oneloop}
\end{figure}

For the NNLO spin-orbit interaction we have 21 diagrams at order $G^3$ with one 
loop, as shown in figure \ref{fig:sonnlog3oneloop}. These diagrams contain both 
cubic self-interaction and two-graviton worldline couplings.

The values of these diagrams are given by
\begin{align}
\text{Fig.~5(a1)}=& 32\frac{G^3 m_2^3}{r^4}
\vec{S}_1\cdot\vec{v}_2\times\vec{n},\\
\text{Fig.~5(a2)}=& 8\frac{G^3 m_1 m_2^2}{r^4}
\vec{S}_1\cdot\vec{v}_2\times\vec{n},\\
\text{Fig.~5(a3)}=& 8\frac{G^3 m_1 m_2^2}{r^4}
\vec{S}_1\cdot\vec{v}_2\times\vec{n},\\
\text{Fig.~5(a4)}=& 8\frac{G^3 m_1 m_2^2}{r^4}
\vec{S}_1\cdot\vec{v}_2\times\vec{n},\\
\text{Fig.~5(a5)}=& 2\frac{G^3 m_1 m_2^2}{r^4}
\vec{S}_1\cdot\vec{v}_2\times\vec{n},\\
\text{Fig.~5(a6)}=& 2\frac{G^3 m_1^2 m_2}{r^4}
\vec{S}_1\cdot\vec{v}_2\times\vec{n},\\
\text{Fig.~5(a7)}=& -8\frac{G^3 m_1 m_2^2}{r^4}
\vec{S}_1\cdot\vec{v}_1\times\vec{n},\\
\text{Fig.~5(a8)}=& -2\frac{G^3 m_1 m_2^2}{r^4}
\vec{S}_1\cdot\vec{v}_1\times\vec{n},\\
\text{Fig.~5(a9)}=& -2\frac{G^3 m_1^2 m_2}{r^4}
\vec{S}_1\cdot\vec{v}_1\times\vec{n},\\
\text{Fig.~5(b1)}=& -\frac{G^3 m_2^3}{r^4}
\vec{S}_1\cdot\vec{v}_2\times\vec{n},\\
\text{Fig.~5(b2)}=& -\frac{G^3 m_2^3}{r^4}
\vec{S}_1\cdot\vec{v}_1\times\vec{n},\\
\text{Fig.~5(b3)}=& \frac{G^3 m_1 m_2^2}{r^4}
\vec{S}_1\cdot\vec{v}_1\times\vec{n},\\
\text{Fig.~5(b4)}=& 0,\\
\text{Fig.~5(b5)}=& 8\frac{G^3 m_1 m_2^2}{r^4}
\vec{S}_1\cdot\vec{v}_1\times\vec{n},\\
\text{Fig.~5(b6)}=& \frac{1}{2}\frac{G^3 m_1 m_2^2}{r^4}
\vec{S}_1\cdot\vec{v}_1\times\vec{n},\\
\text{Fig.~5(b7)}=& \frac{1}{2}\frac{G^3 m_1^2 m_2}{r^4}
\vec{S}_1\cdot\vec{v}_1\times\vec{n},\\
\text{Fig.~5(c1)}=& -2\frac{G^3 m_2^3}{r^4}
\vec{S}_1\cdot\vec{v}_2\times\vec{n},\\
\text{Fig.~5(c2)}=& -\frac{G^3 m_1 m_2^2}{r^4}
\vec{S}_1\cdot\vec{v}_2\times\vec{n},\\
\text{Fig.~5(c3)}=& -2\frac{G^3 m_1 m_2^2}{r^4}
\vec{S}_1\cdot\vec{v}_1\times\vec{n},\\
\text{Fig.~5(c4)}=& -\frac{1}{2}\frac{G^3 m_1 m_2^2}{r^4}
\vec{S}_1\cdot\vec{v}_1\times\vec{n},\\
\text{Fig.~5(c5)}=& -\frac{1}{2}\frac{G^3 m_1^2 m_2}{r^4}
\vec{S}_1\cdot\vec{v}_1\times\vec{n}.
\end{align}
Note that the total value of the diagram in figure 5(b4) equals 0, although it 
does not stand for a short distance contribution. 

\subsubsection{Two-loop interaction}

For the NNLO spin-orbit interaction we have 32 two-loop diagrams at order $G^3$, 
as shown in figure \ref{fig:sonnlog3twoloop}. These diagrams contain two cubic 
vertices or one quartic vertex, and even include cubic vertices with time 
dependence. As explained in \cite{Levi:2011eq}, they contain two-loop Feynman 
integrals of three kinds: Factorizable, nested, and irreducible. The 
factorizable two-loop diagrams do not contribute at the NNLO level, and they 
yield here purely short distance contributions, of the form 
$\delta^{(1)}(\vec{r})$, which are contact interaction terms. For other two-loop 
diagrams calculations should be made, keeping the dimension $d$ general, and the 
limit $d\to3$ is only taken in the end. 

For the irreducible two-loop diagrams, which are the most complicated, 
irreducible two-loop tensor integrals of order 3 are encountered here. These are 
reduced using the integration by parts method to a sum of factorizable and nested 
two-loop integrals, as explained in \cite{Levi:2011eq}, and see appendix~A there. 
In addition to the irreducible two-loop tensor integrals, which were given in 
appendix~A of \cite{Levi:2011eq}, eqs.~(A11), (A12) there, two further irreducible 
tensor integrals are required here, and we provide them in appendix \ref{reduce} 
below. 

The values of the two-loop diagrams are given in the following:

\begin{figure}[p]
	\includegraphics[scale=0.93]{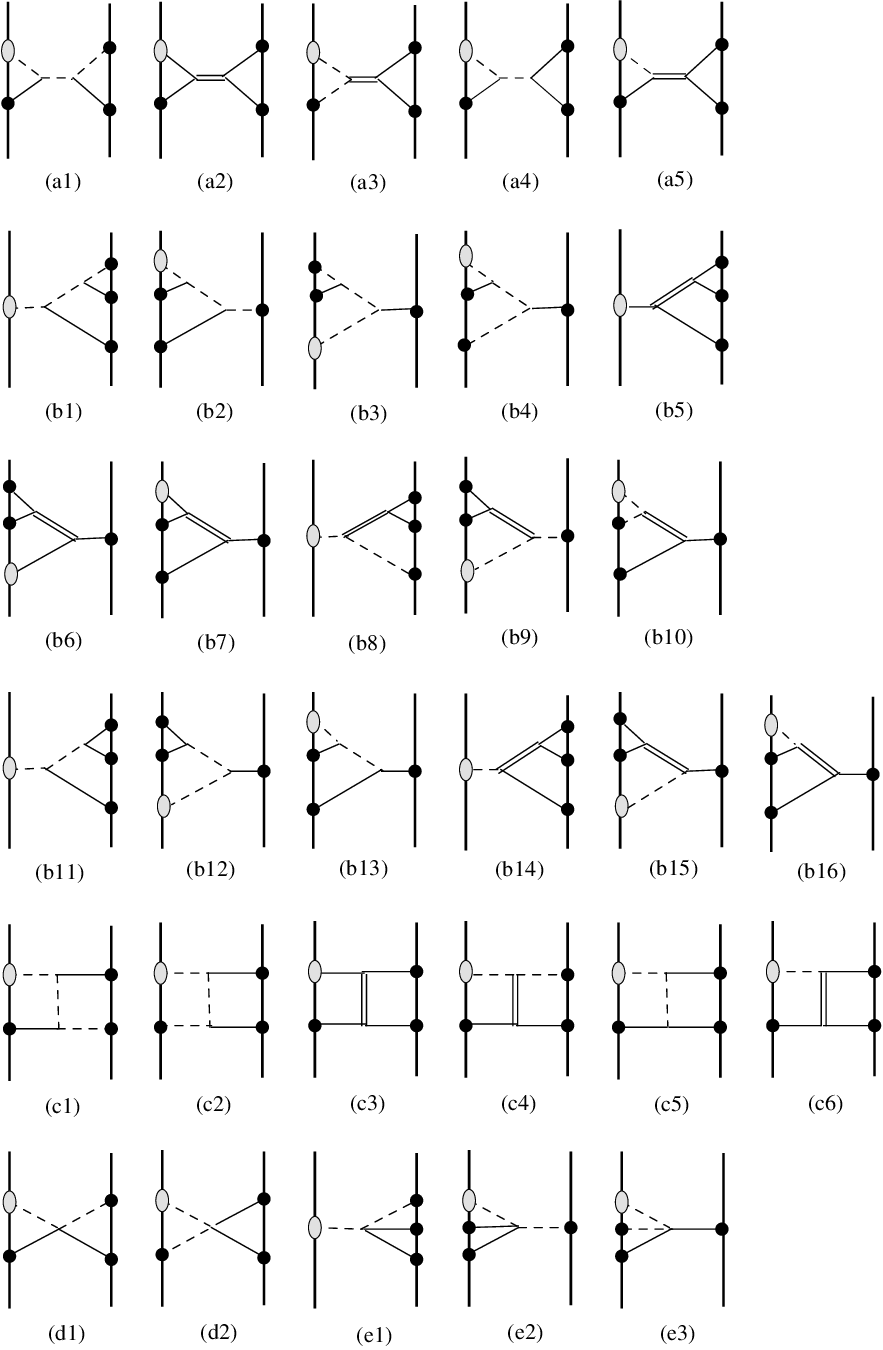}
	\caption{NNLO spin-orbit Feynman diagrams of order $G^3$ with two loops.}
	\label{fig:sonnlog3twoloop}
\end{figure}
\clearpage

\begin{align}
\text{Fig.~6(a1)}=& 0,\\
\text{Fig.~6(a2)}=& 0,\\
\text{Fig.~6(a3)}=& 0,\\
\text{Fig.~6(a4)}=& 0,\\
\text{Fig.~6(a5)}=& 0,\\
\text{Fig.~6(b1)}=& -32\frac{G^3 m_2^3}{r^4}
\vec{S}_1\cdot\vec{v}_2\times\vec{n},\\
\text{Fig.~6(b2)}=& -\frac{24}{5}\frac{G^3 m_1^2 m_2}{r^4}
\vec{S}_1\cdot\vec{v}_2\times\vec{n},\\
\text{Fig.~6(b3)}=& \frac{16}{5}\frac{G^3 m_1^2 m_2}{r^4}
\vec{S}_1\cdot\vec{v}_1\times\vec{n},\\
\text{Fig.~6(b4)}=& \frac{8}{5}\frac{G^3 m_1^2 m_2}{r^4}
\vec{S}_1\cdot\vec{v}_1\times\vec{n},\\
\text{Fig.~6(b5)}=& 2\frac{G^3 m_2^3}{r^4}
\vec{S}_1\cdot\vec{v}_1\times\vec{n},\\
\text{Fig.~6(b6)}=& \frac{6}{5}\frac{G^3 m_1^2 m_2}{r^4}
\vec{S}_1\cdot\vec{v}_1\times\vec{n},\\
\text{Fig.~6(b7)}=& \frac{4}{5}\frac{G^3 m_1^2 m_2}{r^4}
\vec{S}_1\cdot\vec{v}_1\times\vec{n},\\
\text{Fig.~6(b8)}=& -\frac{2}{5}\frac{G^3 m_2^3}{r^4}
\vec{S}_1\cdot\vec{v}_2\times\vec{n},\\
\text{Fig.~6(b9)}=& -\frac{2}{5}\frac{G^3 m_1^2 m_2}{r^4}
\vec{S}_1\cdot\vec{v}_2\times\vec{n},\\
\text{Fig.~6(b10)}=& -2\frac{G^3 m_1^2 m_2}{r^4}
\vec{S}_1\cdot\vec{v}_1\times\vec{n},\\
\text{Fig.~6(b11)}=& 2\frac{G^3 m_2^3}{r^4}
\vec{S}_1\cdot\vec{v}_2\times\vec{n},\\
\text{Fig.~6(b12)}=& -\frac{1}{5}\frac{G^3 m_1^2 m_2}{r^4}
\vec{S}_1\cdot\vec{v}_1\times\vec{n},\\
\text{Fig.~6(b13)}=& \frac{1}{5}\frac{G^3 m_1^2 m_2}{r^4}
\vec{S}_1\cdot\vec{v}_1\times\vec{n},\\
\text{Fig.~6(b14)}=& \frac{2}{5}\frac{G^3 m_2^3}{r^4}
\vec{S}_1\cdot\vec{v}_2\times\vec{n},\\
\text{Fig.~6(b15)}=& -\frac{3}{5}\frac{G^3 m_1^2 m_2}{r^4}
\vec{S}_1\cdot\vec{v}_1\times\vec{n},\\
\text{Fig.~6(b16)}=&\frac{G^3 m_1^2 m_2}{r^4}
\vec{S}_1\cdot\vec{v}_1\times\vec{n},\\
\text{Fig.~6(c1)}=& -4\frac{G^3 m_1 m_2^2}{r^4}
\vec{S}_1\cdot\vec{v}_2\times\vec{n},\\
\text{Fig.~6(c2)}=& 4\frac{G^3 m_1 m_2^2}{r^4}
\vec{S}_1\cdot\vec{v}_1\times\vec{n},\\
\text{Fig.~6(c3)}=& 12\frac{G^3 m_1 m_2^2}{r^4}
\vec{S}_1\cdot\vec{v}_1\times\vec{n},\\
\text{Fig.~6(c4)}=& -16\frac{G^3 m_1 m_2^2}{r^4}
\vec{S}_1\cdot\vec{v}_2\times\vec{n},\\
\text{Fig.~6(c5)}=& \frac{G^3 m_1 m_2^2}{r^4}
\vec{S}_1\cdot\left[2\vec{v}_1\times\vec{n}-2\vec{v}_2\times\vec{n}\right],\\
\text{Fig.~6(c6)}=&\frac{G^3 m_1 m_2^2}{r^4}
\vec{S}_1\cdot\left[-12\vec{v}_1\times\vec{n}+10\vec{v}_2\times\vec{n}\right],\\
\text{Fig.~6(d1)}=& 0,\\
\text{Fig.~6(d2)}=& 0,\\
\text{Fig.~6(e1)}=& 16\frac{G^3 m_2^3}{r^4}
\vec{S}_1\cdot\vec{v}_2\times\vec{n},\\
\text{Fig.~6(e2)}=& \frac{16}{5}\frac{G^3 m_1^2 m_2}{r^4}
\vec{S}_1\cdot\vec{v}_2\times\vec{n},\\
\text{Fig.~6(e3)}=& -\frac{16}{5}\frac{G^3 m_1^2 m_2}{r^4}
\vec{S}_1\cdot\vec{v}_1\times\vec{n}.
\end{align}

\section{Next-to-next-to-leading order spin-orbit potential and Hamiltonian}
\label{VtoH}

Summing up all of the Feynman diagrams from the previous section, we obtain the 
NNLO spin-orbit interaction potential for a binary system of compact spinning 
objects. We split the potential into several pieces according to the number and order of higher-order time derivatives as follows:
\begin{align}
V_{\text{NNLO}}^{\text{SO}} =& \stackrel{(0)}{V}
         + \stackrel{(1)}{V} + \stackrel{(2)}{V} + \stackrel{(3)}{V} + \stackrel{(4)}{V}.
\end{align}
The ordinary part of the potential, which does not contain higher-order time derivatives, reads
\begin{align}
\stackrel{(0)}{V} =&
- \frac{G m_{2}}{4 r^2} \Big[ \vec{S}_{1}\times\vec{n}\cdot\vec{v}_{1} \big( 5 v_{1}^2 \vec{v}_{1}\cdot\vec{v}_{2}
	 - 3 v_{1}^2 v_{2}^2
	 + 4 \vec{v}_{1}\cdot\vec{v}_{2} v_{2}^2
	 - 3 v_{1}^{4}
	 - 3 v_{2}^{4}
	 + 6 \vec{v}_{1}\cdot\vec{n} v_{1}^2 \vec{v}_{2}\cdot\vec{n} \nl
	 - 6 \vec{v}_{1}\cdot\vec{n} \vec{v}_{2}\cdot\vec{n} \vec{v}_{1}\cdot\vec{v}_{2}
	 + 6 \vec{v}_{1}\cdot\vec{n} \vec{v}_{2}\cdot\vec{n} v_{2}^2
	 + 3 v_{2}^2 ( \vec{v}_{1}\cdot\vec{n} )^{2}
	 + 3 v_{1}^2 ( \vec{v}_{2}\cdot\vec{n} )^{2}
                               -15 ( \vec{v}_{1}\cdot\vec{n} )^{2} ( \vec{v}_{2}\cdot\vec{n} )^{2} \big) \nl
	 + \vec{S}_{1}\times\vec{n}\cdot\vec{v}_{2} \big( v_{1}^2 v_{2}^2
	 - 2 \vec{v}_{1}\cdot\vec{v}_{2} v_{2}^2
	 - 2 ( \vec{v}_{1}\cdot\vec{v}_{2} )^{2}
	 + 3 v_{2}^{4}
                               -6 \vec{v}_{1}\cdot\vec{n} \vec{v}_{2}\cdot\vec{n} v_{2}^2
	 - 3 v_{2}^2 ( \vec{v}_{1}\cdot\vec{n} )^{2} \nl
	 - 3 v_{1}^2 ( \vec{v}_{2}\cdot\vec{n} )^{2}
	 + 15 ( \vec{v}_{1}\cdot\vec{n} )^{2} ( \vec{v}_{2}\cdot\vec{n} )^{2} \big)
	 + \vec{S}_{1}\times\vec{v}_{1}\cdot\vec{v}_{2} \big( \vec{v}_{1}\cdot\vec{n} v_{1}^2
	 - 2 \vec{v}_{1}\cdot\vec{n} \vec{v}_{1}\cdot\vec{v}_{2}
	 + 2 \vec{v}_{1}\cdot\vec{n} v_{2}^2 \nl
	 + 2 \vec{v}_{2}\cdot\vec{n} v_{2}^2
         -6 \vec{v}_{1}\cdot\vec{n} ( \vec{v}_{2}\cdot\vec{n} )^{2} \big) \Big] \nl
+ \frac{G^2 m_{2}^2}{4 r^3} \Big[ \vec{S}_{1}\times\vec{n}\cdot\vec{v}_{1} \big( 13 v_{1}^2
	 - 41 \vec{v}_{1}\cdot\vec{v}_{2}
	 + 28 v_{2}^2 -16 \vec{v}_{1}\cdot\vec{n} \vec{v}_{2}\cdot\vec{n}
	 + 8 ( \vec{v}_{1}\cdot\vec{n} )^{2}
	 + 12 ( \vec{v}_{2}\cdot\vec{n} )^{2} \big) \nl
	 - 2 \vec{S}_{1}\times\vec{n}\cdot\vec{v}_{2} \big( v_{1}^2
	 + 7 \vec{v}_{1}\cdot\vec{v}_{2}
	 - 8 v_{2}^2 -56 \vec{v}_{1}\cdot\vec{n} \vec{v}_{2}\cdot\vec{n}
	 - 4 ( \vec{v}_{1}\cdot\vec{n} )^{2}
	 + 62 ( \vec{v}_{2}\cdot\vec{n} )^{2} \big) \nl
	 + \vec{S}_{1}\times\vec{v}_{1}\cdot\vec{v}_{2} \big( 7 \vec{v}_{1}\cdot\vec{n}
	 - 38 \vec{v}_{2}\cdot\vec{n} \big) \Big]
- \frac{2 G^2 m_{1} m_{2}}{r^3} \Big[ \vec{S}_{1}\times\vec{n}\cdot\vec{v}_{1} \big( 2 v_{1}^2
	 + \vec{v}_{1}\cdot\vec{v}_{2}
	 - 3 v_{2}^2 \nl
	 + 12 \vec{v}_{1}\cdot\vec{n} \vec{v}_{2}\cdot\vec{n}
	 - 16 ( \vec{v}_{1}\cdot\vec{n} )^{2}
	 + 4 ( \vec{v}_{2}\cdot\vec{n} )^{2} \big)
	 - 2 \vec{S}_{1}\times\vec{n}\cdot\vec{v}_{2} \big( \vec{v}_{1}\cdot\vec{v}_{2}
	 - v_{2}^2 \nl
	 + 4 \vec{v}_{1}\cdot\vec{n} \vec{v}_{2}\cdot\vec{n}
	 - 4 ( \vec{v}_{1}\cdot\vec{n} )^{2} \big)
	 - \vec{S}_{1}\times\vec{v}_{1}\cdot\vec{v}_{2} \big( 2 \vec{v}_{1}\cdot\vec{n}
	 - 5 \vec{v}_{2}\cdot\vec{n} \big) \Big] \nl
+ \frac{G^3 m_{1}^2 m_{2}}{r^4} \Big[ \vec{S}_{1}\times\vec{n}\cdot\vec{v}_{1}
	 - \vec{S}_{1}\times\vec{n}\cdot\vec{v}_{2} \Big]
+ \frac{5 G^3 m_{1} m_{2}^2}{r^4} \Big[ \vec{S}_{1}\times\vec{n}\cdot\vec{v}_{1}
	 - \vec{S}_{1}\times\vec{n}\cdot\vec{v}_{2} \Big] \nl
+ \frac{G^3 m_{2}^3}{r^4} \Big[ \vec{S}_{1}\times\vec{n}\cdot\vec{v}_{1}
	 - \vec{S}_{1}\times\vec{n}\cdot\vec{v}_{2} \Big]
+ (1 \leftrightarrow 2).
\end{align}
Due to the large number of terms, it makes sense to sort the terms with a single higher-order time derivative into 
\begin{align}
\stackrel{(1)}{V} =& V_{a} + V_{\dot{S}},
\end{align}
where
\begin{align}
V_{a} = &
- \frac{5}{16} \vec{S}_{1}\times\vec{v}_{1}\cdot\vec{a}_{1} v_{1}^{4} \nl
+ \frac{G m_{2}}{4 r} \Big[ \vec{S}_{1}\times\vec{n}\cdot\vec{v}_{1} \big( 4 \vec{v}_{1}\cdot\vec{a}_{1} \vec{v}_{2}\cdot\vec{n}
	 - 4 \vec{v}_{2}\cdot\vec{n} \vec{a}_{1}\cdot\vec{v}_{2}
	 + \vec{a}_{1}\cdot\vec{n} v_{2}^2
	 - v_{1}^2 \vec{a}_{2}\cdot\vec{n}
	 + 4 \vec{v}_{1}\cdot\vec{n} \vec{v}_{1}\cdot\vec{a}_{2} \nl
	 - 4 \vec{v}_{1}\cdot\vec{n} \vec{v}_{2}\cdot\vec{a}_{2}
	 + 3 \vec{a}_{2}\cdot\vec{n} ( \vec{v}_{1}\cdot\vec{n} )^{2}
	 - 3 \vec{a}_{1}\cdot\vec{n} ( \vec{v}_{2}\cdot\vec{n} )^{2} \big)
	 + 2 \vec{S}_{1}\times\vec{n}\cdot\vec{a}_{1} \big( v_{1}^2 \vec{v}_{2}\cdot\vec{n}
	 - \vec{v}_{2}\cdot\vec{n} \vec{v}_{1}\cdot\vec{v}_{2} \nl
	 + \vec{v}_{1}\cdot\vec{n} v_{2}^2
	 + \vec{v}_{2}\cdot\vec{n} v_{2}^2 -3 \vec{v}_{1}\cdot\vec{n} ( \vec{v}_{2}\cdot\vec{n} )^{2} \big)
	 - \vec{S}_{1}\times\vec{v}_{1}\cdot\vec{a}_{1} \big( 12 v_{1}^2 
	 - 12 \vec{v}_{1}\cdot\vec{v}_{2}
	 + 5 v_{2}^2 \nl
         - 4 \vec{v}_{1}\cdot\vec{n} \vec{v}_{2}\cdot\vec{n}
	 - ( \vec{v}_{2}\cdot\vec{n} )^{2} \big)
	 + \vec{S}_{1}\times\vec{n}\cdot\vec{v}_{2} \big( 2 \vec{v}_{2}\cdot\vec{n} \vec{a}_{1}\cdot\vec{v}_{2}
	 - \vec{a}_{1}\cdot\vec{n} v_{2}^2
	 + v_{1}^2 \vec{a}_{2}\cdot\vec{n} \nl
	 - 2 \vec{v}_{1}\cdot\vec{n} \vec{v}_{1}\cdot\vec{a}_{2}
	 + 4 \vec{v}_{1}\cdot\vec{n} \vec{v}_{2}\cdot\vec{a}_{2}
                               -3 \vec{a}_{2}\cdot\vec{n} ( \vec{v}_{1}\cdot\vec{n} )^{2}
	 + 3 \vec{a}_{1}\cdot\vec{n} ( \vec{v}_{2}\cdot\vec{n} )^{2} \big) \nl
	 - 2 \vec{S}_{1}\times\vec{v}_{1}\cdot\vec{v}_{2} \big( 6 \vec{v}_{1}\cdot\vec{a}_{1}
	 - \vec{a}_{1}\cdot\vec{v}_{2}
	 - \vec{v}_{1}\cdot\vec{a}_{2}
	 + 2 \vec{v}_{2}\cdot\vec{a}_{2}
	 - \vec{v}_{1}\cdot\vec{n} \vec{a}_{2}\cdot\vec{n} \big) \nl
	 - \vec{S}_{1}\times\vec{a}_{1}\cdot\vec{v}_{2} \big( 6 v_{1}^2
	 - 4 \vec{v}_{1}\cdot\vec{v}_{2}
	 + 5 v_{2}^2 -6 \vec{v}_{1}\cdot\vec{n} \vec{v}_{2}\cdot\vec{n}
	 + ( \vec{v}_{2}\cdot\vec{n} )^{2} \big)
	 + 2 \vec{S}_{1}\times\vec{n}\cdot\vec{a}_{2} \big( v_{1}^2 \vec{v}_{2}\cdot\vec{n} \nl
	 + \vec{v}_{1}\cdot\vec{n} v_{2}^2
         - 3 \vec{v}_{2}\cdot\vec{n} ( \vec{v}_{1}\cdot\vec{n} )^{2} \big)
	 + \vec{S}_{1}\times\vec{v}_{1}\cdot\vec{a}_{2} \big( v_{1}^2
	 - 2 v_{2}^2
	 + 4 \vec{v}_{1}\cdot\vec{n} \vec{v}_{2}\cdot\vec{n}
	 - ( \vec{v}_{1}\cdot\vec{n} )^{2} \big) \nl
	 - \vec{S}_{1}\times\vec{v}_{2}\cdot\vec{a}_{2} \big( v_{1}^2
	 - ( \vec{v}_{1}\cdot\vec{n} )^{2} \big) \Big] \nl
- \frac{2 G^2 m_{1} m_{2}}{r^2} \Big[ \vec{S}_{1}\times\vec{n}\cdot\vec{v}_{1} \big( 2 \vec{a}_{1}\cdot\vec{n}
	 + \vec{a}_{2}\cdot\vec{n} \big)
	 + \vec{S}_{1}\times\vec{n}\cdot\vec{a}_{1} \big( 3 \vec{v}_{1}\cdot\vec{n}
	 + 2 \vec{v}_{2}\cdot\vec{n} \big) \nl
	 - 2 \vec{S}_{1}\times\vec{n}\cdot\vec{a}_{2} \vec{v}_{1}\cdot\vec{n}
	 + 2 \vec{S}_{1}\times\vec{a}_{1}\cdot\vec{v}_{2}
	 + 2 \vec{S}_{1}\times\vec{v}_{1}\cdot\vec{a}_{2} \Big]
- \frac{G^2 m_{2}^2}{4 r^2} \Big[ 2 \vec{S}_{1}\times\vec{n}\cdot\vec{v}_{1} \vec{a}_{1}\cdot\vec{n} \nl
	 + 4 \vec{S}_{1}\times\vec{n}\cdot\vec{a}_{1} \big( \vec{v}_{1}\cdot\vec{n}
	 - \vec{v}_{2}\cdot\vec{n} \big)
	 + 2 \vec{S}_{1}\times\vec{n}\cdot\vec{v}_{2} \big( \vec{a}_{1}\cdot\vec{n}
	 + 14 \vec{a}_{2}\cdot\vec{n} \big)
	 - 2 \vec{S}_{1}\times\vec{n}\cdot\vec{a}_{2} \big( \vec{v}_{1}\cdot\vec{n} \nl
	 - 18 \vec{v}_{2}\cdot\vec{n} \big)
	 + 14 \vec{S}_{1}\times\vec{v}_{1}\cdot\vec{a}_{1}
	 + 27 \vec{S}_{1}\times\vec{a}_{1}\cdot\vec{v}_{2}
	 + \vec{S}_{1}\times\vec{v}_{1}\cdot\vec{a}_{2}
	 + 12 \vec{S}_{1}\times\vec{v}_{2}\cdot\vec{a}_{2} \Big] \nl
+ (1 \leftrightarrow 2), 
\end{align}
and 
\begin{align}
V_{\dot{S}} = &
\frac{G m_{2}}{2 r} \Big[ \dot{\vec{S}}_{1}\times\vec{n}\cdot\vec{v}_{1} \big( v_{1}^2 \vec{v}_{2}\cdot\vec{n}
	 - \vec{v}_{2}\cdot\vec{n} \vec{v}_{1}\cdot\vec{v}_{2}
	 + \vec{v}_{1}\cdot\vec{n} v_{2}^2
	 + \vec{v}_{2}\cdot\vec{n} v_{2}^2 -3 \vec{v}_{1}\cdot\vec{n} ( \vec{v}_{2}\cdot\vec{n} )^{2} \big) \nl
	 - \dot{\vec{S}}_{1}\times\vec{n}\cdot\vec{v}_{2} \big( \vec{v}_{1}\cdot\vec{n} v_{2}^2
	 + \vec{v}_{2}\cdot\vec{n} v_{2}^2 -3 \vec{v}_{1}\cdot\vec{n} ( \vec{v}_{2}\cdot\vec{n} )^{2} \big)
	 - \dot{\vec{S}}_{1}\times\vec{v}_{1}\cdot\vec{v}_{2} \big( 3 v_{1}^2
	 + 2 v_{2}^2 \nl
                 - 3 \vec{v}_{1}\cdot\vec{n} \vec{v}_{2}\cdot\vec{n}
	 + ( \vec{v}_{2}\cdot\vec{n} )^{2} \big) \Big] \nl
- \frac{G^2 m_{1} m_{2}}{r^2} \Big[ 6 \dot{\vec{S}}_{1}\times\vec{n}\cdot\vec{v}_{1} \big( 3 \vec{v}_{1}\cdot\vec{n}
	 - \vec{v}_{2}\cdot\vec{n} \big)
	 - 2 \dot{\vec{S}}_{1}\times\vec{n}\cdot\vec{v}_{2} \big( 7 \vec{v}_{1}\cdot\vec{n}
	 - 5 \vec{v}_{2}\cdot\vec{n} \big) \nl
	 + 7 \dot{\vec{S}}_{1}\times\vec{v}_{1}\cdot\vec{v}_{2} \Big]
- \frac{G^2 m_{2}^2}{4 r^2} \Big[ \dot{\vec{S}}_{1}\times\vec{n}\cdot\vec{v}_{1} \big( 3 \vec{v}_{1}\cdot\vec{n}
	 - 4 \vec{v}_{2}\cdot\vec{n} \big)
	 + 2 \dot{\vec{S}}_{1}\times\vec{n}\cdot\vec{v}_{2} \big( \vec{v}_{1}\cdot\vec{n} \nl
	 + 15 \vec{v}_{2}\cdot\vec{n} \big)
	 + 27 \dot{\vec{S}}_{1}\times\vec{v}_{1}\cdot\vec{v}_{2} \Big]
+ (1 \leftrightarrow 2) .
\end{align}
The piece with two higher-order time derivatives is given by
\begin{align}
\stackrel{(2)}{V} =&
- \frac{1}{4} G m_{2} \Big[ \vec{S}_{1}\times\vec{n}\cdot\dot{\vec{a}}_{1} \big( v_{2}^2
	 - ( \vec{v}_{2}\cdot\vec{n} )^{2} \big)
	 + 4 \vec{S}_{1}\times\vec{v}_{1}\cdot\dot{\vec{a}}_{1} \vec{v}_{2}\cdot\vec{n}
	 + 6 \vec{S}_{1}\times\dot{\vec{a}}_{1}\cdot\vec{v}_{2} \vec{v}_{2}\cdot\vec{n} \nl
	 - \vec{S}_{1}\times\vec{n}\cdot\dot{\vec{a}}_{2} \big( v_{1}^2
	 - ( \vec{v}_{1}\cdot\vec{n} )^{2} \big)
	 - 2 \vec{S}_{1}\times\vec{v}_{1}\cdot\dot{\vec{a}}_{2} \vec{v}_{1}\cdot\vec{n} \Big] \nl
- \frac{1}{4} G m_{2} \Big[ \ddot{\vec{S}}_{1}\times\vec{n}\cdot\vec{v}_{1} \big( v_{2}^2
	 - ( \vec{v}_{2}\cdot\vec{n} )^{2} \big)
	 - \ddot{\vec{S}}_{1}\times\vec{n}\cdot\vec{v}_{2} \big( v_{2}^2
	 - ( \vec{v}_{2}\cdot\vec{n} )^{2} \big)
	 + 6 \ddot{\vec{S}}_{1}\times\vec{v}_{1}\cdot\vec{v}_{2} \vec{v}_{2}\cdot\vec{n} \Big] \nl
- \frac{1}{4} G m_{2} \Big[ \vec{S}_{1}\times\vec{n}\cdot\vec{v}_{1} \big( 5 \vec{a}_{1}\cdot\vec{a}_{2}
	 + \vec{a}_{1}\cdot\vec{n} \vec{a}_{2}\cdot\vec{n} \big)
	 + 2 \vec{S}_{1}\times\vec{n}\cdot\vec{a}_{1} \big( 2 \vec{v}_{1}\cdot\vec{a}_{2}
	 - 2 \vec{v}_{2}\cdot\vec{a}_{2} \nl
	 + \vec{v}_{1}\cdot\vec{n} \vec{a}_{2}\cdot\vec{n} \big)
	 - \vec{S}_{1}\times\vec{v}_{1}\cdot\vec{a}_{1} \vec{a}_{2}\cdot\vec{n}
	 - \vec{S}_{1}\times\vec{n}\cdot\vec{v}_{2} \big( 3 \vec{a}_{1}\cdot\vec{a}_{2}
	 + \vec{a}_{1}\cdot\vec{n} \vec{a}_{2}\cdot\vec{n} \big) \nl
	 + \vec{S}_{1}\times\vec{a}_{1}\cdot\vec{v}_{2} \vec{a}_{2}\cdot\vec{n}
	 - 2 \vec{S}_{1}\times\vec{n}\cdot\vec{a}_{2} \big( \vec{a}_{1}\cdot\vec{v}_{2}
	 + \vec{a}_{1}\cdot\vec{n} \vec{v}_{2}\cdot\vec{n} \big)
	 - \vec{S}_{1}\times\vec{v}_{1}\cdot\vec{a}_{2} \vec{a}_{1}\cdot\vec{n} \nl
	 - 2 \vec{S}_{1}\times\vec{a}_{1}\cdot\vec{a}_{2} \big( 4 \vec{v}_{1}\cdot\vec{n}
	 - \vec{v}_{2}\cdot\vec{n} \big)
	 + \vec{S}_{1}\times\vec{v}_{2}\cdot\vec{a}_{2} \vec{a}_{1}\cdot\vec{n} \Big] \nl
- \frac{1}{4} G m_{2} \Big[ 2 \dot{\vec{S}}_{1}\times\vec{n}\cdot\vec{v}_{1} \big( 2 \vec{v}_{1}\cdot\vec{a}_{2}
	 - 2 \vec{v}_{2}\cdot\vec{a}_{2}
	 + \vec{v}_{1}\cdot\vec{n} \vec{a}_{2}\cdot\vec{n} \big)
	 + 2 \dot{\vec{S}}_{1}\times\vec{n}\cdot\vec{a}_{1} \big( v_{2}^2
	 - ( \vec{v}_{2}\cdot\vec{n} )^{2} \big) \nl
	 + 4 \dot{\vec{S}}_{1}\times\vec{v}_{1}\cdot\vec{a}_{1} \vec{v}_{2}\cdot\vec{n}
	 - 2 \dot{\vec{S}}_{1}\times\vec{n}\cdot\vec{v}_{2} \big( \vec{v}_{1}\cdot\vec{a}_{2}
	 - 2 \vec{v}_{2}\cdot\vec{a}_{2}
	 + \vec{v}_{1}\cdot\vec{n} \vec{a}_{2}\cdot\vec{n} \big)
	 + 2 \dot{\vec{S}}_{1}\times\vec{v}_{1}\cdot\vec{v}_{2} \vec{a}_{2}\cdot\vec{n} \nl
	 + 12 \dot{\vec{S}}_{1}\times\vec{a}_{1}\cdot\vec{v}_{2} \vec{v}_{2}\cdot\vec{n}
	 + 2 \dot{\vec{S}}_{1}\times\vec{n}\cdot\vec{a}_{2} \big( v_{2}^2 -2 \vec{v}_{1}\cdot\vec{n} \vec{v}_{2}\cdot\vec{n} \big)
	 - 4 \dot{\vec{S}}_{1}\times\vec{v}_{1}\cdot\vec{a}_{2} \big( 2 \vec{v}_{1}\cdot\vec{n}
	 - \vec{v}_{2}\cdot\vec{n} \big) \nl
	 + 2 \dot{\vec{S}}_{1}\times\vec{v}_{2}\cdot\vec{a}_{2} \vec{v}_{1}\cdot\vec{n} \Big] \nl
- \frac{G^2 m_{1} m_{2}}{r} \Big[ \dot{\vec{S}}_{1}\times\vec{n}\cdot\vec{a}_{1}
	 + 6 \dot{\vec{S}}_{1}\times\vec{n}\cdot\vec{a}_{2} \Big]
- \frac{17 G^2 m_{2}^2}{4 r} \dot{\vec{S}}_{1}\times\vec{n}\cdot\vec{a}_{2}
+ (1 \leftrightarrow 2).
\end{align}
The contribution with three higher-order time derivatives reads
\begin{align}
\stackrel{(3)}{V} =&
\frac{1}{4} G m_{2} r \Big[ ( \vec{S}_{1}\times\vec{n}\cdot\dot{\vec{a}}_{1} \vec{a}_{2}\cdot\vec{n}
	 + \vec{S}_{1}\times\vec{n}\cdot\dot{\vec{a}}_{2} \vec{a}_{1}\cdot\vec{n}
	 + 7 \vec{S}_{1}\times\dot{\vec{a}}_{1}\cdot\vec{a}_{2}
	 + \vec{S}_{1}\times\vec{a}_{1}\cdot\dot{\vec{a}}_{2} ) \nl
	 + (2 \dot{\vec{S}}_{1}\times\vec{n}\cdot\dot{\vec{a}}_{2} \vec{v}_{1}\cdot\vec{n}
	 + 2 \dot{\vec{S}}_{1}\times\vec{v}_{1}\cdot\dot{\vec{a}}_{2} )
	 + ( \ddot{\vec{S}}_{1}\times\vec{n}\cdot\vec{v}_{1} \vec{a}_{2}\cdot\vec{n}
	 - \ddot{\vec{S}}_{1}\times\vec{n}\cdot\vec{v}_{2} \vec{a}_{2}\cdot\vec{n} \nl
	 - 2 \ddot{\vec{S}}_{1}\times\vec{n}\cdot\vec{a}_{2} \vec{v}_{2}\cdot\vec{n}
	 + 7 \ddot{\vec{S}}_{1}\times\vec{v}_{1}\cdot\vec{a}_{2}
	 - \ddot{\vec{S}}_{1}\times\vec{v}_{2}\cdot\vec{a}_{2} ) \nl
	 + (2 \dot{\vec{S}}_{1}\times\vec{n}\cdot\vec{a}_{1} \vec{a}_{2}\cdot\vec{n}
	 + 14 \dot{\vec{S}}_{1}\times\vec{a}_{1}\cdot\vec{a}_{2} )
\Big]
+ (1 \leftrightarrow 2),
\end{align}
and finally, even a term of four higher-order time derivatives appears, and reads
\begin{align}
\stackrel{(4)}{V} =& \frac{1}{4} G m_{2} r^2 \ddot{\vec{S}}_{1}\times\vec{n}\cdot\dot{\vec{a}}_{2}
         + (1 \leftrightarrow 2).
\end{align}
It is obvious that the higher order time derivatives of the velocity and spin bloat the potential.
These can be handled at the level of the EOM through a substitution of 
lower order EOM. However, it is often more useful to perform the elimination of 
higher order time derivatives at the level of the potential, and to also 
transform to a Hamiltonian. This will make the result considerably more compact.

For the reduction of higher order time derivatives we follow the procedure 
outlined in \cite{Damour:1990jh, Levi:2014sba}, and its explicit extension for 
spin variables in \cite{Levi:2014sba}. It should be stressed that this procedure 
is in general not equivalent to a substitution of EOM at the level of the 
potential, but rather to a redefinition of variables, which combines with the higher 
order time derivatives to a total time derivative. Yet, as long as this redefinition contributes only 
linearly to the level of approximation, its result is equivalent to an insertion of the lower 
order EOM, up to a total time derivative. This total time derivative is the same
that arises for the derivation of equations of motion through linear variation of variables.
Variation and redefinition of variables are essentially the same.

We are eliminating the higher order time derivatives successively at each PN order,
starting with the LO spin-orbit potential at 1.5PN order. As in \cite{Tulczyjew:1959}, also see \cite{Damour:1982,Damour:1988mr}, 
and also discussed in \cite{Levi:2015msa}, the variable shift reads 
\begin{equation} \label{positionshift}
\vec{y}_1 \rightarrow \vec{y}_1 + \frac{1}{2 m_1}\vec{S}_1\times\vec{v}_1 ,
\end{equation}
and similar for $\vec{y}_2$. 
This shift corresponds to an insertion of EOM into the LO spin-orbit
potential, where the EOM was derived from the complete 3.5PN order potential.
The redefinition in eq.~\eqref{positionshift} is linear in the spin, so that its square does not contribute to the spin-orbit sector.
After this step the potential, and thus the EOM are changed. The higher order time derivatives at the next PN
orders are then eliminated using these modified EOM (the corresponding redefinitions
of variables are omitted here due to their length). The remaining higher order time derivatives
start to appear at 2PN order, so that the redefinitions would contribute quadratically, and thus the insertion of EOM breaks down, only at the 4PN order.
That is, this successive insertion of EOM to remove higher order time derivatives is still valid to linear in spin and 3.5PN order.

It should be noted that this procedure is different from inserting the 
EOM in the complete 3.5PN order potential in a single step. This would correspond to a redefinition of the
position, which contains among others a contribution from the LO spin-orbit and the 2PN order point-mass
potentials. The quadratic contribution of this redefinition therefore leads to NNLO spin-orbit terms.
That is, the method of inserting EOM breaks down in this case, in contrast to the successive
elimination discussed above.
Furthermore, both procedures can lead to different (yet equivalent) results, since the total time derivatives
generated by the successive linear redefinitions and by the corresponding single quadratic redefinition
are in general different.


Next, we can perform a Legendre transformation to obtain a Hamiltonian. For that 
we need to replace the velocity in terms of canonical momenta, which reads
\begin{align}
v_1 =& \hat{p}_1^{i}
	 - \frac{1}{2} \hat{p}_1^{i} \hat{p}_1^2
+ \frac{G m_2}{2 r} \big[ -6 \hat{p}_1^{i}
	 + 7 \hat{p}_2^{i}
	 + n^{i} \vec{n}\cdot\vec{\hat{p}}_2 \big]
- \frac{2 G}{r^2} \epsilon_{ijk} n^{j} S_2^{k}
- \frac{3 G m_2}{2 m_1 r^2} \epsilon_{ikj} n^{k} S_1^{j} \nl
+ \frac{G}{2 r^2} \big[ -5 \epsilon_{ikj} \hat{p}_2^{k} S_2^{j} \vec{n}\cdot\vec{\hat{p}}_1
	 + 4 \epsilon_{ikj} \hat{p}_2^{k} S_2^{j} \vec{n}\cdot\vec{\hat{p}}_2
	 + 6 \epsilon_{ikj} n^{k} S_2^{j} \vec{n}\cdot\vec{\hat{p}}_1 \vec{n}\cdot\vec{\hat{p}}_2 \nl
	 + 2 \epsilon_{ikj} n^{k} S_2^{j} \vec{\hat{p}}_1\cdot\vec{\hat{p}}_2
	 - 2 \hat{p}_2^{i} \vec{S}_2\times\vec{n}\cdot\vec{\hat{p}}_1
	 - 6 n^{i} \vec{n}\cdot\vec{\hat{p}}_2 \vec{S}_2\times\vec{n}\cdot\vec{\hat{p}}_1
	 - 3 n^{i} \vec{n}\cdot\vec{\hat{p}}_1 \vec{S}_2\times\vec{n}\cdot\vec{\hat{p}}_2 \nl
	 + 6 n^{i} \vec{n}\cdot\vec{\hat{p}}_2 \vec{S}_2\times\vec{n}\cdot\vec{\hat{p}}_2
	 - 5 n^{i} \vec{S}_2\times\vec{\hat{p}}_1\cdot\vec{\hat{p}}_2 \big]
+ \frac{35 G^2 m_1}{4 r^3} \epsilon_{ikj} n^{k} S_2^{j}
+ \frac{6 G^2 m_2}{r^3} \epsilon_{ikj} n^{k} S_2^{j} \nl
+ \frac{G m_2}{8 m_1 r^2} \big[ 16 \epsilon_{ijk} \hat{p}_2^{j} S_1^{k} \vec{n}\cdot\vec{\hat{p}}_1
	 + 5 \epsilon_{ijk} n^{j} S_1^{k} \hat{p}_1^2
	 - 20 \epsilon_{ijk} \hat{p}_2^{j} S_1^{k} \vec{n}\cdot\vec{\hat{p}}_2 \nl
	 + 24 \epsilon_{ijk} n^{j} S_1^{k} \vec{n}\cdot\vec{\hat{p}}_1 \vec{n}\cdot\vec{\hat{p}}_2
	 - 10 \hat{p}_1^{i} \vec{S_1}\times\vec{n}\cdot\vec{\hat{p}}_1
	 - 24 n^{i} \vec{n}\cdot\vec{\hat{p}}_2 \vec{S_1}\times\vec{n}\cdot\vec{\hat{p}}_1
	 + 8 \hat{p}_2^{i} \vec{S_1}\times\vec{n}\cdot\vec{\hat{p}}_2 \nl
	 + 24 n^{i} \vec{n}\cdot\vec{\hat{p}}_2 \vec{S_1}\times\vec{n}\cdot\vec{\hat{p}}_2
	 + 16 n^{i} \vec{S_1}\times\vec{\hat{p}}_1\cdot\vec{\hat{p}}_2
	 - 6 \epsilon_{ijk} n^{j} S_1^{k} ( \vec{n}\cdot\vec{\hat{p}}_2 )^{2} \big]
+ \frac{7 G^2 m_2}{2 r^3} \epsilon_{ijk} n^{j} S_1^{k} \nl
+ \frac{5 G^2 m_2^2}{m_1 r^3} \epsilon_{ijk} n^{j} S_1^{k},
\end{align}
where we have used the abbreviation $\vec{\hat{p}}_a\equiv\vec{p}_a/m_a$.
This results in a compact Hamiltonian:
\begin{align} \label{EFTH}
H^{\text{NNLO}}_{\text{SO}} =& \frac{G m_{2}}{16 r^2} \Big[ \vec{S}_{1}\times\vec{n}\cdot\vec{\hat{p}}_{1} \big( 4 \hat{p}_{1}^2 \vec{\hat{p}}_{1}\cdot\vec{\hat{p}}_{2}
	 + 16 \vec{\hat{p}}_{1}\cdot\vec{\hat{p}}_{2} \hat{p}_{2}^2
	 - 16 ( \vec{\hat{p}}_{1}\cdot\vec{\hat{p}}_{2} )^{2}
	 + 7 \hat{p}_{1}^{4}
	 - 4 \hat{p}_{2}^{4} \nl
	 + 24 \vec{\hat{p}}_{1}\cdot\vec{n} \hat{p}_{1}^2 \vec{\hat{p}}_{2}\cdot\vec{n}
	 + 24 \vec{\hat{p}}_{1}\cdot\vec{n} \vec{\hat{p}}_{2}\cdot\vec{n} \vec{\hat{p}}_{1}\cdot\vec{\hat{p}}_{2}
	 - 12 \hat{p}_{2}^2 ( \vec{\hat{p}}_{1}\cdot\vec{n} )^{2}
	 + 3 \hat{p}_{1}^2 ( \vec{\hat{p}}_{2}\cdot\vec{n} )^{2} \nl
	 - 48 \vec{\hat{p}}_{1}\cdot\vec{\hat{p}}_{2} ( \vec{\hat{p}}_{2}\cdot\vec{n} )^{2}
	 + 24 \hat{p}_{2}^2 ( \vec{\hat{p}}_{2}\cdot\vec{n} )^{2}
	 + 60 ( \vec{\hat{p}}_{1}\cdot\vec{n} )^{2} ( \vec{\hat{p}}_{2}\cdot\vec{n} )^{2}
	 - 15 ( \vec{\hat{p}}_{2}\cdot\vec{n} )^{4} \big) \nl
	 - 4 \vec{S}_{1}\times\vec{n}\cdot\vec{\hat{p}}_{2} \big( 2 \hat{p}_{1}^2 \vec{\hat{p}}_{1}\cdot\vec{\hat{p}}_{2}
	 + \hat{p}_{1}^2 \hat{p}_{2}^2
	 + 2 \vec{\hat{p}}_{1}\cdot\vec{\hat{p}}_{2} \hat{p}_{2}^2
	 - 2 ( \vec{\hat{p}}_{1}\cdot\vec{\hat{p}}_{2} )^{2}
	 + 6 \vec{\hat{p}}_{1}\cdot\vec{n} \hat{p}_{1}^2 \vec{\hat{p}}_{2}\cdot\vec{n} \nl
	 + 6 \vec{\hat{p}}_{1}\cdot\vec{n} \vec{\hat{p}}_{2}\cdot\vec{n} \hat{p}_{2}^2
	 - 3 \hat{p}_{2}^2 ( \vec{\hat{p}}_{1}\cdot\vec{n} )^{2}
	 - 3 \hat{p}_{1}^2 ( \vec{\hat{p}}_{2}\cdot\vec{n} )^{2}
	 + 15 ( \vec{\hat{p}}_{1}\cdot\vec{n} )^{2} ( \vec{\hat{p}}_{2}\cdot\vec{n} )^{2} \big) \nl
	 - 2 \vec{S}_{1}\times\vec{\hat{p}}_{1}\cdot\vec{\hat{p}}_{2} \big( 6 \vec{\hat{p}}_{1}\cdot\vec{n} \hat{p}_{1}^2
	 - 5 \hat{p}_{1}^2 \vec{\hat{p}}_{2}\cdot\vec{n}
	 + 12 \vec{\hat{p}}_{1}\cdot\vec{n} \vec{\hat{p}}_{1}\cdot\vec{\hat{p}}_{2}
	 - 16 \vec{\hat{p}}_{2}\cdot\vec{n} \vec{\hat{p}}_{1}\cdot\vec{\hat{p}}_{2} \nl
	 - 4 \vec{\hat{p}}_{1}\cdot\vec{n} \hat{p}_{2}^2
	 + 4 \vec{\hat{p}}_{2}\cdot\vec{n} \hat{p}_{2}^2
	 + 12 \vec{\hat{p}}_{1}\cdot\vec{n} ( \vec{\hat{p}}_{2}\cdot\vec{n} )^{2}
	 - 18 ( \vec{\hat{p}}_{2}\cdot\vec{n} )^{3} \big) \Big] \nl
- \frac{G^2 m_{1} m_{2}}{8 r^3} \Big[ \vec{S}_{1}\times\vec{n}\cdot\vec{\hat{p}}_{1} \big( 25 \hat{p}_{1}^2
	 + 22 \vec{\hat{p}}_{1}\cdot\vec{\hat{p}}_{2}
	 - 45 \hat{p}_{2}^2
	 + 76 \vec{\hat{p}}_{1}\cdot\vec{n} \vec{\hat{p}}_{2}\cdot\vec{n}
	 - 200 ( \vec{\hat{p}}_{1}\cdot\vec{n} )^{2} \nl
	 + 66 ( \vec{\hat{p}}_{2}\cdot\vec{n} )^{2} \big)
	 - 4 \vec{S}_{1}\times\vec{n}\cdot\vec{\hat{p}}_{2} \big( 11 \hat{p}_{1}^2
	 - 14 \vec{\hat{p}}_{1}\cdot\vec{\hat{p}}_{2} -16 \vec{\hat{p}}_{1}\cdot\vec{n} \vec{\hat{p}}_{2}\cdot\vec{n}
	 - 8 ( \vec{\hat{p}}_{1}\cdot\vec{n} )^{2} \big) \nl
	 + \vec{S}_{1}\times\vec{\hat{p}}_{1}\cdot\vec{\hat{p}}_{2} \big( 165 \vec{\hat{p}}_{1}\cdot\vec{n}
	 - 119 \vec{\hat{p}}_{2}\cdot\vec{n} \big) \Big]
+ \frac{G^2 m_{2}^2}{16 r^3} \Big[ \vec{S}_{1}\times\vec{n}\cdot\vec{\hat{p}}_{1} \big( 42 \hat{p}_{1}^2
	 + 10 \vec{\hat{p}}_{1}\cdot\vec{\hat{p}}_{2}
	 - 57 \hat{p}_{2}^2 \nl
	 + 330 \vec{\hat{p}}_{1}\cdot\vec{n} \vec{\hat{p}}_{2}\cdot\vec{n}
	 + 48 ( \vec{\hat{p}}_{1}\cdot\vec{n} )^{2}
	 - 229 ( \vec{\hat{p}}_{2}\cdot\vec{n} )^{2} \big)
	 - \vec{S}_{1}\times\vec{n}\cdot\vec{\hat{p}}_{2} \big( 13 \hat{p}_{1}^2
	 + 228 \vec{\hat{p}}_{1}\cdot\vec{\hat{p}}_{2} \nl
	 - 201 \hat{p}_{2}^2
	 + 18 \vec{\hat{p}}_{1}\cdot\vec{n} \vec{\hat{p}}_{2}\cdot\vec{n}
	 - 32 ( \vec{\hat{p}}_{1}\cdot\vec{n} )^{2}
	 + 285 ( \vec{\hat{p}}_{2}\cdot\vec{n} )^{2} \big)
	 - 4 \vec{S}_{1}\times\vec{\hat{p}}_{1}\cdot\vec{\hat{p}}_{2} \big( 31 \vec{\hat{p}}_{1}\cdot\vec{n} \nl
	 - 6 \vec{\hat{p}}_{2}\cdot\vec{n} \big) \Big]
+ \frac{G^3 m_{1} m_{2}^2}{8 r^4} \Big[ 193 \vec{S}_{1}\times\vec{n}\cdot\vec{\hat{p}}_{1}
	 - 319 \vec{S}_{1}\times\vec{n}\cdot\vec{\hat{p}}_{2} \Big]
+ \frac{G^3 m_{1}^2 m_{2}}{8 r^4} \Big[ 47 \vec{S}_{1}\times\vec{n}\cdot\vec{\hat{p}}_{1} \nl
	 - 96 \vec{S}_{1}\times\vec{n}\cdot\vec{\hat{p}}_{2} \Big]
+ \frac{9 G^3 m_{2}^3}{4 r^4} \Big[ 5 \vec{S}_{1}\times\vec{n}\cdot\vec{\hat{p}}_{1}
	 - 8 \vec{S}_{1}\times\vec{n}\cdot\vec{\hat{p}}_{2} \Big]
	 + (1 \leftrightarrow 2)
\end{align}
The Poisson brackets are the standard canonical ones as shown in 
\cite{Levi:2015msa}.

\subsection{Resolution via canonical transformations} \label{CT} 

If the EFT Hamiltonian obtained here in eq.~\eqref{EFTH} is physically 
equivalent to that of \cite{Hartung:2011te, Hartung:2013dza}, then there exists 
an infinitesimal generator $g$ of a canonical transformation such that 
\bea\label{eq:gH} 
\Delta H&=&\{H,g\}=\{H_{\text{N}}+H_{\text{1PN}}+H^{\text{SO}}_{\text{LO}},\,\,\,
g_{\text{NNLO}}^{\text{SO}}+g_{\text{NLO}}^{\text{SO}}+g_{\text{2PN}}^{}\}
\nn\\
&=&
\Delta H_{\text{2PN}}+\Delta H_{\text{3PN}}+\Delta H_{\text{SO}}^{\text{NLO}}
+\Delta H_{\text{SO}}^{\text{NNLO}}, 
\eea
where here we have dropped contributions to sectors beyond linear in spin and 
beyond NNLO, and where
\be \label{hdiff}
\Delta H=H_{\text{EFT}}-H_{\text{ADM}}.
\ee
Thus, the contribution to the NNLO spin-orbit sector comprises   
\be \label{ctHnnloso}
\Delta H_{\text{SO}}^{\text{NNLO}}=\{H_N,g_{\text{NNLO}}^{\text{SO}}\}+
\{H_{\text{1PN}},g_{\text{NLO}}^{\text{SO}}\}
+\{H^{\text{SO}}_{\text{LO}},g_{\text{2PN}}^{}\},
\ee
and we also have here contributions to lower orders, given by 
\begin{align} \label{nloctHso}
\Delta H_{\text{SO}}^{\text{NLO}}&=\{H_{\text{N}},g_{\text{NLO}}^{\text{SO}}\},
\nn\\
\Delta H_{\text{2PN}}&=\{H_{\text{N}},g_{\text{2PN}}^{}\},
\end{align}
so we also require that the canonical transformation is consistent with the 
equivalence at NLO of the spin-orbit, and 2PN non spinning Hamiltonians.

Similarly to the construction considerations in \cite{Levi:2010zu,Levi:2014sba}, 
we find for the infinitesimal generator of PN canonical transformations for the 
NNLO spin-orbit sector, $g_{\text{NNLO}}^{\text{SO}}$, the following general form: 
\begin{align} \label{gnnloso}
g_{\text{NNLO}}^{\text{SO}}=&\frac{G m_2}{r}\vec{S}_1\cdot\left[
\vec{\hat{p}}_1\times\vec{\hat{p}}_2 \left(g_1\hat{p}_1^2
+g_2\vec{\hat{p}}_1\cdot\vec{\hat{p}}_2+g_3\hat{p}_2^2
+g_4\left(\vec{\hat{p}}_1\cdot\vec{n}\right)^2
+g_5\vec{\hat{p}}_1\cdot\vec{n} \vec{\hat{p}}_2\cdot\vec{n}
\right.\right.\nn\\
&\left.+g_6\left(\vec{\hat{p}}_2\cdot\vec{n}\right)^2\right)
+\vec{\hat{p}}_1\times\vec{n}\left(
g_7\hat{p}_1^2 \vec{\hat{p}}_1\cdot\vec{n}
+g_8\hat{p}_1^2 \vec{\hat{p}}_2\cdot\vec{n}
+g_9\vec{\hat{p}}_1\cdot\vec{\hat{p}}_2
\vec{\hat{p}}_1\cdot\vec{n}\right.\nn\\
&+g_{10}\vec{\hat{p}}_1\cdot\vec{\hat{p}}_2 \vec{\hat{p}}_2\cdot\vec{n}
+g_{11}\hat{p}_2^2 \vec{\hat{p}}_1\cdot\vec{n}
+g_{12}\hat{p}_2^2 \vec{\hat{p}}_2\cdot\vec{n}
+g_{13}\vec{\hat{p}}_1\cdot\vec{n}
\left(\vec{\hat{p}}_2\cdot\vec{n}\right)^2\nn\\
&\left.+g_{14}\left(\vec{\hat{p}}_1\cdot\vec{n}\right)^2
\vec{\hat{p}}_2\cdot\vec{n}
+g_{15}\left(\vec{\hat{p}}_1\cdot\vec{n}\right)^3
+g_{16}\left(\vec{\hat{p}}_2\cdot\vec{n}\right)^3\right)
+\vec{\hat{p}}_2\times\vec{n}\left(
g_{17}\hat{p}_1^2 \vec{\hat{p}}_1\cdot\vec{n}\right.\nn\\
&+g_{18}\hat{p}_1^2 \vec{\hat{p}}_2\cdot\vec{n}
+g_{19}\vec{\hat{p}}_1\cdot\vec{\hat{p}}_2
\vec{\hat{p}}_1\cdot\vec{n}
+g_{20}\vec{\hat{p}}_1\cdot\vec{\hat{p}}_2
\vec{\hat{p}}_2\cdot\vec{n}
+g_{21}\hat{p}_2^2 \vec{\hat{p}}_1\cdot\vec{n}
+g_{22}\hat{p}_2^2 \vec{\hat{p}}_2\cdot\vec{n}\nn\\
&\left.\left.+g_{23}\vec{\hat{p}}_1\cdot\vec{n}
\left(\vec{\hat{p}}_2\cdot\vec{n}\right)^2
+g_{24}\left(\vec{\hat{p}}_1\cdot\vec{n}\right)^2
\vec{\hat{p}}_2\cdot\vec{n}
+g_{25}\left(\vec{\hat{p}}_1\cdot\vec{n}\right)^3
+g_{26}\left(\vec{\hat{p}}_2\cdot\vec{n}\right)^3\right)
\right]\nn\\
&+\frac{G^2 m_2}{r^2}\vec{S}_1\cdot\left[\vec{\hat{p}}_1\times\vec{\hat{p}}_2
\left(g_{27}m_1+g_{28}m_2\right)\right.\nn\\
&+\vec{\hat{p}}_1\times\vec{n}\left(\vec{\hat{p}}_1\cdot\vec{n}
\left(g_{29}m_1+g_{30}m_2\right)+\vec{\hat{p}}_2\cdot\vec{n}
\left(g_{31}m_1+g_{32}m_2\right)\right)
\nn\\
&\left.+\vec{\hat{p}}_2\times\vec{n}\left(\vec{\hat{p}}_1\cdot\vec{n}
\left(g_{33}m_1+g_{34}m_2\right)+\vec{\hat{p}}_2\cdot\vec{n}
\left(g_{35}m_1+g_{36}m_2\right)\right)\right].
\end{align}

We should also have the generators contributing first to lower orders, as noted 
in eq.~(\ref{ctHnnloso}), so that their coefficients are already set from 
eq.~(\ref{nloctHso}).
For the NLO spin-orbit sector we have the generator from eq.~(7.8) in 
\cite{Levi:2014sba, Levi:2010zu} with its coefficients set to the values
\begin{align}\label{gnlosofix}
g_1&=-\frac{1}{2},\,\,g_2=0,\,\,g_3=\frac{1}{2},\,\,g_4=0,\,\,g_5=0.
\end{align} 
We should also take into account the generator, which contributes 
first at the 2PN non spinning sector, from eq.~(7.10) in \cite{Levi:2014sba}
with its coefficients set to 
\begin{align} \label{g2pnfix}
g_1&=0,\,\,g_2=-\frac{1}{2},\,\,g_3=0,\,\,g_4=0,\,\,g_5=0,\,\,g_6=0,
\,\,g_7=-\frac{1}{4}.
\end{align} 

Thus, we plug in eq.~(\ref{ctHnnloso}) our ansatz for $g_{\text{NNLO}}^{\text{SO}}$ 
from eq.~(\ref{gnnloso}), together with the fixed generators in 
eqs.~(\ref{gnlosofix}), (\ref{g2pnfix}), and we 
compare that to eq.~(\ref{hdiff}). Comparing $O(G)$ terms fixes the $O(G)$ 
coefficients of $g_{\text{NNLO}}^{\text{SO}}$ to the values
\begin{align}
g_{1}&= \frac{7}{16}, & g_{2}&= -\frac{3}{2}, & g_{3}&= \frac{9}{16}, & g_{4}&= 0, 
& g_{5}&= 0, & g_{6}&= -\frac{13}{16}, & g_{7}&= 0, \nn\\
g_{8}&= \frac{1}{16}, & g_{9}&= 0, & g_{10}&= -1, & g_{11}&= \frac{1}{4}, 
& g_{12}& = \frac{7}{16}, & g_{13}&= -\frac{3}{4}, & g_{14}&= 0, \nn\\ 
g_{15}&= 0, & g_{16}&= -\frac{3}{16}, & g_{17}&= 0, & g_{18}&= 0, & g_{19}&= 0, & g_{20}&= 0, & 
g_{21}&= 0,  \nn\\ 
g_{22}&= 0, & g_{23}&= 0, & g_{24}&= 0, & g_{25}&= 0, & g_{26}&= 0.
\end{align}
This eliminates all of the $O(G)$ terms in the difference. Comparing the remaining 
$O(G^2)$ terms in the difference fixes the $O(G^2)$ coefficients of 
$g_{\text{NNLO}}^{\text{SO}}$ to the values
\begin{align}
g_{27}&= -\frac{19}{16}, & g_{28}&= -\frac{1}{2}, & g_{29}&= -\frac{53}{8}, 
& g_{30}&= -\frac{3}{4},
& g_{31}&= -3, \nn\\
g_{32}&= -\frac{17}{4}, & g_{33}&= 2, & g_{34}&= -\frac{1}{2}, & g_{35}&= 0, 
& g_{36}&= -\frac{11}{2}.
\end{align}
This eliminates all of the $O(G^2)$ terms, as well as all terms at $O(G^3)$ 
in the difference. Hence, we have shown that the ADM Hamiltonian and the EFT 
Potential at NNLO spin-orbit are completely equivalent.

\subsection{Hamiltonians in the center of mass frame}

Since the center of mass of a binary moves uniformly along a straight line,
it makes sense to separate this motion and reduce the number of orbital variables.
This is achieved by a specialization of the Hamiltonian to the center of mass frame,
where the total linear momentum vanishes. Here the total linear momentum is just the
sum of the individual momenta, $\vec{p}_1 + \vec{p}_2 = 0$, since we are considering
the conservative sector where the recoil due to gravitational waves vanishes.
The orbital dynamics is then governed by a single canonical momentum
$\vec{p} \equiv \vec{p}_1 = - \vec{p}_2$ and its conjugate $\vec{r} = \vec{y}_1 - \vec{y}_2$,
the binary separation vector.

We further introduce the tetrad basis $\vec{n}$, $\vec{\lambda}$, $\vec{l}$, where
$\vec{l} = \vec{L} / L$, $\vec{\lambda} = \vec{l} \times \vec{n}$, and the orbital
angular momentum is $\vec{L} = r \vec{n} \times \vec{p}$, where a similar tetrad was used in
\cite{Blanchet:2011zv} for an approximate solution of the EOM with spin. This is a sensible choice for the description of a spinning binary, as it allows, for instance, a separation of $\vec{p}$ into circular and radial components,
\begin{equation}
\vec{p} = p_r \vec{n} + \frac{L}{r} \vec{\lambda} ,
\end{equation}
where $p_r = \vec{p} \cdot \vec{n}$. However, we will not expand the spin products as in
\begin{align}\label{SSproduct}
\vec{S}_A \cdot \vec{S}_B =
      \vec{S}_A \cdot \vec{n} \, \vec{n} \cdot \vec{S}_B
      + \vec{S}_A \cdot \vec{\lambda} \, \vec{\lambda} \cdot \vec{S}_B
      + \vec{S}_A \cdot \vec{l} \, \vec{l} \cdot \vec{S}_B ,
\end{align}
since these spin products better highlight the structure of the
interaction and keep expressions more compact. Furthermore, the products
$\vec{S}_A \cdot \vec{S}_A$ are constants in this work. As a convention, we then replace multiple products
involving $\vec{L}$ by the products $\vec{S}_A \cdot \vec{S}_B$, so with this prescription
at most one scalar product containing $\vec{L}$ remains in each term.

As in \cite{Levi:2014sba} we are expressing the results in terms of dimensionless variables, where
dimensions of length are rescaled by the total mass $G m = G(m_1 + m_2)$ and dimensions of mass are rescaled by the
reduced mass $\mu = m_1 m_2 / m$. This rescaling is denoted by a tilde, e.g., $\tilde{L} \equiv \frac{L}{G m \mu}$.
The dependence on the masses is then reduced to the mass ratio $q \equiv m_1 / m_2$, or the symmetric
mass ratio $\nu \equiv \mu / m$.

We use here the Hamiltonians arising from the formalism recently set up in \cite{Levi:2015msa}, and
hence within the corresponding gauge choices.
We include also the results from the cubic order in spin from \cite{Levi:2014gsa}, see also \cite{Hergt:2007ha, Hergt:2008jn} for the black hole case, and \cite{Marsat:2014xea}.
All expressions here are then complete to the 3.5PN order with spins.
Before we present the spin-dependent Hamiltonians, we provide for completeness the
point-mass Hamiltonians to 2PN order in the same formalism \cite{Gilmore:2008gq}
and center of mass frame, reading
\begin{align}
H_\text{N} =&
         \frac{\tilde{L}^2}{2 \tilde{r}^2}
	 - \frac{1}{\tilde{r}}
         + \frac{\tilde{p}_r^2}{2} , \\
H_\text{1PN} =&
         \frac{1}{8} \bigg[ \frac{\tilde{L}^2}{\tilde{r}^2} + \tilde{p}_r^2 \bigg]^2 (3 \nu - 1)
	 - \frac{\tilde{L}^2}{2 \tilde{r}^3} (3 + \nu)
	 + \frac{1}{2 \tilde{r}^2}
	 - \frac{\tilde{p}_r^2}{2 \tilde{r}} \left( 3 + 2 \nu \right) , \\
H_\text{2PN} =&
	 \frac{\tilde{L}^6}{16 \tilde{r}^6} (1 - 5 \nu + 5 \nu^2)
	 + \frac{\tilde{L}^4}{8 \tilde{r}^5} (5 - 16 \nu - 3 \nu^2)
	 + \frac{\tilde{L}^2}{4 \tilde{r}^4} (11 + 12 \nu)
	 - \frac{1}{4 \tilde{r}^3} (2 + \nu) \nl
	 + \frac{3 \tilde{L}^4 \tilde{p}_r^2}{16 \tilde{r}^4} (1 - 5 \nu + 5 \nu^2)
	 + \frac{\tilde{L}^2 \tilde{p}_r^2}{4 \tilde{r}^3} (5 - 18 \nu - 4 \nu^2)
	 + \frac{9 \tilde{p}_r^2}{4 \tilde{r}^2} (1 + 2 \nu) \nl
	 + \frac{3 \tilde{L}^2 \tilde{p}_r^4}{16 \tilde{r}^2} (1 - 5 \nu + 5 \nu^2)
	 + \frac{\tilde{p}_r^4}{8 \tilde{r}} ( 5 - 20 \nu - 8 \nu^2)
         + \frac{\tilde{p}_r^6}{16} (1 - 5 \nu + 5 \nu^2).
\end{align}
The 2PN order in the point-mass sector is sufficient for deriving the spin-dependent part of the binding energy to 4PN order,
or to transform the spin-dependent Hamiltonians to EOB gauge up to 4PN order.

The center of mass spin-orbit Hamiltonians to 3.5PN order read
\begin{align}
H_\text{LO}^\text{SO} =& \frac{\nu \tilde{\vec{L}}\cdot\tilde{\vec{S}}_{1}}{2 \tilde{r}^3} (4 + 3 q^{-1} ) + [1 \leftrightarrow 2] , \\
H_\text{NLO}^\text{SO} =& \bigg[ \frac{13 \nu^2 \tilde{L}^2}{8 \tilde{r}^5}
	 - \frac{\nu}{4 \tilde{r}^4} (24 + 5 \nu)
	 + \frac{43 \nu^2 \tilde{p}_r^2}{8 \tilde{r}^3} \bigg] \tilde{\vec{L}}\cdot\tilde{\vec{S}}_{1} \nl
+ \bigg[ \frac{5 \nu \tilde{L}^2}{8 \tilde{r}^5} ( 2 \nu - 1 )
	 - \frac{5 \nu}{4 \tilde{r}^4} (4 + \nu)
	 + \frac{\nu \tilde{p}_r^2}{8 \tilde{r}^3} (34 \nu - 5) \bigg] \frac{\tilde{\vec{L}}\cdot\tilde{\vec{S}}_{1}}{q} + [1 \leftrightarrow 2] , \\
H_\text{NNLO}^\text{SO} =& \bigg[ 
	 \frac{\nu^2 \tilde{L}^4}{16 \tilde{r}^7} (22 \nu - 19)
	 - \frac{\nu^2 \tilde{L}^2}{16 \tilde{r}^6} (265 + 17 \nu)
	 + \frac{3 \nu}{2 \tilde{r}^5} (8 + 7 \nu)
         + \frac{\nu^2 \tilde{L}^2 \tilde{p}_r^2}{8 \tilde{r}^5} (46 \nu - 19) \nl
         - \frac{9 \nu^2 \tilde{p}_r^2}{16 \tilde{r}^4} (32 + 7 \nu)
	 + \frac{\nu^2 \tilde{p}_r^4}{8 \tilde{r}^3} (80 \nu - 17)
\bigg] \vec{\tilde{L}}\cdot\vec{\tilde{S}}_{1}
+ \bigg[
	 \frac{\nu \tilde{L}^4}{16 \tilde{r}^7} (7 - 32 \nu + 17 \nu^2) \nl
	 + \frac{\nu \tilde{L}^2}{16 \tilde{r}^6} (42 - 173 \nu - 17 \nu^2)
	 + \frac{\nu}{8 \tilde{r}^5} (90 + 61 \nu)
         + \frac{\nu \tilde{L}^2 \tilde{p}_r^2}{16 \tilde{r}^5} (14 - 88 \nu + 73 \nu^2) \nl
         + \frac{9 \nu \tilde{p}_r^2}{16 \tilde{r}^4} (10 - 31 \nu - 7 \nu^2)
	 + \frac{\nu \tilde{p}_r^4}{16 \tilde{r}^3} (7 - 56 \nu + 131 \nu^2)
\bigg] \frac{\vec{\tilde{L}}\cdot\vec{\tilde{S}}_{1}}{q}
+ [1 \leftrightarrow 2] ,
\end{align}
where the NNLO Hamiltonian was derived in the present paper and the lower order Hamiltonians
are taken from \cite{Levi:2015msa}. The relevant Hamiltonians at quadratic order in spin
are also given in \cite{Levi:2015msa}. After transformation to the center of mass form,
they read
\begin{align}
H_\text{LO}^\text{S$_1$S$_2$} =& \frac{\nu}{\tilde{r}^3}
         \bigl[ 3 \vec{n}\cdot\vec{\tilde{S}}_{1} \vec{n}\cdot\vec{\tilde{S}}_{2}
	 - \vec{\tilde{S}}_{1}\cdot\vec{\tilde{S}}_{2} \bigr] , \\
H_\text{NLO}^\text{S$_1$S$_2$} =&
- \bigg[ \frac{\nu \tilde{L}^2}{2 \tilde{r}^5} (5 + 2 \nu)
	 - \frac{7 \nu}{\tilde{r}^4}
	 + \frac{7 \nu \tilde{p}_r^2}{4 \tilde{r}^3} (-2 + \nu)
\bigg] \vec{\tilde{S}}_{1}\cdot\vec{\tilde{S}}_{2}
+ \frac{\nu \tilde{L}^2}{2 \tilde{r}^5}
         (5 + \nu) \vec{\tilde{S}}_{1}\cdot\vec{\lambda} \vec{\tilde{S}}_{2}\cdot\vec{\lambda} \nl
+ \bigg[ \frac{3 \nu \tilde{L}^2}{4 \tilde{r}^5} (2 + 3 \nu)
	 - \frac{13 \nu}{\tilde{r}^4}
	 + \frac{\nu \tilde{p}_r^2}{4 \tilde{r}^3} (-14 + 23 \nu)
\bigg] \vec{n}\cdot\vec{\tilde{S}}_{1} \vec{n}\cdot\vec{\tilde{S}}_{2}
+ \frac{\nu \tilde{L} \tilde{p}_r}{4 \tilde{r}^4} (4 - 25 \nu) \nl
         \times \big[ \vec{n}\cdot\vec{\tilde{S}}_{2} \vec{\tilde{S}}_{1}\cdot\vec{\lambda}
         + \vec{n}\cdot\vec{\tilde{S}}_{1} \vec{\tilde{S}}_{2}\cdot\vec{\lambda} \big]
- \frac{9 \nu^2 \tilde{L} \tilde{p}_r}{2 \tilde{r}^4}
         \bigg[ q \vec{n}\cdot\vec{\tilde{S}}_{2} \vec{\tilde{S}}_{1}\cdot\vec{\lambda}
         + \frac{1}{q} \vec{n}\cdot\vec{\tilde{S}}_{1} \vec{\tilde{S}}_{2}\cdot\vec{\lambda} \bigg] , \\
H_\text{LO}^\text{S$^2$} =& \frac{\nu C_{1(\text{ES}^2)}}{2 q \tilde{r}^3}
	 \big[ 3 (\vec{n}\cdot\vec{\tilde{S}}_{1} )^{2} - \tilde{S}_{1}^2 \big]
          + [1 \leftrightarrow 2] , \\
H_\text{NLO}^\text{S$^2$} =&
\nu^2 \bigg\{
- \bigg[ \frac{5 \tilde{L}^2}{4 \tilde{r}^5}
	 - \frac{1}{\tilde{r}^4}
	 + \frac{\tilde{p}_r^2}{8 \tilde{r}^3}
\bigg] \tilde{S}_{1}^2
+ \bigg[ \frac{21 \tilde{L}^2}{8 \tilde{r}^5}
	 - \frac{2}{\tilde{r}^4}
	 + \frac{\tilde{p}_r^2}{8 \tilde{r}^3}
\bigg] ( \vec{n}\cdot\vec{\tilde{S}}_{1} )^{2} \nl
+ \frac{5 \tilde{L}^2}{4 \tilde{r}^5} \vec{\tilde{S}}_{1}\cdot\vec{\lambda}
         \bigg[ \vec{\tilde{S}}_{1}\cdot\vec{\lambda}
         - \frac{\tilde{r} \tilde{p}_r}{\tilde{L}} \vec{n}\cdot\vec{\tilde{S}}_{1} \bigg]
\bigg\}
+ \nu^2 C_{1(\text{ES}^2)} \bigg\{
\bigg[ \frac{\tilde{L}^2}{2 \tilde{r}^5}
	 - \frac{3}{2 \tilde{r}^4}
	 - \frac{\tilde{p}_r^2}{\tilde{r}^3}
\bigg] \tilde{S}_{1}^2 \nl
+ \bigg[ \frac{7}{2 \tilde{r}^4} + \frac{\tilde{p}_r^2}{\tilde{r}^3} \bigg]
         ( \vec{n}\cdot\vec{\tilde{S}}_{1} )^{2}
- \frac{\tilde{L}^2}{2 \tilde{r}^5} \vec{\tilde{S}}_{1}\cdot\vec{\lambda}
         \bigg[ \vec{\tilde{S}}_{1}\cdot\vec{\lambda} 
         - \frac{\tilde{r} \tilde{p}_r}{\tilde{L}} \vec{n}\cdot\vec{\tilde{S}}_{1} \bigg]
\bigg\} \nl
+ \frac{\nu}{q} \bigg\{
\bigg[ \frac{1}{\tilde{r}^4} (1 + \nu)
	 + \frac{\tilde{p}_r^2}{8 \tilde{r}^3} (1 - 2 \nu)
\bigg] \tilde{S}_{1}^2
+ \frac{\tilde{L}^2}{4 \tilde{r}^5} (5 - 4 \nu)
         \bigg[ \tilde{S}_{1}^2
         - ( \vec{\tilde{S}}_{1}\cdot\vec{\lambda} )^{2}
         + \frac{\tilde{r} \tilde{p}_r}{\tilde{L}} \vec{n}\cdot\vec{\tilde{S}}_{1} \vec{\tilde{S}}_{1}\cdot\vec{\lambda} \bigg] \nl
- \bigg[ \frac{3 \tilde{L}^2}{8 \tilde{r}^5} (7 - 6 \nu)
	 + \frac{1}{\tilde{r}^4} (1 + 2 \nu)
	 + \frac{\tilde{p}_r^2}{8 \tilde{r}^3} (1 - 2 \nu)
\bigg] ( \vec{n}\cdot\vec{\tilde{S}}_{1} )^{2}
\bigg\}
+ \frac{\nu}{q} C_{1(\text{ES}^2)} \bigg\{
\bigg[ \frac{\tilde{L}^2}{4 \tilde{r}^5} (\nu - 5) \nl
	 + \frac{1}{2 \tilde{r}^4} (4 - 3 \nu)
	 + \frac{\tilde{p}_r^2}{4 \tilde{r}^3} (1 - 8 \nu)
\bigg] \tilde{S}_{1}^2
- \frac{\tilde{L} \tilde{p}_r}{2 \tilde{r}^4}
         (1 + 2 \nu) \vec{n}\cdot\vec{\tilde{S}}_{1} \vec{\tilde{S}}_{1}\cdot\vec{\lambda}
+ \frac{\tilde{L}^2}{2 \tilde{r}^5} (1 - \nu) ( \vec{\tilde{S}}_{1}\cdot\vec{\lambda} )^2 \nl
+ \bigg[ \frac{3 \tilde{L}^2}{4 \tilde{r}^5} (3 + \nu)
	 + \frac{1}{2 \tilde{r}^4} (7 \nu - 10)
	 + \frac{\tilde{p}_r^2}{4 \tilde{r}^3} (5 + 16 \nu)
\bigg] ( \vec{n}\cdot\vec{\tilde{S}}_{1} )^{2}
\bigg\} + [1 \leftrightarrow 2] .
\end{align}
The LO cubic in spin Hamiltonian at 3.5PN order from \cite{Levi:2014gsa} reads in the center of mass frame 
\begin{align}
H_\text{LO}^\text{S$^3$} =& \frac{3 \nu^2}{2 \tilde{r}^5}
\vec{\tilde{L}}\cdot\vec{\tilde{S}}_{1} \bigg\{
\frac{1}{6 q^2}
          \big[ \tilde{S}_{1}^2 - 5 ( \vec{n}\cdot\vec{\tilde{S}}_{1} )^{2} \big]
          \bigl[ 4 C_{1(\text{BS}^3)} (1 + q) - 3 C_{1(\text{ES}^2)} \bigr]
+ q \big[ \tilde{S}_{2}^2
         - ( \vec{\tilde{S}}_{2}\cdot\vec{\lambda} )^{2} \big] \nl
- \frac{1}{q} \big[ 2 \vec{\tilde{S}}_{1}\cdot\vec{\tilde{S}}_{2}
         - 5 \vec{n}\cdot\vec{\tilde{S}}_{1} \vec{n}\cdot\vec{\tilde{S}}_{2}
	 - \vec{\tilde{S}}_{1}\cdot\vec{\lambda} \vec{\tilde{S}}_{2}\cdot\vec{\lambda} \big]
+ \frac{C_{2(\text{ES}^2)}}{2} (3 + 4 q) \bigg[
	 3 \tilde{S}_{2}^2 - 5 ( \vec{n}\cdot\vec{\tilde{S}}_{2} )^{2} \nl
	         - 2 ( \vec{\tilde{S}}_{2}\cdot\vec{\lambda} )^{2}
         - \frac{2 \tilde{r} \tilde{p}_r}{\tilde{L}}
                 \vec{n}\cdot\vec{\tilde{S}}_{2} \vec{\tilde{S}}_{2}\cdot\vec{\lambda} \bigg]
- C_{1(\text{ES}^2)} (3 + 4 {q^{-1}}) \bigg[
	 \vec{\tilde{S}}_{1}\cdot\vec{\tilde{S}}_{2} - \vec{\tilde{S}}_{1}\cdot\vec{\lambda} \vec{\tilde{S}}_{2}\cdot\vec{\lambda} \nl
         - \frac{\tilde{r} \tilde{p}_r}{\tilde{L}} \vec{n}\cdot\vec{\tilde{S}}_{1} \vec{\tilde{S}}_{2}\cdot\vec{\lambda} \bigg]
+ \frac{\tilde{r} \tilde{p}_r}{\tilde{L}} \vec{\tilde{S}}_{2}\cdot\vec{\lambda}
         \vec{n}\cdot \bigg[ \frac{\vec{\tilde{S}}_{1}}{q} - q \vec{\tilde{S}}_{2} \bigg]
\bigg\} + [1 \leftrightarrow 2] .
\end{align}
Recall that in the black hole case it holds $C_{A(\text{ES}^2)} = 1 = C_{A(\text{BS}^3)}$,
and that $[1 \leftrightarrow 2]$ implies $q \leftrightarrow q^{-1}$. It is nice to note
that in the representation that we picked, the Hamiltonian $H_\text{LO}^\text{S$^3$}$
appears as a correction to the spin-orbit interaction at 3.5PN order, which is nonlinear in the spins.
It should also be noted though that the representation is not unique. One can expand
a scalar product of spins with eq.~\eqref{SSproduct}, which produces two products involving
$\vec{L}$, and then absorb two different products involving $\vec{L}$ by rewriting
them in terms of a spin product. 

All of the results in the present section are still gauge dependent. We are going to work out gauge invariant
relations in the following section, which however require restrictions on the
orbit and spin orientations.

\section{Complete gauge invariant relations to 3.5PN order with spins}
\label{GI}

Starting from the Hamiltonians given in the last section, it is straightforward to obtain
gauge invariant relations between the binding energy and the orbital frequency or orbital
angular momentum for circular orbits.
Circular orbits are defined by $\tilde{r}=\text{const}$. This holds if the radial
momentum vanishes, $\tilde{p}_r = 0$.
Since there is no obvious way to define spin orientations in a gauge invariant manner,
we further restrict to aligned spins, $\vec{S}_A \cdot \vec{n} = 0 = \vec{S}_A \cdot \vec{\lambda}$,
and $\vec{S}_A \cdot \vec{L} = S_A L$.
The relations presented here were given in part already in \cite{Levi:2014sba}.
Here we complete these relations to the 3.5PN order with spins by including
recent results for the cubic order in spin \cite{Levi:2014gsa}. The computation follows that in \cite{Levi:2014sba}.

The circular orbit relation between $\tilde{L}$ and $\tilde{r}$ takes 
the form
\begin{align}
\frac{1}{\tilde{r}} =&
\frac{1}{\tilde{L}^2}
+ \frac{4}{\tilde{L}^4}
+ \frac{1}{\tilde{L}^6} \bigg[ 25 - \frac{43 \nu}{8} \bigg]
+ \bigg[ \frac{3 \nu}{\tilde{L}^6} + \frac{\nu}{2 \tilde{L}^8} (290 + 19 \nu) \bigg] \tilde{S}_1 \tilde{S}_2 \nl
+ \bigg\{ - \bigg[ \frac{6 \nu}{\tilde{L}^5}
	 + \frac{\nu}{8 \tilde{L}^7} ( 648 - 47 \nu)
         + \frac{\nu}{16 \tilde{L}^9} ( 15348 - 4508 \nu + 13 \nu^2)
 \bigg] \tilde{S}_1 \nl
- \bigg[ \frac{9 \nu}{2 \tilde{L}^5}
	 + \frac{\nu}{8 \tilde{L}^7} (445 - 44 \nu)
	 + \frac{\nu}{16 \tilde{L}^9} (10029 - 3614 \nu + 14 \nu^2) \bigg] \frac{\tilde{S}_1}{q} \nl
+ \bigg[ \frac{3 \nu}{2 \tilde{L}^6} C_{1(\text{ES}^2)}
         + \frac{\nu}{4 \tilde{L}^8} \big[ 121 + 112 \nu + 2 (49 + 5 \nu) C_{1(\text{ES}^2)} \big]
	  \bigg] \frac{\tilde{S}_1^2}{q}
+ \frac{\nu^2}{4 \tilde{L}^8} (135 + 14 C_{1(\text{ES}^2)}) \tilde{S}_1^2 \nl
- \frac{\nu^2}{\tilde{L}^9} \bigg[
\bigg( 5 q (1 + q) C_{2(\text{BS}^3)}
         + \frac{3}{4} q (48 + 31 q) C_{2(\text{ES}^2)} \bigg) \tilde{S}_2^3 \nl
+ \bigg( 72 + \frac{93}{2} q
	 + \frac{51}{4} (3 + 4 q) C_{2(\text{ES}^2)} \bigg) \tilde{S}_1 \tilde{S}_2^2
\bigg]
+ [1 \leftrightarrow 2]
\bigg\} .
\end{align}
This relation is gauge dependent and therefore differs from the result in ADM gauge
given by eq.~(8.22) in \cite{Levi:2014sba}.
A gauge invariant quantity is the orbital frequency $\omega$, which is commonly
expressed in terms of $x = \tilde{\omega}^{2/3}$ in the binding energy.
The gauge invariant relation between $\tilde{L}$ and $x$ has the following addition
compared to eq.~(8.24) in \cite{Levi:2014sba}:
\begin{align}
\frac{1}{\tilde{L}^2} =& x + \dots + x^{9/2} \nu^2 \bigg[
\big[ 6 C_{2(\text{BS}^3)} - 26 C_{2(\text{ES}^2)}
	 + 6 q (C_{2(\text{BS}^3)} - 4 C_{2(\text{ES}^2)})
\big] q \tilde{S}_2^3 \nl
- \big[ 52 + 6 C_{2(\text{ES}^2)}
	 + 8 q (6 + C_{2(\text{ES}^2)})
\big] \tilde{S}_1 \tilde{S}_2^2
+ [1 \leftrightarrow 2] \bigg] .
\end{align}

Finally, the gauge invariant relation for the binding energy as a function of the
orbital angular momentum reads
\def\LADM{\tilde{L}}
\def\Sf{\tilde{S}_1}
\def\Ss{\tilde{S}_2}
\def\nnl{\nn\\ & \quad}
\begin{align} \label{EJ}
e_{\text{spin}}(\LADM) &=
\left(\Sf+\Ss\right) \frac{\nu}{\LADM^5} \left[ 2
        + \frac{1}{\LADM^2} \left( \frac{3 \nu}{8}+18 \right)
        + \frac{1}{\LADM^4} \left( \frac{5 \nu^2}{16}-27 \nu+162 \right) \right] \nnl
+ \left(\Sf / q + \Ss q\right) \frac{\nu}{\LADM^5} \left[ \frac{3}{2}
        + \frac{99}{8 \LADM^2}
        - \frac{1}{\LADM^4} \left(\frac{195 \nu}{8}-\frac{1701}{16}\right) \right] \nnl
- \frac{\nu  \Sf \Ss}{\LADM^6} \left[ 1
        + \frac{1}{\LADM^2} \left(\frac{13 \nu}{4}+\frac{69}{2}\right)
        \right]
- \bigg\{ \frac{\nu \Ss^2}{\LADM^6} \bigg[
        q \bigg( \frac{C_{2(\text{ES}^2)}}{2}
        +\frac{1}{8\LADM^2} \left(54\nu+63\right) \nnl
        +\frac{C_{2(\text{ES}^2)}}{4\LADM^2} \left(5\nu+21\right) \bigg)
        + \frac{\nu}{\LADM^2} \left( \frac{65}{8} + C_{2(\text{ES}^2)} \right) \bigg]
- \frac{\nu^2}{\LADM^9} \bigg[ q \Ss^3 \big[ q ( 6 C_{2(\text{ES}^2)} + C_{2(\text{BS}^3)} ) \nnl
        + 9 C_{2(\text{ES}^2)} + C_{2(\text{BS}^3)} \big]
+ \Sf \Ss^2 \big[ 12 (1+C_{2(\text{ES}^2)}) q + 9 (2+C_{2(\text{ES}^2)}) \big] \bigg] + [1 \leftrightarrow 2] \bigg\} ,
\end{align}
and as a function of $x$ it reads
\begin{align}
e_{\text{spin}}(x) &=
\nu x^{5/2} \left(\Sf+\Ss\right) \left[ - \frac{4}{3}
        + x \left( \frac{31 \nu}{18} - 4 \right)
        - x^2 \left( \frac{7 \nu^2}{12}-\frac{211 \nu}{8}+\frac{27}{2} \right) \right] \nnl
+ \nu x^{5/2} \left(\Sf / q + \Ss q\right) \left[ - 1
        + x \left( \frac{5 \nu}{3}-\frac{3}{2} \right)
        - x^2 \left( \frac{5 \nu^2}{8}-\frac{39 \nu}{2}+\frac{27}{8} \right) \right] \nnl
+ \nu x^3 \Sf \Ss \left[ 1
        + x \left(\frac{5 \nu}{18}+\frac{5}{6}\right)
        \right]
+ \bigg\{ \nu x^3 \Ss^2 \bigg[ q \bigg( \frac{C_{2(\text{ES}^2)}}{2}
        + \frac{5 x}{6} \left(\nu-3\right) \nnl
        + \frac{5 x C_{2(\text{ES}^2)}}{4} \left(\nu+1\right) \bigg)
        + \nu x \left(\frac{25}{18}+\frac{5 C_{2(\text{ES}^2)}}{3}\right) \bigg]
+ \nu^2 x^{9/2} \bigg[ q \Ss^3 \big[
        2 ( C_{2(\text{ES}^2)} - C_{2(\text{BS}^3)} ) \nnl
        + q ( 3 C_{2(\text{ES}^2)} - 2 C_{2(\text{BS}^3)} ) \big]
+ \Sf \Ss^2 \big[ q (6-4 C_{2(\text{ES}^2)}) + ( 4 - 3 C_{2(\text{ES}^2)} ) \big]
\bigg]
+ [1 \leftrightarrow 2] \bigg\} .
\end{align}
The point-mass gauge invariant relations to 4PN order can be found in \cite{Damour:2014jta}.

\section{Conclusions} \label{theendmyfriend}

In this work we implemented the EFT for gravitating spinning objects in the PN 
scheme \cite{Levi:2015msa} at the NNLO level, which was first treated in 
\cite{Levi:2011eq}. We derived the NNLO spin-orbit interaction potential at the 
3.5PN order for rapidly rotating compact objects. Such high PN orders are 
required for the successful detection of gravitational radiation, as the EOB 
Hamiltonian, e.g., requires parameters for the even higher 5PN and 6PN orders in 
the point-mass case, in order to produce good waveforms. From the NNLO spin-orbit 
interaction potential, which we obtain here in a Lagrangian form for the first 
time, we directly derived the corresponding Hamiltonian. We then compared our 
result to the ADM Hamiltonian result \cite{Hartung:2011te}, and arrived at a 
complete agreement between the ADM and EFT results. Therefore, in order to 
complete the spin dependent conservative sector to 4PN order, it remains to apply 
the EFT for gravitating spinning objects \cite{Levi:2015msa} at NNLO to quadratic 
level in the spin, and for higher order in spin finite size effects, as was already done in 
\cite{Levi:2014gsa} for cubic and quartic orders in spin. 
The NNLO spin-squared result is indeed presented in another recent paper \cite{Levi:2015ixa}, which then completes the conservative sector to 4PN order. Finally, we provide the relevant Hamiltonians in the center of mass frame, and the complete gauge invariant relations among the binding energy, angular momentum, and orbital frequency of an inspiralling binary with generic compact spinning components to 3.5PN order.  

The spin-orbit sector constitutes the most elaborate spin dependent sector at
each order, and accordingly we encountered here a proliferation of the
relevant Feynman diagrams, where there are 132 diagrams contributing to this
sector, and a significant increase of the computational complexity, e.g.~there 
are 32 two-loop diagrams here.
We also recall that as the spin is derivative-coupled, higher-order tensor 
expressions are required for all integrals involved in
the calculations, compared to the non spinning case. However, the computation
is made efficient through the use of the ``NRG'' fields, which are advantageous
also in the spin dependent sectors, together with the various gauge choices
included in the EFT for gravitating spinning objects \cite{Levi:2015msa}. In
addition, we automatized the EFT computations here, and carried out the
automated computations in parallel. Hence, it is clear that for higher order
corrections automated EFT computations, utilizing the ``NRG'' fields, should be
implemented, and are most powerful and efficient. It should be stressed that
in order to obtain such higher order results, all lower order results are 
required consistently within one formalism, and so also for that the derivation 
presented in this work is essential. This work then paves the way for the 
obtainment of the next-to-NNLO spin-orbit interaction potential at 4.5PN order 
for rapidly rotating compact objects, once this level of accuracy would be 
approached.

\acknowledgments

This work has been done within the Labex ILP (reference ANR-10-LABX-63) part 
of the Idex SUPER, and received financial French state aid managed by the 
Agence Nationale de la Recherche, as part of the programme Investissements 
d'Avenir under the reference ANR-11-IDEX-0004-02. ML is grateful for the kind 
hospitality at the Max Planck Institute for Gravitational Physics (AEI) during 
the final stages of this work.

\appendix

\section{Irreducible two-loop tensor integrals} \label{reduce}

In the evaluation of the irreducible two-loop diagrams we encounter irreducible 
two-loop tensor integrals up to order 3. These are reduced using the integration 
by parts method to a sum of factorizable and nested two-loop integrals, as 
explained in \cite{Levi:2011eq}, and see appendix~A there. In addition to the 
irreducible two-loop tensor integrals, which were given in appendix~A of 
\cite{Levi:2011eq}, the two following reductions are also required here:
\begin{align}
\int_{\vec{k}_1\vec{k}_2} \frac{k_1^i}{k_1^2 \left(p-k_1\right)^2 
	k_2^2 \left(p-k_2\right)^2 \left(k_1-k_2\right)^2}&=\frac{p^i}{d-4} 
	\int_{\vec{k}_1\vec{k}_2}\left[\frac{1}{k_1^2 \left(p-k_1\right)^2 k_2^2 
		\left(p-k_2\right)^4}\right.\nn\\
&\left.\qquad\qquad\qquad-\frac{1}{k_1^2 \left(k_1-k_2\right)^2 k_2^2 
	\left(p-k_2\right)^4 }\right],\\
\int_{\vec{k}_1\vec{k}_2} \frac{k_1^i k_1^j}{k_1^2
	\left(p-k_1\right)^2 
	k_2^2 \left(p-k_2\right)^2 \left(k_1-k_2\right)^2}&=\frac{1}{d-4} 
\int_{\vec{k}_1\vec{k}_2}\left[\frac{p^i p^j -p^i k_1^j-p^j k_1^i +2k_1^ik_1^j}
     {k_1^2 \left(p-k_1\right)^2 k_2^2 \left(p-k_2\right)^4}\right.\nn\\
&\left.\qquad\qquad\quad	
-\frac{p^i p^j -p^i k_1^j -p^j k_1^i +2k_1^ik_1^j}{k_1^2 \left(k_1-k_2\right)^2 
	k_2^2 \left(p-k_2\right)^4}\right].
\end{align}

\bibliographystyle{jhep}
\bibliography{gwbibtex}

\end{document}